\documentclass[11pt]{article}
\usepackage[dvips]{graphicx}
\usepackage{float}
\usepackage{epsfig}
\usepackage{latexsym,amsmath,amsfonts,amssymb}
\usepackage{rotating}
\usepackage[american]{babel}
\usepackage[dvips]{graphicx}
\usepackage{bbm}
\usepackage{color}
\usepackage{slashed}
\usepackage[unicode]{hyperref}
\usepackage{lscape}
\usepackage{bigints}
\usepackage{enumerate}
\usepackage[shortlabels]{enumitem}
\usepackage{subcaption}
\usepackage{tikz}
\usetikzlibrary{decorations.pathreplacing}
\usetikzlibrary{shapes}
\usepackage{graphicx}
\usepackage{epsfig}
\usepackage{rotating}
\usepackage{amssymb}
\usepackage{dsfont}
\usepackage{psfrag}
\usepackage{amsmath,euscript,array,mathrsfs,amsfonts}
\usepackage{slashed}
\usepackage{array}
\usepackage{youngtab}
\usepackage{color}
\usepackage{bbold}
\usepackage{hyperref}
\usepackage[thinc]{esdiff}
\hypersetup{
colorlinks = true,
linkcolor = red,
linktocpage = true,
citecolor = blue
}

\usepackage{tikz}
\usetikzlibrary{calc}
 \usetikzlibrary{decorations.text}
 \usetikzlibrary{shapes}

 \usetikzlibrary{decorations.pathmorphing}
\usetikzlibrary{decorations.pathreplacing}
\usetikzlibrary{arrows.meta}
\tikzset{
  >={To[length=5pt]}
  }
\usetikzlibrary{shapes, shapes.geometric, shapes.symbols, shapes.arrows, shapes.multipart, shapes.callouts, shapes.misc}
\tikzset{snake it/.style={decorate, decoration=snake}}
\tikzset{7brane/.style={circle, draw=black, fill=black,ultra thick,inner sep=1.5 pt, minimum size=1 pt,}, c/.default={4pt}}
\tikzset{cross/.style={cross out, draw=black,thick, minimum size=2*(#1-\pgflinewidth), inner sep=0pt, outer sep=0pt}, cross/.default={5pt}}
\tikzset{big7brane/.style={circle, draw=black, fill=black,ultra thick,inner sep=2.5 pt, minimum size=1 pt,}, c/.default={4pt}}
\tikzset{u/.style={circle, draw=black, fill=white,inner sep=2 pt, minimum size=2 pt,},f/.style={square, draw=black, fill=white,ultra thick,inner sep=4 pt, minimum size=2 pt,}}
\tikzset{so/.style={circle, draw=black, fill=red,inner sep=2 pt, minimum size=2 pt,},f/.style={square, draw=black, fill=white,ultra thick,inner sep=4 pt, minimum size=2 pt,}}
\tikzset{sp/.style={circle, draw=black, fill=blue,inner sep=2 pt, minimum size=2 pt,},f/.style={square, draw=black, fill=white,ultra thick,inner sep=4 pt, minimum size=2 pt,}}
\tikzset{uf/.style={rectangle, draw=black, fill=white,inner sep=3 pt, minimum size=4 pt,}}
\tikzset{spf/.style={rectangle, draw=black, fill=blue, thick,inner sep=3 pt, minimum size=4 pt, circle, draw=black, fill=blue,thick,inner sep=2 pt, minimum size=2 pt,},f/.style={square, draw=black, fill=white,ultra thick,inner sep=4 pt, minimum size=2 pt,}}
\tikzset{sof/.style={rectangle, draw=black, fill=red, thick,inner sep=3 pt, minimum size=4 pt,}}
\usetikzlibrary{positioning}
\usetikzlibrary{arrows}
\usetikzlibrary{decorations.pathreplacing}
\usetikzlibrary{shapes}

\makeatletter\def\l@subsubsection#1#2{}%
\makeatother

\renewcommand\theequation{\arabic{section}.\arabic{equation}} 

\def\cA{{\cal A}}

\def\cF{{\cal F}}

\def\CC{\ensuremath{\mathds C}}
\def\RR{\ensuremath{\mathds R}}
\def\ZZ{\ensuremath{\mathds Z}}

\DeclareMathOperator{\vol}{vol}

\DeclareMathOperator{\sech}{sech}
\DeclareMathOperator{\tr}{tr}

\DeclareMathOperator{\csch}{csch}

\newcommand{\be}{\begin{equation}}
\newcommand{\ee}{\end{equation}}
\newcommand{\ba}{\begin{eqnarray}}
\newcommand{\ea}{\end{eqnarray}}

\def\im{Invent. Math.}

\def\hat{\widehat}
\def\a{\alpha}
\def\b{\beta}
\def\c{\gamma}
\def\d{\delta}
\def\f{\phi}               
\def\vf{\varphi}  
\def\tvf{\tilde{\varphi}}
\def\vp{\varphi}
\def\g{\gamma}
\def\h{\eta}
\def\j{\psi}
\def\k{\kappa}                    
\def\l{\lambda}
\def\m{\mu}
\def\n{\nu}
\def\o{\omega}  \def\w{\omega}
\def\p{\partial} 
\def\q{\theta}  \def\th{\theta}                  
\def\r{\rho}                                     
\def\s{\sigma}                                   
\def\t{\tau}
\def\u{\upsilon}
\def\x{\xi}
\def\z{\zeta}
\def\pt{\tilde{\varphi}}
\def\tt{\tilde{\theta}}
\def\lab{\label}
\def\wg{\wedge}
\def\bpsi{\bar{\psi}}
\def\bt{\bar{\theta}}
\def\bvf{\bar{\varphi}}
\def\W{\Omega}

\newcommand{\nb}{\nonumber}
\newcommand{\td}{\mathrm{d}}

\DeclareMathOperator{\str}{str}

\newcommand{\beq}{\begin{equation}}
\newcommand{\eeq}{\end{equation}}
\newcommand{\bea}{\begin{eqnarray}}
\newcommand{\eea}{\end{eqnarray}}

\newcommand{\beqs}{\begin{eqnarray}}
\newcommand{\eeqs}{\end{eqnarray}}
\newcommand{\bal}{\begin{aligned}}
\newcommand{\eal}{\end{aligned}}
\makeatletter
\newcommand\setItemnumber[1]{\setcounter{enum\romannumeral\@enumdepth}{\numexpr#1-1\relax}}
\makeatother
\begin{document}
\baselineskip=15.5pt
\pagestyle{plain}
\setcounter{page}{1}

\def\del{{\partial}}
\def\vev#1{\left\langle #1 \right\rangle}
\def\cn{{\cal N}}
\def\co{{\cal O}}


\def\IC{{\mathbb C}}
\def\IR{{\mathbb R}}
\def\IZ{{\mathbb Z}}
\def\RP{{\bf RP}}
\def\CP{{\bf CP}}
\def\Poincaré{{Poincar\'e }}
\def\tr{{\rm tr}}
\def\tp{{\tilde \Phi}}

\def\TL{\hfil$\displaystyle{##}$}
\def\TR{$\displaystyle{{}##}$\hfil}
\def\TC{\hfil$\displaystyle{##}$\hfil}
\def\TT{\hbox{##}}
\def\HLINE{\noalign{\vskip1\jot}\hline\noalign{\vskip1\jot}}
\def\seqalign#1#2{\vcenter{\openup1\jot
   \halign{\strut #1\cr #2 \cr}}}
\def\lbldef#1#2{\expandafter\gdef\csname #1\endcsname {#2}}
\def\eqn#1#2{\lbldef{#1}{(\ref{#1})}%
\begin{equation} #2 \label{#1} \end{equation}}
\def\eqalign#1{\vcenter{\openup1\jot
     \halign{\strut\span\TL & \span\TR\cr #1 \cr
    }}}

\def\eno#1{(\ref{#1})}
\def\href#1#2{#2}
\def\half{\frac{1}{2}}



\def\ads{{\it AdS}}
\def\adsp{{\it AdS}$_{p+2}$}
\def\cft{{\it CFT}}

\newcommand{\ber}{\begin{eqnarray}}
\newcommand{\eer}{\end{eqnarray}}

\newcommand{\beqar}{\begin{eqnarray}}
\newcommand{\cO}{{\cal O}}
\newcommand{\cT}{{\cal T}}
\newcommand{\cR}{{\cal R}}
\newcommand{\eeqar}{\end{eqnarray}}
\newcommand{\tht}{\thteta}
\newcommand{\lm}{\lambda}\newcommand{\Lm}{\Lambda}


\newcommand{\nonu}{\nonumber}
\newcommand{\oh}{\displaystyle{\frac{1}{2}}}
\newcommand{\dsl}
   {\kern.06em\hbox{\raise.15ex\hbox{$/$}\kern-.56em\hbox{$\partial$}}}
\newcommand{\as}{\not\!\! A}
\newcommand{\ps}{\not\! p}
\newcommand{\ks}{\not\! k}
\newcommand{\D}{{\cal{D}}}
\newcommand{\dv}{d^2x}
\newcommand{\Z}{{\cal Z}}
\newcommand{\N}{{\cal N}}
\newcommand{\Dsl}{\not\!\! D}
\newcommand{\Bsl}{\not\!\! B}
\newcommand{\Psl}{\not\!\! P}

\newcommand{\eeqarr}{\end{eqnarray}}


\def\del{{\delta^{\hbox{\sevenrm B}}}} \def\ex{{\hbox{\rm e}}}
\def\azb{A_{\bar z}} \def\az{A_z} \def\bzb{B_{\bar z}} \def\bz{B_z}
\def\czb{C_{\bar z}} \def\cz{C_z} \def\dzb{D_{\bar z}} \def\dz{D_z}
\def\im{{\hbox{\rm Im}}} \def\mod{{\hbox{\rm mod}}} \def\tr{{\hbox{\rm Tr}}}
\def\ch{{\hbox{\rm ch}}} \def\imp{{\hbox{\sevenrm Im}}}
\def\trp{{\hbox{\sevenrm Tr}}} \def\vol{{\hbox{\rm Vol}}}
\def\rl{\Lambda_{\hbox{\sevenrm R}}} \def\wl{\Lambda_{\hbox{\sevenrm W}}}
\def\fc{{\cal F}_{k+\cox}} \def\vev{vacuum expectation value}
\def\nodiv{\mid{\hbox{\hskip-7.8pt/}}}
\def\ie{{\em i.e.}}
\def\ie{\hbox{\it i.e.}}

\def\CC{{\mathchoice
{\rm C\mkern-8mu\vrule height1.45ex depth-.05ex
width.05em\mkern9mu\kern-.05em}
{\rm C\mkern-8mu\vrule height1.45ex depth-.05ex
width.05em\mkern9mu\kern-.05em}
{\rm C\mkern-8mu\vrule height1ex depth-.07ex
width.035em\mkern9mu\kern-.035em}
{\rm C\mkern-8mu\vrule height.65ex depth-.1ex
width.025em\mkern8mu\kern-.025em}}}

\def\RR{{\rm I\kern-1.6pt {\rm R}}}
\def\NN{{\rm I\!N}}
\def\ZZ{{\rm Z}\kern-3.8pt {\rm Z} \kern2pt}
\def\IB{\relax{\rm I\kern-.18em B}}
\def\ID{\relax{\rm I\kern-.18em D}}
\def\II{\relax{\rm I\kern-.18em I}}
\def\IP{\relax{\rm I\kern-.18em P}}
\newcommand{\CS}{{\scriptstyle {\rm CS}}}
\newcommand{\CSs}{{\scriptscriptstyle {\rm CS}}}
\newcommand{\rc}{\nonumber\\}
\newcommand{\bear}{\begin{eqnarray}}
\newcommand{\eear}{\end{eqnarray}}

\newcommand{\LL}{{\cal L}}

\def\mani{{\cal M}}
\def\calo{{\cal O}}
\def\calb{{\cal B}}
\def\calw{{\cal W}}
\def\calz{{\cal Z}}
\def\cald{{\cal D}}
\def\calc{{\cal C}}
\newcommand{\gt}{\tilde{g}}

\def\to{\rightarrow}
\def\ele{{\hbox{\sevenrm L}}}
\def\ere{{\hbox{\sevenrm R}}}
\def\zb{{\bar z}}
\def\wb{{\bar w}}
\def\nodiv{\mid{\hbox{\hskip-7.8pt/}}}
\def\menos{\hbox{\hskip-2.9pt}}
\def\dr{\dot R_}
\def\drr{\dot r_}
\def\ds{\dot s_}
\def\da{\dot A_}
\def\dga{\dot \gamma_}
\def\ga{\gamma_}
\def\dal{\dot\alpha_}
\def\al{\alpha_}
\def\cl{{closed}}
\def\cls{{closing}}
\def\vev{vacuum expectation value}
\def\tr{{\rm Tr}}
\def\to{\rightarrow}
\def\too{\longrightarrow}

\newcommand{\dd}{\mathrm{d}}

\def\a{\alpha}
\def\b{\beta}
\def\c{\gamma}
\def\d{\delta}
\def\e{\epsilon}           
\def\F{\Phi}
\def\f{\phi}               
\def\vf{\varphi}  \def\tvf{\tilde{\varphi}}
\def\vp{\varphi}
\def\g{\gamma}
\def\h{\eta}
\def\j{\psi}
\def\k{\kappa}                    
\def\l{\lambda}
\def\m{\mu}
\def\n{\nu}
\def\o{\omega}  \def\w{\wedge}
\def\q{\theta}  \def\th{\theta}                  
\def\r{\rho}                                     
\def\s{\sigma}                                   
\def\t{\tau}
\def\u{\upsilon}
\def\x{\xi}
\def\X{\Xi}
\def\z{\zeta}
\def\pt{\tilde{\varphi}}
\def\tt{\tilde{\theta}}
\def\lab{\label}
\def\6{\partial}
\def\wg{\wedge}
\def\atanh{{\rm arctanh}}
\def\bpsi{\bar{\psi}}
\def\bt{\bar{\theta}}
\def\bvf{\bar{\varphi}}

\def\ft#1#2{{\textstyle{{\scriptstyle #1}\over {\scriptstyle #2}}}}
\def\fft#1#2{{#1 \over #2}}
\def\del{\partial}
\def\sst#1{{\scriptscriptstyle #1}}

\def\dalemb#1#2{{\vbox{\hrule height .#2pt
        \hbox{\vrule width.#2pt height#1pt \kern#1pt
                \vrule width.#2pt}
        \hrule height.#2pt}}}
\def\square{\mathord{\dalemb{6.8}{7}\hbox{\hskip1pt}}}
\def\hF{\hat F}
\def\tA{\widetilde A}
\def\tcA{{\widetilde{\cal A}}}
\def\tcF{{\widetilde{\cal F}}}
\def\hA{\hat{\cal A}}
\def\cF{{\cal F}}
\def\cA{{\cal A}}
\def\wdg{{\sst \wedge}}

\def\0{{\sst{(0)}}}
\def\1{{\sst{(1)}}}
\def\2{{\sst{(2)}}}
\def\3{{\sst{(3)}}}
\def\4{{\sst{(4)}}}
\def\5{{\sst{(5)}}}
\def\6{{\sst{(6)}}}
\def\7{{\sst{(7)}}}
\def\8{{\sst{(8)}}}
\def\n{{\sst{(n)}}}
\def\tV{\widetilde V}
\def\tW{\widetilde W}
\def\tH{\widetilde H}
\def\tE{\widetilde E}
\def\tF{\widetilde F}
\def\tA{\widetilde A}
\def\tP{{\widetilde P}}
\def\tD{\widetilde D}
\def\bA{\bar{\cal A}}
\def\bF{\bar{\cal F}}
\def\tG{\widetilde G}
\def\tT{\widetilde T}
\def\Z{\rlap{\sf Z}\mkern3mu{\sf Z}}
\def\R{\rlap{\rm I}\mkern3mu{\rm R}}
\def\G{{\Gamma}}
\def\gg{\bf g}
\def\CS{{\cal S}}
\def\S{{\cal S}}
\def\P{{\cal P}}
\def\ep{\epsilon}
\def\td{\tilde}
\def\wtd{\widetilde}
\def\half{{\textstyle{1\over2}}}
\def\Qw{{Q_{\rm wave}}}
\def\Qnut{{Q_{\sst{\rm NUT}}}}
\def\mun{{\mu_{\sst{\rm NUT}}}}
\def\muw{{\mu_{\rm wave}}}
\let\a=\alpha \let\b=\beta \let\g=\gamma \let\d=\delta \let\e=\epsilon
\let\z=\zeta \let\h=\eta \let\q=\theta \let\i=\iota \let\k=\kappa
\let\l=\lambda \let\m=\mu \let\n=\nu \let\x=\xi 
\let\s=\sigma \let\t=\tau \let\u=\upsilon \let\f=\phi \let\c=\chi \let\y=\psi
\let\w=\omega  \let\D=\Delta \let\Q=\Theta \let\L=\Lambda
\let\X=\Xi  \let\U=\Upsilon \let\F=\Phi \let\Y=\Psi
\let\C=\Chi \let\W=\Omega     
\let\la=\label \let\ci=\cite \let\re=\ref
\let\se=\section \let\sse=\subsection \let\ssse=\subsubsection 
\def\bd{\begin{document}} \def\ed{\end{document}}
\def\ds{\documentstyle} \let\fr=\frac \let\bl=\bigl \let\br=\bigr
\let\Br=\Bigr \let\Bl=\Bigl 
\let\bm=\bibitem
\let\na=\nabla
\let\pa=\partial \let\ov=\overline 
\def\ba{\begin{eqnarray}}
\def\ea{\end{eqnarray}}
\def\ft#1#2{{\textstyle{{\scriptstyle #1}\over {\scriptstyle #2}}}}
\def\fft#1#2{{#1 \over #2}}
\def\del{\partial}
\def\sst#1{{\scriptscriptstyle #1}}
\def\oneone{\rlap 1\mkern4mu{\rm l}}
\def\ie{{\it i.e.\ }}
\def\via{{\it via}}
\def\semi{{\ltimes}}
\def\str{{\rm str}}
\def\jm{{\rm j}}
\def\im{{\rm i}}
\def\mapright#1{\smash{\mathop{-\!\!\!-\!\!\!-\!\!\!-\!\!\!-\!\!\!
             \longrightarrow}\limits^{#1}}}
\def\maprightt#1#2{\smash{\mathop{-\!\!\!-\!\!\!-\!\!\!-\!\!\!-\!\!\!
             \longrightarrow}\limits^{#1}_{#2}}}

\newcommand{\ho}[1]{$\, ^{#1}$}
\newcommand{\hoch}[1]{$\, ^{#1}$}
\newcommand{\ra}{\rightarrow}
\newcommand{\lra}{\longrightarrow}
\newcommand{\Lra}{\Leftrightarrow}
\newcommand{\bp}{\tilde \beta^\prime}
\newcommand{\Tr}{{\rm Tr} } 
\def\rme{{\rm e}}


\newfont{\namefont}{cmr10}
\newfont{\addfont}{cmti7 scaled 1440}
\newfont{\boldmathfont}{cmbx10}
\newfont{\headfontb}{cmbx10 scaled 1728}





\newcommand{\hyph}[1]{$#1$\nobreakdash-\hspace{0pt}}
\providecommand{\abs}[1]{\lvert#1\rvert}
\newcommand{\Nugual}[1]{$\mathcal{N}= #1 $}
\newcommand{\sub}[2]{#1_\text{#2}}
\newcommand{\partfrac}[2]{\frac{\partial #1}{\partial #2}}
\newcommand{\bsp}[1]{\begin{equation} \begin{split} #1 \end{split} \end{equation}}
\newcommand{\calF}{\mathcal{F}}
\newcommand{\calO}{\mathcal{O}}
\newcommand{\calM}{\mathcal{M}}
\newcommand{\calV}{\mathcal{V}}
\newcommand{\bbZ}{\mathbb{Z}}
\newcommand{\bbC}{\mathbb{C}}
\newcommand{\cK}{{\cal K}}

\newcommand{\Thq}{\Theta\left(\r-\r_q\right)}
\newcommand{\Dq}{\d\left(\r-\r_q\right)}
\newcommand{\kten}{\kappa^2_{\left(10\right)}}
\newcommand{\pbi}[1]{\imath^*\left(#1\right)}
\newcommand{\tth}{\tilde{\th}}
\newcommand{\tf}{\tilde{\f}}
\newcommand{\tj}{\tilde{\j}}
\newcommand{\tw}{\tilde{\omega}}
\newcommand{\tz}{\tilde{z}}
\newcommand{\prj}[2]{(\partial_r{#1})(\partial_{\j}{#2})-(\partial_r{#2})(\partial_{\j}{#1})}
\def\atanh{{\rm arctanh}}
\def\sech{{\rm sech}}
\def\csch{{\rm csch}}
\allowdisplaybreaks[1]

\def\red{\textcolor[rgb]{0.98,0.00,0.00}}

\newcommand{\Dan}[1] {{\textcolor{blue}{#1}}}

\numberwithin{equation}{section}



%

%
\setcounter{footnote}{0}
\renewcommand{\theequation}{{\rm\thesection.\arabic{equation}}}

\begin{titlepage}

\begin{center}

\vskip .5in 
\noindent

{\Large \bf{ Holography for Confined and Deformed Theories: TsT-Generated Solutions in type IIB Supergravity} }
\bigskip\medskip

Federico Castellani$^\dagger$\footnote{federico.castellani@unifi.it} and Carlos Nunez$^*$\footnote{c.nunez@swansea.ac.uk}  \\

\bigskip\medskip
{\small 
$^\dagger$INFN, Sezione di Firenze and
Dipartimento di Fisica e Astronomia, Universit\'a di Firenze, 

Via G. Sansone 1, I-50019 Sesto Fiorentino (Firenze), Italy.
\\
$^*$ Department of Physics, Swansea University, Swansea SA2 8PP, United Kingdom
}

\vskip .5cm 
\vskip .9cm 
     	{\bf Abstract }\vskip .1in
\end{center}

\noindent
In this paper, we present a series of  solutions of type IIB Supergravity, obtained through the TsT deformation of a seed background that is holographically dual to ${\cal N}=4$ Super Yang-Mills theory compactified on a circle (with a twist). We explore various holographic observables, focusing on disentangling the contributions of confinement, marginal deformations, and their interplay. Additionally, we provide a detailed analysis of how certain observables are influenced by the dynamics of the Kaluza-Klein (KK) modes arising from the circle compactification. Our results offer an improved understanding of the connections between the type IIB deformed backgrounds and the dual QFTs.

 \noindent
\vskip .5cm
\vskip .5cm
\vfill
\eject

\end{titlepage}

\setcounter{footnote}{0}

\small{
\tableofcontents}

\normalsize

\newpage
\renewcommand{\theequation}{{\rm\thesection.\arabic{equation}}}
%
\section{Introduction and general idea of this work}
The $AdS$/CFT conjecture and its refinements \cite{Maldacena:1997re, Gubser:1998bc, Witten:1998qj} have led to applications of gauge-string duality to the study of Renormalization Group (RG) flows in quantum field theories (QFTs). This duality have been applied to a wide variety of systems, describing the RG flows of maximally supersymmetric (SUSY) Yang-Mills theories in different dimensions \cite{Itzhaki:1998dd, Boonstra:1998mp, Girardello:1998pd, Girardello:1999bd}.

One  extension of the gauge-string duality is via wrapped brane constructions. The idea is to begin with a higher-dimensional brane system (typically D4, D5, or D6 branes) and wrap these branes on a shrinking $q$-cycle $\Sigma_q$, often while preserving some amount of supersymmetry to ensure the stability of the supergravity background. The resulting system is proposed to be dual to the RG flow between a higher-dimensional QFT and a lower-dimensional, often minimally supersymmetric, QFT in $(1+1)$, $(2+1)$, or $(3+1)$ dimensions. The dimensional reduction is characterised by the difference between the brane and wrapped cycle dimensionalities, $(p+1-q)$, the amount of SUSY preserved, etc. Numerous examples of such constructions have been developed (see early works such as \cite{Witten:1998zw, Maldacena:2000yy, Nieder:2000kc, Acharya:2000mu, Gauntlett:2000ng, Papadopoulos:2000gj, Edelstein:2001pu, Gomis:2001aa, Nunez:2001pt, Petrini:2001fk, Maldacena:2001pb, Schvellinger:2001ib, Gauntlett:2001ps, Gursoy:2002tx, Naka:2002jz, Edelstein:2002zy, Gomis:2001vg, Gomis:2001vk,Kruczenski:2003pv}). For comprehensive reviews, see \cite{Bigazzi:2002gyi, Bertolini:2003iv, Aharony:2002up}. These developments have yielded  connections between gauge-string duality and QFT observables, including the addition of dynamical quarks (fundamental fields of the gauge group), as seen in works such as \cite{Casero:2006pt, Aharony:2006da, Paredes:2006wb, Benini:2006hh, Benini:2007gx, Conde:2011sw, Casero:2007jj, Hoyos-Badajoz:2008znk, Bigazzi:2008qq, Bigazzi:2008zt, Bigazzi:2009bk, Gaillard:2010qg}. For more detailed reviews, see \cite{Erdmenger:2007cm, Nunez:2010sf}.

Despite the significant progress, two key challenges afflict wrapped brane systems:

\begin{itemize} \item{ \underline{UV Behavior}: the UV limit of these systems (usually $r \to \infty$ in the dual string description) often describes a higher-dimensional dynamics that is not a well-defined QFT and requires a string-theoretic UV completion. The non-$AdS$ asymptotics of these backgrounds complicate the application of holographic renormalization techniques \cite{Papadimitriou:2004ap, Kanitscheider:2008kd}, limiting the computational power of the dual descriptions.}
\item{ \underline{KK-Modes Interactions}: The wrapping of D$_p$ branes on a $\Sigma_q$ cycle corresponds to compactifying a maximally SUSY $(p+1)$-dimensional QFT on $\Sigma_q$. This introduces an infinite tower of massive Kaluza-Klein (KK) modes that interact with the massless particles (gauge fields, scalars, and fermions), whose dynamics are typically of primary interest. In the supergravity approximation, there is no hierarchy between the KK-mode masses and the QFT's strong coupling scale, making it challenging to disentangle the effects of these modes from the low-energy dynamics. A clean separation would require working in the string sigma model on the background (see \cite{Hori:2002cd} for an example).} 
\end{itemize}
In this paper, we address these two issues above. To tackle the first, we focus on a system of D3 branes wrapped on a supersymmetric (SUSY) circle. The presence of some preserved supersymmetry guarantees the stability of the background. Importantly, the asymptotic geometry is $AdS_5$. This offers a significant advantage over other wrapped brane systems by allowing the use of well-developed Holographic Renormalisation. The specific background we use was introduced by Anabal\'on and Ross \cite{Anabalon:2021tua}, with similar backgrounds discussed in \cite{Anabalon:2022aig, Anabalon:2023lnk, Nunez:2023xgl, Nunez:2023nnl, Fatemiabhari:2024aua, Anabalon:2024qhf, Anabalon:2024che, Giliberti:2024eii}. Some of these solutions have already been explored within the context of holography.

Regarding the second issue—the KK-mode interaction with the massless sector—we know that achieving a hierarchical separation between the strong coupling scale and the mass of the KK modes is only possible at finite $\alpha'$, which introduces numerous technical challenges. Instead, we adopt a more practical and modest approach aimed at identifying whether an observable computed via the string dual receives contributions from the KK modes. The paper \cite{Gursoy:2005cn} suggested a method to achieve this by comparing an observable calculated in the original (seed) background with the same observable in a “deformed” background. The deformed background is obtained by performing a TsT transformation (T-duality, shift of coordinates, and T-duality) on the seed geometry.

The TsT transformation was proposed by Lunin and Maldacena \cite{Lunin:2005jy} as a way of producing backgrounds dual to marginal deformations of SCFTs. The paper \cite{Gursoy:2005cn} used the fact that the KK-modes are typically charged under internal symmetries of the QFT, associated with non-R-isometries of the dual background.

If the calculation of the observable differs when performed in the seed and the deformed background, we argue that such observable receives contribution from the KK-modes.

In our case, we use the Anabal\'on-Ross background \cite{Anabalon:2021tua} as the seed solution, which can be understood as a vacuum expectation value (VEV) deformation of ${\cal N}=4$ $SU(N)$ SYM. In the IR, this theory flows to a ${\cal N}=2$ Chern-Simons theory with level $N$. A detailed exploration of the associated QFT has been carried out in \cite{Kumar:2024pcz} and \cite{Cassani:2021fyv}.

We perform two distinct TsT transformations to generate what we term the “$\beta$-deformed”\footnote{We refer to this case also as “beta-deformed” interchangeably.} and “dipole-deformed” backgrounds. These new backgrounds help disentangle the effects of KK modes arising from the twisted compactification of ${\cal N}=4$ SYM on a circle. The backgrounds are characterized by two parameters, $Q$ and $\gamma$. The parameter $Q$ is associated with the circle compactification, the confinement of the theory, and the appearance of a mass gap. The parameter $\gamma$, on the other hand, originates from the TsT transformation along certain isometries of the background. In the QFT, this is related to the introduction of a $\star$-product in the interactions, modifying how KK modes interact with the low-energy fields (gauge fields, fermions, and scalars) and among themselves.

Observables computed holographically may or may not depend on these parameters, although independence is rare and non-generic. Observables that do depend on both $Q$ and $\gamma$ can be interpreted as being affected by both confinement and marginal deformations. Please, notice that the first (confinement) is a low energy effect, whilst the marginal deformation occurs close to the UV fixed point.

This paper is long and detailed. We aim to present the material pedagogically, providing extensive explanations and calculations to benefit researchers working in this area. Below, we outline the contents of the paper.

\subsection{Outline of this paper}\label{QFTapproach}
The paper is divided in two main parts: Geometry is the first part, whilst the second is Quantum Field Theory and Observables.

Section \ref{section-geometry} (on the geometries) presents in detail the seed background written by Anabal\'on and Ross. After that, the new solutions obtained by the application of TsT are given. One of the solutions uses the $U(1)'s$ (the torus isometries) to be inside the $S^5$ (they are R-symmetries), we call this the $\beta$-deformed solution. The second solution uses a $U(1)$ inside the five sphere and the second is chosen to be that on which the UV-QFT is compactified. We call this “dipole background” following the notation of \cite{Bergman:2001rw}. 
{All backgrounds are presented with details, quantised charges are discussed and geometric invariants written. Details of the calculations are fully spelled in Appendix \ref{TsTdetails}. In Appendix \ref{sec-non-commutative}, we present the solution dual to the non-commutative version of the QFT and a last solution dual to a dipole-version of the non-commutative background. We do not study further  the last two new solutions.}

Section \ref{QFT-section} presents details about the QFT dual to the first two new backgrounds above mentioned. We start with a careful account of the perturbative analysis of the QFT, once it is compactified on $S^1$, with an R-symmetry twist. We discuss the $\star$-product in the beta-defomed
and in the dipole deformed QFTs. We clarify why any observable involving the KK-modes part of the dynamics is likely to be weighted by a factor of the parameter $\gamma$. We study Coulomb branches before and after the deformations and comment briefly about Chern-Simons terms from the holographic viewpoint. In this same section we calculate a number of observable quantities that could have not been obtained were not for the holographic dual solution. Among these observables, we discuss: Wilson, 't Hooft and  Polyakov loops, making the point that confinement of electric quarks works in our systems and the breaking of a
$\mathbb{Z}_N$ symmetry is at work.
The Entanglement Entropy is calculated and a discussion is made around the behaviour in the different backgrounds. The observables above
make a good case for the confining properties of the background, as we explain.

An interesting quantity, called the flow central charge is computed. This can be argued to be indicative of the number of degrees of freedom in our QFT. This quantity is energy-dependent and in our dual QFTs, that represent a flow across dimensions, interpolates between zero degrees of freedom at low energy (indicating a gapped system) and the degrees of freedom corresponding to the UV CFT. We analyse the dipole-deformed QFT with some care, for which the number of degrees of freedom diverges in the UV, suggesting a non-local, non-field theoretical description. Semiclassical strings are carefully studied, showing that some of these heavy operators in the QFT feel the effects of both confinement and marginal deformations.

Section \ref{concl} gives some brief conclusions and closing comments, proposing some possible future work.
\section{Geometry}\label{section-geometry}

Let us start this section briefly reviewing the relevant features of a set of smooth type II backgrounds presented in \textit{e.g.} \cite{Anabalon:2021tua, Kumar:2024pcz, Chatzis:2024top,Chatzis:2024kdu}. These are type IIB supergravity solutions constructed by a lifting procedure of the five-dimensional minimal gauged supergravity solution of \cite{Anabalon:2021tua} to different type II backgrounds of the form
\begin{eqnarray}
   \widehat{AdS_5}\times \widehat{M^{5}}\,,
\end{eqnarray}
the hats refer to deformations of both the $AdS_5$ and the five-dimensional internal manifold $M^5$.
More precisely, the deformation algorithm can be described as  follows. \\
Let us start by performing a compactification on $S^1$ circle of radius $R$ of one of the Poincaré directions of $AdS_5$, let us call it $\phi$. We deform further the $AdS_5$ part of the background by introducing a cigar-like geometry in the subspace spanned by the compact $\phi$ and the holographic radial direction $r$. In doing so,  we introduce a warping function $f(r)$ at whose zero the $S^1_\phi$ circle smoothly shrinks to zero size (in the case in which $f(r)$ displays multiple positive roots, the tip of the cigar is placed at the bigger root). Finally, introducing 
a finely-tuned fibration for the internal manifold $M^5$ over the $S^1_\phi$ circle direction, we can ask for the preservation of four supercharges for the type IIB supergravity solution. We refer to \cite{Anabalon:2021tua, Chatzis:2024top,Chatzis:2024kdu} for a more detailed analysis of this family of new backgrounds and generalizations to massive IIA and eleven dimensional supergravity solutions. \\
In this work, we focus mainly on the case of the $\widehat{AdS_5}\times \widehat{S^{5}}$ background. In \cite{Castellani:2024pmx} an extension to the $\widehat{AdS_5}\times \widehat{T^{1,1}}$ solution \cite{Chatzis:2024top,Chatzis:2024kdu}.\\
Following the steps of the algorithm sketched above, we can modify the $AdS_5 \times S^5$ solution to obtain a metric for the deformed background that reads \cite{Anabalon:2021tua}\footnote{Notice that in general the warping function is defined as $f(r) = 1-\frac{\m^2}{r^4}-\frac{Q^6}{r^6}$, but only the solution with $\m=0$ is SUSY preserving. We focus mostly on the $\mu=0$ case in what follows. We also set $g_s=\alpha'=1$.}
 \begin{eqnarray}
 \label{metric-ARxS5}
& & \mathrm{d}s^2 _{10}  = r^2 (-\mathrm{d}t^2+\mathrm{d}x_1^2 + \mathrm{d}x_2^2 +  f(r)\mathrm{d}\phi^2) + \frac{\mathrm{d}r^2}{ r^2 f(r)}+ \sum_{i=1}^3 \mathrm{d}\mu_i^2 + \mu_i^2 \left( \mathrm{d}\phi_i + \mathcal{A} \right) ^2\,,\nonumber\\
&&f(r) = 1-\frac{Q^6}{r^6}\,,\quad \mathcal{A} =  Q^3\left( \frac{1}{r^2}- \frac{1}{Q^2}\right)\mathrm{d}\phi\, .
\\
%
%
 & & F_5= ( 1+ {\star_{10} }) G_5, \qquad 
 G_5=  -4 r^3 \mathrm{d}t\wg \mathrm{d}x_1 \wg\mathrm{d}x_2\wg\mathrm{d}r\wg \mathrm{d}\phi -  2Q^3 J_2\wedge \mathrm{d}t\wedge \mathrm{d}x_1 \wedge \mathrm{d}x_2\,, 
 \nonumber\\
 & & J_2=  \sum_{i=1}^3 \mu_i \mathrm{d}\mu_i \wedge \left(\mathrm{d}\phi_i +\mathcal{A}\right)\,.\label{RR-S5}
\end{eqnarray}
For a lighter notation, we have fixed the $AdS$ radius $l= (4\pi g_s N)^{1/4} =1$, and we have defined
\begin{equation}
\label{mu_i}
\mu_1= \sin\theta \sin\varphi,~\mu_2= \sin\theta \cos\varphi, ~\mu_3= \cos\theta\,, \quad \sum_{i=1}^3 \m_i^2 = 1\,.
\end{equation}
The holographic radial direction $r$ ranges between its minimal value at $r =Q$, namely the zero of the warp factor $f(r)$, and the conformal boundary at $r\to \infty$. The compact coordinate $\phi$ is set to be periodic with period
\bea
\label{RQ}
\phi \sim \phi + 2\pi R\,, \quad R = \frac{2}{r^2f^\prime(r)}\bigg|_Q = \frac{1}{3Q}\,,
\eea
in order to avoid conical singularities at the tip of the cigar. The range of the other angles is,
\bea
\theta\in[0,\pi/2]\,,\quad \varphi\in[0,\pi/2]\,,\quad  \phi_1,\phi_2,\phi_3\in[0,2\pi]\,. 
\eea
Notice that in eq.(\ref{metric-ARxS5}) the five-sphere is written in such a way that all the three $U(1)$ isometric directions $\phi_i$ are uniformly fibered over the compactification $S^1_\phi$ circle, through the one form $\mathcal{A}$. Note also that eq.(\ref{RQ}) sets a precise relation between the compactification circle radius $R$ and the magnitude of the gauge field $Q$. 

The five-form in eq.(\ref{RR-S5}) is sourced by a stack of D3-branes. The quantization condition for the flux is (restoring momentarily the parameters $g_s, \alpha'$),
\bea
\label{QD3}
Q_{D3} =\frac{1}{(2\pi)^4g_s\a^{\prime\,2}}\int \star_{10} G_5 = \frac{l^4}{4\pi g_s\a^{\prime\,2}} = N\,,
\eea
%
We close here the presentation of the seed-background in eqs.(\ref{metric-ARxS5})-(\ref{RR-S5}). In what follows, we apply different TsT transformations, generating new backgrounds.

\subsection{The (marginal) beta-deformation}\label{sec_beta_deformation}
In this subsection we follow \cite{Lunin:2005jy} and perform a so-called $\beta$-transformation of the $\widehat{AdS_5}\times \widehat{S^5}$ background presented in eqs.(\ref{metric-ARxS5})-(\ref{RR-S5}). The  transformation, a solution generating technique,  can be schematically summarized by the following steps:
\begin{itemize}
    \item Identify an appropriate $U(1)\times U(1)$ symmetry inside the isometry group of a string theory background. In particular, taking two directions, say $\Theta_1$ and $\Theta_2$, the two $U(1)$ symmetries can be seen as acting as shifts on these two coordinates;
    \item Perform a T-duality transformation along the first circle, \textit{e.g.} $U(1)_{\Theta_1}$;
    \item Implement a shift in the second $U(1)_{\Theta_2}$ isometry direction, $\Theta_2\to \Theta_2 + \g \Theta_1$, with $\g$ a real number parametrizing the transformation;
    \item $T$-dualize back along $U(1)_{\Theta_1}$.
\end{itemize}
It is custom to refer to the the latter background generating procedure also as a TsT transformation.  In Appendix \ref{TsTdetails_LM}, a detailed computation of the original Lunin-Maldacena TsT-transformation in $AdS_5\times S^5$ is provided. \\

Let us consider the deformed $\widehat{AdS_5}\times \widehat{S^5}$ metric in eq. (\ref{metric-ARxS5}) and rewrite the internal $\widehat{S^5}$ as  follows
\ba
\label{ds2S5}
 \mathrm{d}s^2 _{S^5}
 &=&\sum_{i=1}^3 \mathrm{d}\mu_i^2 +\m_1^2 \left( D\psi -\mathrm{d}\varphi_1\right)^2+ \m_2^2 \left( D\psi +\mathrm{d}\varphi_1+\mathrm{d}\varphi_2\right) ^2+\m_3^2 \left( D\psi -\mathrm{d}\varphi_2\right)^2\,,
\ea
where we have introduced a new basis for the three $U(1)$ isometry directions given by
\ba
\label{anglephi}
\phi_1 = \psi -\varphi_1\,, \quad \phi_2 = \psi +\varphi_1+\varphi_2\,, \quad \phi_3 = \psi -\varphi_2\,, 
\ea
and subsequently defined
\ba
D\psi = \mathrm{d}\psi +\mathcal{A}\,.
\ea
Referring to \cite{Lunin:2005jy}, we  express eq.(\ref{ds2S5}) to easily recognize a two-dimensional torus subspace, realising the $U(1)\times U(1)$ isometry needed for the TsT transformation.  In particular, choosing the two $U(1)$ acting as shifts along the $\varphi_1$ and $\varphi_2$ directions, it is convenient to rewrite the five-sphere metric (\ref{ds2S5}) in the form
\ba
\label{ds2LM}
 \mathrm{d}s^2 _{S^5}
 &=&\sum_{i=1}^3 \mathrm{d}\mu_i^2 + 9\frac{\m_1^2\m_2^2\m_3^2}{g_0}D\psi^2+(\m_1^2+\m_2^2)\left( D\tilde\varphi_1 + \frac{\m_2^2}{\m_1^2+\m_2^2}D\tilde\varphi_2\right)^2+\frac{g_0}{\m_1^2+\m_2^2} D\tilde\varphi_2^2\,,\nonumber\\
\ea
where the following objects have been introduced
\ba
g_0 &=& \m_1^2\m_2^2+ \m_1^2\m_3^2+\m_2^2\m_3^2\,,\nonumber\\
 D\tilde\varphi_1 &=& \mathrm{d}\varphi_1 + \left(3\frac{\m_2^2\m_3^2}{g_0}-1\right)D\psi  =  D\varphi_1 +\left(3\frac{\m_2^2\m_3^2}{g_0}-1\right)\mathcal{A}\,,\nonumber\\
 D\tilde\varphi_2 &=&\mathrm{d}\varphi_2 + \left(3\frac{\m_2^2\m_1^2}{g_0}-1\right)D\psi  =  D\varphi_2 +\left(3\frac{\m_2^2\m_1^2}{g_0}-1\right)\mathcal{A}\,.
\ea
Following the prescription summarised above, we can perform the  transformation along $\varphi_1$ and $\varphi_2$ in eq.(\ref{ds2LM}), providing a new type IIB  supergravity solution, which reads
\bea
\label{sol1}
 \mathrm{d}s^2 _{10}  &=& r^2 (-\mathrm{d}t^2+\mathrm{d}x_1^2 + \mathrm{d}x_2^2 +  f(r)\mathrm{d}\phi^2) + \frac{\mathrm{d}r^2}{ r^2 f(r)} +  \mathrm{d}s^2_{S^5_\b}\,,\nonumber\\
  \mathrm{d}s^2_{S^5_\b}&=& \sum_{i=1}^3 \mathrm{d}\mu_i^2 + 9\frac{\m_1^2\m_2^2\m_3^2}{g_0}D\psi^2+G(\m_1^2+\m_2^2)\left( D\tilde\varphi_1 + \frac{\m_2^2}{\m_1^2+\m_2^2}D\tilde\varphi_2\right)^2 +\frac{g_0 G}{\m_1^2+\m_2^2} D\tilde\varphi_2^2\,,\nonumber\\
B&=& \g g_0 \,G \, D\tilde\varphi_1\wg D\tilde\varphi_2\,, \quad e^{2\Phi} = G\,,\nonumber\\
F_3 &=& \frac{\gamma}{G}\, i_{\varphi_2}i_{\varphi_1}\star_{10\beta}G_5\,,\quad F_5= (1 + \star_{10\beta}) G_5\,,\quad
F_7 =  G_5 \wedge B\,.
\eea
Here the $G$ factor has been defined as
\ba
G^{-1} &=& 1+ \g^2 g_0\,,
\ea
and the ${\star_{10\beta}}$ symbol represents the Hodge-dual respect with the deformed metric in eq.(\ref{sol1}). Moreover, notice that the $G_5$ form is still given as in eq.(\ref{RR-S5}). 
Since the $AdS$ part of the background is not involved in the TsT transformation, then the asymptotic behavior of the solution is unchanged respect with the $\widehat{AdS_5}\times \widehat{S^5}$ solution (\ref{metric-ARxS5}). We refer to Appendix \ref{TsTdetails} for more details on the derivation of (\ref{sol1}). \\
We can verify that the $\beta$-deformation does not modify the quantization flux condition (\ref{QD3}), once we involve the Page five-form flux \cite{Benini:2007gx},\cite{Hamilton:2016ito}, namely \footnote{Here we are back to set $g_s= \alpha^\prime=1$.}
\begin{multline}
\label{QD3prime}
Q_{D3}^\prime =
\frac{1}{(2\pi)^4}\int_{\Sigma_5}  \star_{10\beta} G_5 + B\wg F_3  =\frac{1}{(2\pi)^4}\int_{\Sigma_5} \star_{10\beta} G_5 \left(1+\g^2 g_0\right)= \frac{1}{(2\pi)^4}\int_{\Sigma_5}  \star_{10} G_5 =  N\,,
\end{multline}
where we have taken the cycle $\Sigma_5$ as spanned by the deformed five sphere directions.\footnote{The D3-brane Maxwell charge is simply given by the integral of $\star_{10\beta} G_5$ over the cycle $\Sigma_5$. In this case, the $G$ factor in the five-form is not canceled out by the $B\wg F_3$ term. The charge is not trivial and not quantised. It does not acquire an $r$-dependence,  so it does not change between UV and IR.}\\
Moreover, we can verify if the $F_3$ flux is supported by the presence of  $D5$-branes, generated in the deformed background. To do that, let us evaluate the following charge
\bea
\label{QD5prime}
Q_{D5}^\prime =
\frac{1}{(2\pi)^2}\int_{\Sigma_3}  F_3 =\frac{1}{(2\pi)^2}\frac{\g}{G}\int_{\Sigma_3} i_{\varphi_2}i_{\varphi_1} \star_{10\beta} G_5 =\frac{\g}{(2\pi)^4}\int_{\Sigma_5}  \star_{10} G_5 = \g N\,,
\eea
with $\Sigma_3 = \left[\theta,\varphi,\psi\right]$. 
We observe that in the special case in which
\bea
\g = \frac{m}{n}\,,  \quad N = n\kappa\,,\quad m\,,n\,,\kappa \in \mathbb{N}\,,
\eea
we have that eq.(\ref{QD5prime}) provides a charge quantization condition for a new stack of $N_{D5} =m\kappa$ D5-branes. In what follows, and more precisely, in discussing the Coulomb branches of the dual field theory in Section \ref{sec:Coulomb}, we come back to this interesting feature.
On the other hand, we can check that the present transformation does not introduce additional D1-branes sourcing the $F_3$ and $F_7$ fluxes, for any $\g$ values. Indeed, we have 
\bea
\label{QD1prime}
Q_{D1}^\prime =
\frac{1}{(2\pi)^6}\int_{\Sigma_7}  \star_{10\beta} F_3 + B\wg G_5=\frac{1}{(2\pi)^6}\int_{\Sigma_7}   B\wg G_5 - F_7 = 0\,,
\eea
for every seven cycle $\Sigma_7$ supporting the relevant fluxes involved above.

Notice that, expressing back the deformed solution (\ref{sol1}) in terms of the angles $\phi_i$ basis in (\ref{anglephi}) it is possible to check that (\ref{sol1}) is in agreement with the ones found in \cite{Lunin:2005jy} and \cite{Liu:2019cea}.\footnote{In \cite{Liu:2019cea} authors consider a consistent truncation of the TsT transformed $AdS_5\times S^5$ containing a graviphoton gauge field $A$. In the IIB supergravity background the latter is identified as the gauge field associated the $\psi$ shift isometry. Then the ten-dimensional metric is obtained by the $AdS_5\times S^5$ by replacing $\mathrm{d}\phi_i \to \mathrm{d}\phi_i + A$.}

For the present $\beta$-transformed background in eq.(\ref{sol1}), we have been able to verify on Mathematica the Bianchi identity for the $B$ field as well as the Maxwell equations for $F_3$, $F_5$.
On the other hand, the dilaton and Einstein equations are computationally too demanding. We expect them to work well.

\subsection{Dipole deformation}\label{sec:dipole}
In this subsection, we perform an alternative TsT transformation on the $\widehat{AdS_5}\times \widehat{S^5}$ background of eq.(\ref{metric-ARxS5}). We choose the $U(1)\times U(1)$ isometries to be an internal $S^5$ angle and the compact $\phi$ direction. 
Since the SUSY spinor depends on the $\phi$-direction, the T-duality along it breaks SUSY. We call this a dipole-deformation, following the notation in \cite{Bergman:2001rw}.

In order to generate the new background, it is convenient to write the metric of the five-sphere $\widehat{S^5}$ in eq.(\ref{metric-ARxS5}) as a (deformed) $U(1)$ fibration over $\mathbb{C}P^2$.
 With this parametrisation, the $\widehat{AdS_5}\times \widehat{S^5}$ metric reads
\bea
\label{S5CP2}
& &\mathrm{d}s^2_{10}=\mathrm{d}s^2_{8}+ r^2f(r)\mathrm{d}\phi^2+ (\mathcal{D}\varphi+\mathcal{A} )^2\,,\nonumber\\
&& \mathrm{d}s^2_{8} = r^2 (-\mathrm{d}t^2+\mathrm{d}x_1^2 + \mathrm{d}x_2^2 ) + \frac{\mathrm{d}r^2}{ r^2 f(r)} + \mathrm{d}s^2_{\mathbb{C}P^2} \,,\quad  \mathcal{D}\varphi = \mathrm{d}\varphi +\eta\,,
\eea
where the $\mathrm{d}s^2_{\mathbb{C}P^2}$ is the Fubini-Study metric  \cite{Pope:1980ub,Herrero:2011bk}
\bea
\label{CP2metric}
\mathrm{d}s^2_{\mathbb{C}P^2} = \mathrm{d}\a^2 + \frac{1}{4} \sin^2\a \left[ \cos^2\a (\mathrm{d}\psi-\cos\theta_1\mathrm{d}\theta_2) +\mathrm{d}\theta_1^2 +\sin^2\theta_1\mathrm{d}\theta_2^2\right]\,,
\label{metric_S5_as_CP2}
\eea
along with the K\"ahler potential 
\bea
\eta = \frac12 \sin^2\a(\mathrm{d}\psi-\cos\theta_1\mathrm{d}\theta_2)\,.
\label{oneformcp2}
\eea
Here the one-form $\mathcal{A}$ is still given as in eq.(\ref{metric-ARxS5}). Using the parametrisation of eq.(\ref{metric_S5_as_CP2}), the forms in eq.(\ref{RR-S5}) can be recast as,
\bea
\label{RR-S5_CP2}
 && F_5= ( 1+ {\star_{10} }) G_5\,, \nonumber\\
 && G_5=  -4 r^3 \mathrm{d}t\wg \mathrm{d}x_1 \wg\mathrm{d}x_2\wg\mathrm{d}r\wg \mathrm{d}\phi -  2Q^3 J_2\wedge \mathrm{d}t\wedge \mathrm{d}x_1 \wedge \mathrm{d}x_2\,, \quad J_2=  \frac 12 \mathrm{d}\eta\,.
\eea
We proceed with the Lunin-Maldacena TsT deformation along the directions $\Theta_1 = \varphi$ and $\Theta_2 = \phi$, obtaining the  new type IIB supergravity solution,
\bea
\label{S5dip}
&&\mathrm{d}s^2_{10}=\mathrm{d}s^2_{8} + G r^2f(r)\mathrm{d}\phi^2+  G\left(\mathcal{D}\varphi +\mathcal{A}\right)^2\,,\nonumber\\
&&B =\gamma r^2 f(r) \,G\,\mathcal{D}\varphi \wg \mathrm{d}\phi\,, \qquad  \quad e^{2\Phi} = G\,,\nonumber\\
&&F_5  = \left(1 + \star_{10\beta}\right)G_5\,,\quad F_3 =  \frac{\g}{G} \,\,i_{\phi}i_{\varphi}\star_{10\beta}G_5\,,\quad F_7 = G_5 \wg B\,. 
\eea
We have defined 
\ba
\label{GDipole}
&& G^{-1} = 1 + \g^2r^2f(r)\,.
\ea
As above, the $\star_{10\beta}$ symbol refers to the Hodge-dual operator of the $\beta$-deformed metric in eq.(\ref{S5dip}) and the five-form $G_5$ is still written in terms of the original solution as in eq.(\ref{RR-S5_CP2}). \\
The 2-form potential $C_2$ and the associated $F_3$ can be written as,
\bea
\label{C2}
&& C_2 = -\gamma \frac{Q^3}{r^2}J_2\,,\quad F_3 = \mathrm{d}C_2 = 2\g\frac{Q^3}{r^3}\mathrm{d}r\wg J_2\,.
\eea
The quantization of Page fluxes  follows closely the observations in Subsection \ref{sec_beta_deformation}. Indeed, using the expressions for the fluxes in eq.(\ref{S5dip}), we have \footnote{Notice that  $B\wg F_3 $ term does not contribute to eq.(\ref{QD3prime_dipole}), the Page charge equals the Maxwell charge.}
\bea
\label{QD3prime_dipole}
Q_{D3}^\prime =\frac{1}{(2\pi)^4}\int_{\Sigma_5}  \star_{10\beta} G_5 + B\wg F_3  = \frac{1}{(2\pi)^4}\int  \star_{10} G_5 =  N\,,
\eea
with $\Sigma_5 = \left[\mathbb{C}P^2,\varphi\right]$. For D1 branes we find, 
\bea
\label{QD1prime_dipole}
Q_{D1}^\prime =\frac{1}{(2\pi)^6}\int_{\Sigma_7}  \star_{10\beta} F_3 + B\wg G_5 =\frac{1}{(2\pi)^6}\int_{\Sigma_7}   B\wg G_5 - F_7 = 0\,.
\eea
%
We do not impose a charge quantization condition for D5-branes, since F$_3$ has no magnetic components on a compact cycle, we conclude that in the present dipole deformation, we do not have D5-branes in the background (also for rational $\g$).\\
Finally, let us analyze the NS5-brane charge by considering a three cycle 
\bea
\label{NS5_dipole}
Q_{\text{NS}5}^\prime =\frac{1}{(2\pi)^2}\int_{M_3}  H_3\,. 
\eea
Since $H_3$ has 'legs' only on non-compact cycles, we do not impose the quantizations condition.

For the background in eqs.(\ref{S5dip})-(\ref{GDipole}), we checked with Mathematica the Bianchi identities for $H_3$, $F_3$  and $F_5$, as well as the Maxwell, dilaton and Einstein equations. We also verified  that the seven and three forms are Hodge dual to each other.\\
Let us present the results for  geometric invariants.  
The Ricci scalar for the metric in eq. (\ref{S5dip}) reads
\bea
R =\frac{2 \gamma ^2 \left(Q^{12} \left(26 \gamma ^2 r^2+28\right)+Q^6 \left(2 r^6-34 \gamma ^2 r^8\right)+r^{12} \left(8 \gamma ^2 r^2+15\right)\right)}{\left(-\gamma ^2 Q^6 r+\gamma ^2 r^7+r^5\right)^2}\,,
\eea
which in the IR ($r\to Q$) and UV ($r\to \infty$) displays the following asymptotic behaviour, 
\bea
\label{Rr0}
R\vert_{r=Q} = 90 Q^2 \g^2\,, \quad R\vert_{r\to \infty} \sim 16-\frac{2}{\gamma ^2 r^2}+O\left(r^{-3}\right)\,.
\eea
This suggests the absence of singularities for the background at the two ends of the geometry ($r=Q$), ($r\to\infty$), and also along the radial coordinate. Let us present the asymptotic form of other geometric invariants,
\bea
&& R_{\m\n}R^{\m\n}\vert_{r=Q} = 36 \left(81 \gamma ^4 Q^4-2 \gamma ^2 Q^2+6\right)\,,\quad R_{\m\n}R^{\m\n}\vert_{r\to \infty} \sim 176-\frac{96}{\gamma ^2 r^2}+O\left(r^{-3}\right)\,,\nonumber\\
&& R_{\m\n\r\s}R^{\m\n\r\s}\vert_{r=Q} = 36 \left(99 \gamma ^4 Q^4+40 \gamma ^2 Q^2+14\right)\,,\quad  R_{\m\n\r\s}R^{\m\n\r\s}\vert_{r\to \infty}\sim 232-\frac{248}{\gamma ^2 r^2}+O\left(r^{-3}\right)\,,\nonumber\\
\eea
%
The above invariants in Einstein frame instead have asymptotics
\begin{eqnarray}
& & R\vert_{r\to \infty}\sim \frac{5}{2\sqrt{\g r}} +O\left(r^{-\frac{5}{2}}\right)\,,\\
& & R_{\mu\nu}R^{\mu\nu}\vert_{r\to \infty}\sim \frac{821}{4\g r} +O\left(r^{-3}\right)\,,\\
& & R_{\mu\nu\rho\sigma}R^{\mu\nu\rho\sigma}\vert_{r\to \infty}\sim \frac{4057}{16\g r} +O\left(r^{-3}\right)\,.\label{ivarinatseinstein}
\end{eqnarray}
Also in this frame we have that the curvature scalars, being in particular vanishing, are regular in the UV limit.\\
Let us conclude by commenting on the UV asymptotic behavior of the background metric in eq.(\ref{S5dip}). Notice from eq.(\ref{GDipole}) that in the large $r$ limit, the $G$ factor scales as
\bea
G\big|_{r\to \infty} \sim \frac{1}{\g^2r^2}\,,\label{xxy}
\eea
from which, we have that the spacetime described by the metric in eq.(\ref{S5dip}) is no longer asymptotically $AdS_5$.


\section{Field theory and observables}\label{QFT-section}
{In this section we discuss details about the perturbative QFTs and study holographic observables for the strongly coupled QFTs dual to the backgrounds in Section \ref{section-geometry}. }

\subsection{Comments on the dual QFTs}
We start giving some  details of the quantum field theories dual to the TsT deformed type IIB backgrounds presented in Section \ref{section-geometry}. We display the main features of the marginal and dipole deformed theories, including their perturbative Lagrangians, the compactification  and SUSY preserving mechanisms. In Subsection \ref{sec:Coulomb}, we comment about the expected Coulomb branches of the  theories we analyse.

\subsubsection{Twisted circle compactification of \texorpdfstring{$\mathcal{N}=4$}{N=4} SYM}
\label{sec:twisted_N_4}
Let us begin by displaying some details on the 3D $\mathcal{N}=2$ SUSY theory derived from a twisted circle compactification of $\mathcal{N}=4$ $SU(N)$ SYM in four dimension. In doing this perturbative analysis, we refer mainly to \cite{Kumar:2024pcz}. Our starting point is the Lagrangian for the $\mathcal{N}=4$ SYM theory on $\mathbb{R}^{1,3}$ (in the $\mathcal{N}=1$ formalism), that using the conventions in \cite{Kovacs:1999fx} is given by \footnote{Here we are using the normalization conventions of $\Tr(T^aT^b)=\frac{1}{2}\delta^{ab}$, and $[T^a,T^b]=if^{abc}T^c$, with $f^{abc}$ structure constant of $SU(N)$. The indices $M\,,N$ run over the $\mathbb{R}^{1,3}$ directions, while after the compactification along the $\phi$ circle, the Greek indices run over the $\mathbb{R}^{1,2}$ directions}
\ba\label{SYMlagr}
{\mathcal L}  &=&  \Tr\bigg\{-\frac{1}{2}
F_{MN}^{2}-2\overline{D_{M }Z^i }D^{M }Z^i
+ g^{2}[\bar{Z}^i, Z^i]^{2}+  2g^2[\bar Z^j,\bar Z^k][Z^k,Z^j] + \nb\\
&&+i\bar{\lambda}\slashed{D}\lambda +i\bar{\psi}^i\slashed{D}\psi^i+2\sqrt{2} g\left(\bar{\lambda}\left[Z^i,
\psi^i\right]- \left[\bar{\psi}^i,\bar{Z}^i\right]\lambda\right)-\sqrt{2}\e_{ijk}\left(\left[\bar{\psi}^i,
             Z^j\right]\psi^k- \left[\bar{\psi}^i,
             \bar{Z}^j\right]\psi^k\right) \bigg\}\,.   \nb\\
\ea
In the Lagrangian of eq. (\ref{SYMlagr}), all the fields are in the adjoint representation of  $SU(N)$,  $\Phi = \Phi^a T^a$. Moreover notice that the gauge field $A_\m$ and the Weyl fermion $\lambda$ are the propagating fields of the $\mathcal{N}=1$ vector multiplet, while the complex scalars $Z_i$ and the Weyl fermions $\psi_i$ correspond to the three $\mathcal{N}=1$ chiral multiplets of $\mathcal{N}=4$. The covariant derivative is defined as
\ba
D_M = \p_M -ig\left[A_M,\cdot\right]\,.
\ea

We  analyze the case in which the Lagrangian in eq. (\ref{SYMlagr}) is  reduced on a $S^1$ cycle of radius $R$ in the $\phi$ direction, imposing periodic and anti-periodic boundary conditions for bosons and fermions, respectively. The Fourier decomposition of the fields content of the $\mathcal{N}=1$ vector and chiral multiplets in  eq. (\ref{SYMlagr}) reads \footnote{The reality condition for $A_\m$ requires the identifications $A^{(-n)}_\m(x) = A^{(n)\,\dagger}_\m(x)$. Analogously for $\Theta^{(n)}(x)$.}
\ba
&&A_\m(x,\phi) = \sum_{n\in \mathbb{Z}} e^{i \frac{n\phi}{R}}A^{(n)}_\m(x)\,,\nb\\
&&A_\phi(x,\phi) \equiv \Theta(x,\phi) = \sum_{n\in \mathbb{Z}} e^{i \frac{n\phi}{R}}\Theta^{(n)}(x)\,,\nb\\
&&Z^j(x,\phi) = \sum_{n\in \mathbb{Z}} e^{i \frac{n\phi}{R}}Z^{j\, (n)}(x)\,,~~j:1,2,3.\nb\\
&&\chi(x,\phi) = \sum_{n\in \mathbb{Z}} e^{i \frac{(n+1/2)}{R}\phi}\chi^{ (n)}(x)\,,\quad \chi = \l\,,\psi^j\,,~~j:1,2,3.
\ea
In the light of this decomposition, we can analyze some of terms in  the Lagrangian of eq. (\ref{SYMlagr}). We start looking at the gaugino kinetic term
\ba
i\bar{\lambda} \slashed{D}\lambda &=& i\bar\l(x,\phi) \left[\g^\m D_\m \l(x,\phi)+ \g^\phi \p_\phi\l(x,\phi) -i\g^\phi g\left[\Theta(x,\phi), \l(x,\phi)\right]\right]\nb\\
&=& \sum_{n,m,p \in \mathbb{Z}} e^{i \frac{(m+1/2)}{R}\phi}e^{-i \frac{(n+1/2)}{R}\phi} e^{i \frac{p}{R}\phi} i\bar\l^{(n)}(x)\bigg\{\d_{p,0}\g^\m\p_\m \l^{(m)}(x)+\nb\\
&&-i\g^\m g\left[A_\m^{(p)}(x), \l^{(m)}(x)\right]+i\frac{(n+1/2)}{R} \d_{p,0}\g^\phi \l^{(m)}(x)-i\g^\phi g\left[\Theta^{(p)}(x), \l^{(m)}(x)\right]\bigg\}\,,\nb\\
\ea
which after the integration over the $S^1_\phi$ cycle reduces to 
\ba
\frac{1}{2\pi R}\int_{S^1_\phi}\mathrm{d}\phi\,\,\Tr\left[i\bar{\lambda} \slashed{D}\lambda\right]
&=&  \Tr\bigg[\sum_{n,m \in \mathbb{Z}} \bar\l^{(n)}(x)\bigg\{\d_{m,0}i\g^\m\p_\m \l^{(n)}(x) +\g^\m g\left[A_\m^{(n-m)}(x), \l^{(m)}(x)\right]+\nb\\&&-\frac{(n+1/2)}{R}\d_{m,0}\g^\phi\l^{(n)}(x)+\g^\phi g\left[\Theta^{(n-m)}(x), \l^{(m)}(x)\right]\bigg\}\bigg]\,.
\ea
Let us focus on the bi-linear terms for a certain $\l^{(n)}(x)$ mode, namely
\ba
 \bar\l^{(n)}(x)\left[i\g^\m\p_\m - \frac{(n+1/2)}{R}\g^\phi\right]\l^{(n)}(x)\,,
\ea
from which, using the fact that $\{\g^\m, \g^\phi\} =0$, we can recognize a Kaluza-Klein mass term for the gaugino mode,
\ba
\label{massl}
m_{n,\l} = \frac{|n+1/2|}{R} \neq 0\,,\quad \forall n \in \mathbb{Z}\,.
\ea
We  find the same results for the Weyl fermions $\psi^j$. We  focus on the gauge field (integrated over $S^1_\phi$) kinetic term that reads
\ba
\label{KKA}
 && \frac{1}{2\pi R}\int_{S^1_\phi} d\phi \frac{1}{2} \Tr\left[F_{MN}^2\right]=\frac{1}{2\pi R}\int_{S^1_\phi}\mathrm{d}\phi\,\, \frac{1}{2}\Tr\left[ F_{\m\n}^2\right] +\sum_{n \in \mathbb{Z}}\Tr\left[\p_\m \Theta^{(-n)}\p^\m\Theta^{(n)} - \frac{n^2}{R^2} A_\m^{(-n)}A^{(n)\,\m}\right]+\nb\\
 && -ig\sum_{n,m \in \mathbb{Z}} \Tr\left[\left(\p_\m \Theta^{(-n-m)}+ i\frac{n+m}{R}A_\m^{(-n-m)}\right)\left[A^{(n)\,\m}, \Theta^{(m)}\right]\right]+\nb\\
 && - \frac{g^2}{2} \sum_{n,m,p,s \in \mathbb{Z}}\d_{n+m+p+s,0} \Tr \left[\left[A^{(n)\,\m}, \Theta^{(m)}\right]\left[A^{(p)}_\m ,\Theta^{(s)}\right]\right]\,,
\ea
where, for a lighter expression we have left implicit the $\mathbb{R}^{1,2}$ gauge field strength terms. Moreover from eq. (\ref{KKA}) we can recognize a mass term for the KK modes $A^{(n)}_\m(x)$, \textit{i.e.}
\ba
\label{massA}
m_{n, A} =\frac{|n|}{R}\,.\label{diegoleo}
\ea
The mass of each mode of the scalar field $\Theta$ is the same as the one in eq.(\ref{diegoleo}), namely $ m_{n,\Theta}=\frac{|n|}{R}$. This can be seen by decomposing in modes the equations of motion.
\\
Finally, we look at  the reduced scalar kinetic terms 
\ba
&&\int d\phi \Tr (D_MZ^j)^2=\frac{1}{2\pi R}\int_{S^1_\phi}\mathrm{d}\phi\,\Tr \left[\overline{D_{\mu }Z^i }D^{\mu }Z^i\right] +\sum_{n \in \mathbb{Z}}  \frac{n^2}{R^2}\Tr\left[ \bar Z^{i\,(n)}Z^{i\,(n)}\right] +\nb\\
&&-g \sum_{n,m \in \mathbb{Z}} \frac{n}{R}\Tr\left[\bar Z^{i\,(n+m)}\left[\Theta^{(m)}\,,Z^{i\,(n)}\right]+\left[\bar Z^{i\,(n+m)}\,,\Theta^{(m)}\right]Z^{i\,(n)}\right]+\nb\\
&&+g^2\sum_{n,m,p,s \in \mathbb{Z}}\d_{n,m+p+s} \Tr\left[\left[\bar Z^{i\,(n)}\,,\Theta^{(m)}\right]\,\left[\Theta^{(p)}\,,Z^{i\,(s)}\right]\right]\,,
\ea
that provide a mass term for the modes of the scalar fields, given by
\ba
\label{massZ}
m_{n, Z^i} =\frac{|n|}{R}\,.
\ea
 We consider the theory to be coupled to an external gauge field ${\cal A}_{\mu,\phi}$ for the diagonal combination (Cartan) of the  $R$-symmetry $U(1)^3$, such that the covariant derivative components are modified as follows 
 \begin{equation}
 \label{N4DC}
        D_{\mu}\Phi \rightarrow D_{\mu}\Phi\,, \quad D_{\phi}\rightarrow D_\phi\Phi - i R_{\Phi}\, \mathcal{A}_{\phi}\Phi,\quad\quad
        \mathcal{A}_\phi = Q\,\,.
    \end{equation}
We have considered the covariant derivative acting on a general field $\Phi$, and defined $R_\Phi$ as the charge of the $\Phi$ field under the diagonal $U(1)^3$ part of the $R$-symmetry. More precisely, it is possible to define the normalized charge of the complex scalar as
    \begin{equation}
        R_{Z^i} = 1\,,\label{rscalar}    
    \end{equation}
while for the fermions $\lambda$ and $\psi^i$
    \begin{equation}
        R_\l = \frac{3}{2} \,,\quad R_{\psi^i} = -\frac{1}{2}\,,\label{rfermion}
    \end{equation}
    Note that the R-charges in eqs.(\ref{rscalar})-(\ref{rfermion}) are scaled by a factor of $\frac{3}{2}$ of the canonical R-charge. 
Also, remind that the charges for $\bar{Z}_i\,,\bar{\lambda}\,,\bar \psi^i$ have opposite signs.
The rules for the covariant derivatives in eq. (\ref{N4DC}), along with the charge values in eqs. (\ref{rscalar}) and (\ref{rfermion}) yield a shift in the field masses in eqs. (\ref{massl}), (\ref{massA}) and (\ref{massZ}), as 
\ba
\label{massN4}
&&m_{n,Z^i}\rightarrow \left| \frac{n}{R} - Q\right|\,,\quad m_{n,A}\rightarrow \left| \frac{n}{R}\right|\, 
,\nb\\
&& m_{n,\psi^i}\rightarrow \left| \frac{n+1/2}{R} + \frac{Q}{2}\right| \,,\quad m_{n,\l}\rightarrow \left| \frac{n+1/2}{R} - \frac{3Q}{2}\right|\,.
\ea
Thus, by fixing the value of the background field to be 
\ba
Q = \frac{1}{3R}\,,
\ea
we can allow the 3D theory gaugino $\l$ to have massless zero modes. Indeed in this case the spectrum in eq. (\ref{massN4}) reduces
\ba
\label{Bspecturm}
&&m_{n,Z^i}\rightarrow \frac{|n-1/3|}{R}\,,\quad m_{n,A}\rightarrow  \frac{|n|}{R}\,,
\nb\\
&& m_{n,\psi^i}\rightarrow \frac{|n+2/3|}{R} \,,\quad m_{n,\l}\rightarrow\frac{|n|}{R}\,.
\ea
Let us observe how the spectra for the complex scalar $Z^i$ and their Weyl fermion partner $\psi^i$ massive modes are settle such that 
\bea
m_{n+1,Z^i} = m_{n,\psi^i}\,.
\eea
This feature thus ensures the perturbative spectrum of the 3D theory to be supersymmetric.
\subsubsection{Marginal beta deformation of twisted \texorpdfstring{$\mathcal{N}=4$}{N=4} on \texorpdfstring{$\mathbb{R}^{1,2}\times S^1$}{R1,2xS1}}
\label{sec:marginal_N_4}
Here, referring mainly to the Lunin-Maldacena approach in \cite{Lunin:2005jy}, we display the analysis of marginal beta-deformation of the twisted $S^1_\phi$ circle compactification of 4D $\mathcal{N}=4$ theory presented in the previous Section \ref{sec:twisted_N_4}.\\
Let us start by taking into account the $SU(N)$ $\mathcal{N}=4$ theory, whose Lagrangian is given in eq. (\ref{SYMlagr}). Then, considering a $U(1)\times U(1)$ global symmetry of the latter, a class of deformed theories can be introduced. Indeed, identifying $\hat{Q}^1$ and $\hat{Q}^2$ with the generators of the two global $U(1)$ and describing their action on a general field $\Phi$ of the theory as
\bea
\label{hat_Q}
\hat{Q}^i \, \Phi = Q^i_{\Phi} \Phi\,, \quad i = 1,2\,,
\eea
we deform the Lagrangian in eq. (\ref{SYMlagr}) by defining an alternative field product (denoted as $\star$) given by \cite{Lunin:2005jy}
\bea
\label{beta_star_product}
\Phi_A \Phi_B \longrightarrow \Phi_A \star \Phi_B \equiv e^{i\pi\g\left(Q^1_{\Phi_A}Q^2_{\Phi_B}-Q^2_{\Phi_A}Q^1_{\Phi_B}\right)}\Phi_A \Phi_B\,, \quad \g \in \mathbb{R}\,.
\eea
In the case in which the $U(1)\times U(1)$ is taken inside the R-symmetry of the original theory, the introduction of the product in eq. (\ref{beta_star_product}) defines a marginal deformation of 4D $\mathcal{N}=4$: the so-called \textit{beta}-deformation \cite{Lunin:2005jy,Leigh:1995ep}. This is the case of interest in this subsection.\\
The family of deformations following from the prescription in eq. (\ref{beta_star_product}) are conjectured to correspond in the dual gravity theories to TsT-transformations \cite{Lunin:2005jy}. In particular, in this picture, the two global $U(1)$ symmetries in eq. (\ref{beta_star_product}) are related with the two $U(1)$ metric isometries, involved in performing the TsT-transformation of the supergravity solution.  

Let us now focus on the marginal beta-deformation of the 4D $\mathcal{N}=4$ theory and its twisted circle compactification. The beta-deformation modifies the Lagrangian in eq.(\ref{SYMlagr}) by the introduction of the new product in eq. (\ref{beta_star_product}) for the terms involving fields that are charged under the  $U(1)\times U(1)$ R-symmetry. Then, it is useful to look closely at the action of the $U(1)$ generators on the fields as in eq. (\ref{hat_Q}), namely
\bea
\label{Q12_Phi}
&&\hat{Q}^{1,2}\, A_M = 0\,, \quad \hat{Q}^{1,2} \, Z^i = -\hat{Q}^{1,2} \, \bar{Z}^i = Q^{1,2}_{Z^i} Z^i\,,\nb\\
&& \hat{Q}^{1,2} \, \l = -\hat{Q}^{1,2} \, \bar{\l} = Q^{1,2}_{\l} \l\,, \quad \quad \hat{Q}^{1,2} \, \psi^i = -\hat{Q}^{1,2} \, \bar{\psi}^i = Q^{1,2}_{\psi^i} \psi^i\,.
\eea
Notice that since the gauge field is uncharged under the R-symmetry, thus considering a generic field $\Phi$ and eq. (\ref{beta_star_product}), we observe that
\bea
\label{AM_star_product}
\Phi A_M \longrightarrow \Phi \star A_M =e^{i\pi\g\left(Q^1_{\Phi}Q^2_{A}- Q^2_{\Phi}Q^1_{A}\right)}\Phi A_M = \Phi A_M   \,.
\eea
This implies that the terms in the Lagragian in eq. (\ref{SYMlagr}) involving either $A_\m$ or $\Theta$ are unchanged by the introduction of the modified product. We  conclude that the spectrum in eq. (\ref{Bspecturm}) of the 3D $\mathcal{N}=2$ SUSY theory is invariant under the present marginal deformation. By the same reasoning, the interaction terms involving two R-symmetry charged fields in the Lagrangian in eq. (\ref{SYMlagr}) are typically affected by the beta deformation. In particular, the terms that will actually feel the effect of the deformation, are the ones that involve fields belonging to different chiral multiplets. This feature can be interpreted as follows:
the two $U(1)$ global symmetries used to define the product in eq. (\ref{beta_star_product}) can be seen as acting on the three $\mathcal{N}=1$ chiral multiplets of $\mathcal{N}=4$ as \cite{Lunin:2005jy}
\bea
\label{beta_star_product_chiral}
&&\left(\Phi_1,\Phi_2,\Phi_3\right) \longrightarrow \left(\Phi_1,e^{i\varphi_A}\Phi_2,e^{-i\varphi_A}\Phi_3\right)\,,\nb\\
&&\left(\Phi_1,\Phi_2,\Phi_3\right) \longrightarrow \left(e^{-i\varphi_B}\Phi_1,e^{i\varphi_B}\Phi_2,\Phi_3\right)\,.
\eea
From these transformation rules, we  observe that fields belonging to the same chiral multiplets are clearly equally charged under $\hat{Q}^{1,2}$, namely
\bea
\label{QZ_Qpsi}
Q^{1,2}_{Z^i} = Q^{1,2}_{\psi^i}\,.
\eea
Using eq. (\ref{QZ_Qpsi}) we  notice that
\bea
\label{lambda_Z_psi}
\bar{\lambda} \left[\bar{Z}^i, \psi \right] \longrightarrow \bar{\lambda}\star \left[\bar{Z}^i, \psi^i \right]_\star = \bar{\lambda}\star \left[\bar{Z}^i, \psi^i \right]= \bar{\lambda} \left[\bar{Z}^i, \psi^i \right]\,.
\eea
Since all the terms in the Lagrangian in eq. (\ref{SYMlagr}) that involve the vector multiplet fields $(A_M, \lambda)$ are invariant under the introduction of the product in eq. (\ref{beta_star_product}), and given that after the compactification the zero modes belong only to the spectrum of the latter field (see eq. (\ref{Bspecturm})), we conclude that if an observable (operator) of the beta-deformed theory depends on the $\gamma$-parameter of eq. (\ref{beta_star_product}), then the operator (observable) in 'made-out' by the Kaluza-Klein modes.\\ Hence, the beta-deformation give us a useful tool in discerning between observables of the twisted compactified $\mathcal{N}=4$ theory that depend on the IR theory only or also in its UV completion.  In this direction, we provide the analysis of some relevant observables in Subsection \ref{sec:observables}.

\subsubsection{Dual Dipole QFT}
Starting  from the Lagrangian in eq. (\ref{SYMlagr}), we  analyze the dual QFT of the dipole TsT transformation in Subsection \ref{sec:dipole}. Before doing that, let us briefly review the symmetry properties of the solution. As the $D3$-branes in this case wrap the $S^1_\phi$ cycle, then the Lorentz symmetry along the branes decompose to $SO(1,2)\times U(1)_\phi$. Moreover, in the transverse space is present a $SU(4)_R$ symmetry, in which we can identify the diagonal $U(1)$, related to the $\varphi$ isometry in the dual supergravity metric in eq. (\ref{S5dip}), \textit{i.e.} the circle isometry respect to which the dipole TsT transformation is performed. Here, we will refer to the latter as $U(1)_\varphi$.\\
After the compactification on the $\phi$ direction and the introduction of the external gauge field $\mathcal{A}_\phi$, we can label the fields in the Lagrangian of eq. (\ref{SYMlagr}) by their quantum numbers under the two $U(1)_\phi$ and $U(1)_\varphi$ symmetries, as in eq. (\ref{Q12_Phi}). In particular we will use the following notation 
\ba
\hat{R}\,\Phi = R_\Phi \Phi\,,\quad \hat{L}\,\Phi = L_\Phi \Phi\,,
\ea
where we have assigned each field in the theory $\Phi$ a dipole length $L_\Phi$, according to its charge under the $U(1)_\phi$ isometry, while $R_\Phi$ represent its charge under the R-symmetry diagonal $U(1)$. We can further define the (unit) dipole operator as \cite{Gursoy:2005cn,Dasgupta:2001zu,Guica:2017mtd}
\be
\label{DipoleFTL}
\hat L\equiv-i \, \p_\phi\,.
\ee 
Then, following the prescription for the deformations in eq. (\ref{beta_star_product}), we can replace the field products in the $\mathcal{N}=4$ Lagrangian in eq. (\ref{SYMlagr}) with the  modified one
\ba
\label{starprod}
 \Phi _A \Phi _B\rightarrow\Phi _A \star \Phi _B \equiv e\,^{i\pi \g \left( R_{\Phi_A} L_{\Phi_B}- R_{\Phi_B}L_{\Phi_A}\right)}\Phi _A\Phi _B\,.
\ea
Notice that the operators $\hat R$ and $\hat L$ have mass dimensions zero and one, respectively, as it should be clear from eq. (\ref{DipoleFTL}). Hence, modifying the field products via eq. (\ref{starprod}) is equivalent to introducing an irrelevant deformation in the field theory.
Then, using eqs. (\ref{rscalar}) and (\ref{rfermion}) and recalling that the gauge field is invariant under R-symmetry, we can start analyzing the bosonic part of the (not-compactified) Lagrangian in eq. (\ref{SYMlagr}) after the dipole-deformation, that reads
\ba\label{DipoleLag}
&& \mathcal{L}_B=  \Tr\left[-\frac{1}{2}
F_{MN}\star F^{MN}-2\overline{D_{M }Z^i }\star D^{M }Z^i
+ g^{2}[\bar{Z}^i, Z^i]_\star^{2}+  2g^2[\bar Z^j,\bar Z^k]_\star \star [Z^k,Z^j]_\star\right]\,.\nb\\
\ea
Here, the covariant derivative acting on the R-charged field (as the scalar $Z^i$) is defined with the modified product in eq. (\ref{beta_star_product}) \cite{Gursoy:2005cn,Guica:2017mtd,Meier:2023lku}
\ba
\label{DipoleLagr}
 &&D_\m \Phi =\partial _\m \Phi-i g\left(A_\m \star \Phi -\Phi \star A_\m\right)\,,\nb\\
 &&D_\phi \Phi =\partial _\phi \Phi-i g\left(\Theta \star \Phi -\Phi \star \Theta\right) -i R_{\Phi}\, \mathcal{A}_{\phi}\Phi \,.
\ea
Since the gauge field $A_M$ (and so the 3D gauge field $A_\m$ and scalar $\Theta$) is uncharged under R-symmetry, the first term in eq. (\ref{DipoleLag}) is invariant under this deformation and after the compactification takes again the form of eq. (\ref{KKA}). This leaves untouched the spectrum for the KK modes from the  3D gauge field $A_\m$ and $\Theta$ in eq. (\ref{Bspecturm}).

The scalars  part of the Lagrangian ($Z^i$), is modified by the deformation. This follows from the fact that these fields display non-vanishing $R_\Phi$ charges as in eqs. (\ref{rscalar})-(\ref{rfermion}). Then, it is useful to consider again the (integrated over the $S^1_\phi$ cycle) kinetic term for the scalars $Z^i$, but this time with the dipole deformed field product  in eqs.(\ref{DipoleLagr}):
\bea
\label{DmZDmZ}
\frac{1}{2\pi R}\int_{S^1_\phi}\mathrm{d}\phi\, \Tr \overline{D_{M }Z^i } \star D^{M }Z^i = \frac{1}{2\pi R}\int_{S^1_\phi}\mathrm{d}\phi\,  \Tr \left[\overline{D_{\m }Z^i }\star D^{\m }Z^i + \overline{D_{\phi }Z^i }\star D^{\phi }Z^i\right]\,.
\eea
Let us focus on the first term in the above equation, namely
\begin{multline}
\label{DZDZ}
 \frac{1}{2\pi R}\int_{S^1_\phi}\mathrm{d}\phi\,  \Tr \overline{D_{\m }Z^i }\star D^{\m }Z^i =  \frac{1}{2\pi R}\int_{S^1_\phi}\mathrm{d}\phi\,  \Tr \bigg[\p_\m \bar Z^i \star \p^\m Z^i-i g\,\p_\m \bar Z^i \star\left[A^\m, Z^i\right]_{\star}+\\-i g\left[A_\m, \bar Z^i\right]_{\star}\star \p^\m Z^i - g^2\left[A_\m, \bar Z^i\right]_{\star}\star\left[A^\m, Z^i\right]_{\star}\,.
\end{multline}
It is helpful to notice that 
\ba
\label{rule1}
\bar Z^i \star Z^i  = e\,^{i\pi\g \left( R_{\bar Z^i} L_{Z^i}- R_{Z^i} L_{\bar Z^i}\right)} \sum_{n,m \in \mathbb{Z}} e^{-i\frac{n}{R}\phi} e^{i\frac{m}{R}\phi} \bar Z^{i\, (n)} Z^{i\, (m)}\nb\\
=  \sum_{n,m \in \mathbb{Z}} e\,^{i\pi\g\frac{n-m}{R}}e^{-i\frac{n}{R}\phi} e^{i\frac{m}{R}\phi} \bar Z^{i\, (n)} Z^{i\, (m)}\,,
\ea
and 
\ba
\label{rule2}
A_\m \star Z^i  = e\,^{i\pi\g \left( R_{A} L_{Z^i}- R_{Z^i} L_{A}\right)} \sum_{n,m \in \mathbb{Z}} e^{i\frac{n}{R}\phi} e^{i\frac{m}{R}\phi} A_\m^{(n)} Z^{i\, (m)}\nb\\
=  \sum_{n,m \in \mathbb{Z}} e\,^{-i\pi\g\frac{n}{R}}e^{i\frac{n}{R}\phi} e^{i\frac{m}{R}\phi} A_\m^{(n)} Z^{i\, (m)}\,.
\ea
Moreover we have that 
\ba
\label{rule3}
&&\left[A_\m, Z^i\right]_{\star} =  \sum_{n,m \in \mathbb{Z}} e^{i\frac{n+m}{R}\phi} \left[e\,^{-i\pi\g\frac{n}{R}}A_\m^{(n)} Z^{i\,(m)}- e\,^{i\pi\g\frac{n}{R}}Z^{i\, (m)}A_\m^{(n)}\right]\,,\nb\\
&&\left[A_\m, \bar Z^i\right]_{\star} =  \sum_{n,m \in \mathbb{Z}} e^{i\frac{n-m}{R}\phi} \left[e\,^{i\pi\g\frac{n}{R}}A_\m^{(n)} \bar Z^{i\,(m)}- e\,^{-i\pi\g\frac{n}{R}}\bar Z^{i\, (m)}A_\m^{(n)}\right]\,,
\ea
from which eq. (\ref{DZDZ}) reduces to 
\begin{multline}
\label{DmZ}
\frac{1}{2\pi R}\int_{S^1_\phi}\mathrm{d}\phi\, \Tr \p_\m \bar Z^i \p^\m Z^i -ig \sum_{n,m \in \mathbb{Z}}\Tr\bigg\{\p^\m\bar Z^{i \, (n+m)} \left[e\,^{-i\pi\g\frac{n}{R}}A_\m^{(n)} Z^{i\,(m)}- e\,^{i\pi\g\frac{n}{R}}Z^{i\, (m)}A_\m^{(n)}\right]+\\
+ \left[e\,^{i\pi\g\frac{n}{R}}A_\m^{(n)} \bar Z^{i\,(n+m)}- e\,^{-i\pi\g\frac{n}{R}}\bar Z^{i\, (n+m)}A_\m^{(n)}\right]\p^\m Z^{i \, (m)}\bigg\}+\\
-g^2\sum_{n,m,s \in \mathbb{Z}}\left[e\,^{i\pi\g\frac{n}{R}}A_\m^{(n)} \bar Z^{i\,(n+m+s)}- e\,^{-i\pi\g\frac{n}{R}}\bar Z^{i\, (n+m+s)}A_\m^{(n)}\right]\times\\\times\left[e\,^{-i\pi\g\frac{m}{R}}A^{(m)\,\m} Z^{i\,(s)}- e\,^{i\pi\g\frac{m}{R}}Z^{i\, (s)}A^{(m)\,\m}\right]\bigg]\,.
\end{multline}
Let us now analyze the second term in eq (\ref{DmZDmZ}), namely
\ba
\label{DphiZ}
 \frac{1}{2\pi R}\int_{S^1_\phi}\mathrm{d}\phi\,  \Tr \,\overline{D_{\phi }Z^i }\star D^{\phi }Z^i\,,
\ea
then using similar rules as in eqs. (\ref{rule1}),(\ref{rule2}), and (\ref{rule3}) (notice that $\Theta$ has the same quantum number under $\hat R$ and $\hat L$ as $A_\m$), we can write eq. (\ref{DphiZ}) as 
\begin{multline}
\label{DphiZDphiZ}
\Tr\bigg[\sum_{n\in \mathbb{Z}} \left(\frac{n}{R}-Q\right)^2\, \bar Z^{i\,(n)}Z^{i\,(n)}+\\-g \sum_{n,m \in \mathbb{Z}}\bigg\{\left(\frac{(n+m)}{R}-Q\right)\bar Z^{i \, (n+m)} \left[e\,^{-i\pi\g\frac{n}{R}}\Theta^{(n)} Z^{i\,(m)}- e\,^{i\pi\g\frac{n}{R}}Z^{i\, (m)}\Theta^{(n)}\right]+\\-\left(\frac{m}{R}-Q\right) \left[e\,^{i\pi\g\frac{n}{R}}\Theta^{(n)} \bar Z^{i\,(n+m)}- e\,^{-i\pi\g\frac{n}{R}}\bar Z^{i\, (n+m)}\Theta^{(n)}\right]Z^{i \, (m)}\bigg\}+\\
-g^2\sum_{n,m,s \in \mathbb{Z}}\left[e\,^{i\pi\g\frac{n}{R}}\Theta^{(n)} \bar Z^{i\,(n+m+s)}- e\,^{-i\pi\g\frac{n}{R}}\bar Z^{i\, (n+m+s)}\Theta^{(n)}\right]\left[e\,^{-i\pi\g\frac{m}{R}}\Theta^{(m)} Z^{i\,(s)}- e\,^{i\pi\g\frac{m}{R}}Z^{i\, (s)}\Theta^{(m)}\right]\bigg]\,.
\end{multline}
Hence, from eqs.(\ref{DmZ}), (\ref{DphiZDphiZ}), and the fact that the YM kinetic term is invariant under the deformation in eq. (\ref{starprod}), we have that the spectrum of the bosonic sector of the effective 3D theory is not modified in response of the dipole deformation. Indeed,  both $A_\m$ and $\Theta$ are uncharged under $U(1)_\varphi$.  On the other hand, the interactions between the various massless and KK fields are  affected. A similar analysis can be performed regarding the fermion sector of the theory, leading to the same conclusion: in the dipole case the  spectrum in eq. (\ref{beta_star_product}) is untouched after the deformation.\\
In analogy with the observation in eq.(\ref{lambda_Z_psi}), we can analyze the following (integrated) interaction term 
\begin{multline}
\label{lambdaZpsi}
 \frac{1}{2\pi R}\int_{S^1_\phi}\mathrm{d}\phi\,  \Tr \,    \bar \lambda \star \left[\bar Z^i\,,\psi^i\right]_\star\,=\\
 =\Tr\bigg[ \sum_{n,m,s \in \mathbb{Z}}\delta_{m,n+s}  e^{i\pi\g \frac{s(R_\psi-R_Z) -(m-n)R_\lambda}{R}} \bar\lambda^{(s)} \left[e^{i\pi\g\frac{n R_Z- m R_\psi }{R}}\bar Z^{i\, (n)}\psi^{i\,(m)}-e^{-i\pi\g\frac{n R_Z- m R_\psi }{R}}\psi^{i\,(m)}\bar Z^{i\, (n)}\right]\,,
\end{multline}
which this time depends in a non-trivial way on the transformation parameter $\g$. In other words, the interactions between the KK-modes do change under the dipole deformation. Note, as we anticipated, SUSY is broken by this transformation. A T-duality along the R-symmetry direction is performed.\\
This closes the presentation of the Lagrangian (perturbative) study of the QFTs dual to the backgrounds in eqs.(\ref{sol1}) and (\ref{S5dip}).  Let us now discuss the space of vacua of these theories.


\subsubsection{Coulomb branches}
\label{sec:Coulomb}
In this section we study the Coulomb branches that appear (or not) for the marginal beta deformed
twisted $\mathcal{N}=4$ $SU(N)$ theory on $\mathbb{R}^{1,2}\times S^1$ presented in Subsection \ref{sec:marginal_N_4}. We perform this analysis using both the point of view of  the field theory and that of the TsT-transformed dual supergravity background in Subsection \ref{sec_beta_deformation}.

\paragraph{Field theory point of view}\,\\
In Subsection \ref{sec:marginal_N_4} we have seen that a TsT-transformation along two isometries of the internal $S^5$ of $AdS_5\times S^5$ is conjectured to correspond to a marginal deformation in the dual $\mathcal{N}=4$ SYM theory. More precisely the deformation in eq. (\ref{beta_star_product}) (or analogously in eq. (\ref{beta_star_product_chiral})) turns out to be a change of the superpotential (in the $\mathcal{N}=1$ formalism) of the form 
\ba
\label{betadefW}
\Tr\left(\Phi_i[\Phi_j,\Phi_k]\right) \longrightarrow \Tr\left(\Phi_i\left[e^{i\pi\g}\Phi_j \Phi_k- e^{-i\pi\g} \Phi_k \Phi_j\right]\right)\,.
\ea
Here $\Phi_i$ ($i=1,2,3$) are the three chiral multiplets in the adjoint representation of the gauge group, the scalar components of which are $Z_i$. This feature suggests that the Coulomb branch configurations are affected by the beta deformation as well.
For early analysis of the Coulomb branch in the marginal beta deformed 4D $\mathcal{N}=4$ SYM theory we refer to \textit{e.g} \cite{Benini:2004nn,Dorey:2004xm}.\\
Before the deformation, the F and D terms constrains read
\begin{equation} 
[Z_{i},Z_{j}]=0\,,\quad \sum_{i=1}^{3}\,[Z_{i},Z_{i}^{\dagger}]=0\,,
\label{FDcons}
\end{equation}
and the vacuum equations are solved by taking the three chirals as diagonal matrices,
\begin{equation}
\langle Z_{i} \rangle =  {\rm diag}\left[ x^{(i)}_{1},
x^{(i)}_{2},\ldots,x^{(i)}_{N}\right]\,,
\label{arb}
\end{equation}
where, clearly the eigenvalues 
$x^{(i)}_{a}$ (here $a=1,2,\ldots,N$) are unconstrained. With this choice, the 
$SU(N)$  gauge symmetry of $\mathcal{N}=4$ SYM is spontaneously broken down to  
$U(1)^{N-1}$. This is the classical Coulomb branch of the theory.
On the other hand, turning on the deformation (\ref{betadefW}), the constraints in eq. (\ref{FDcons}) modify to
\begin{equation} 
[Z_{i},Z_{j}]_{\star}=0\,,\quad \sum_{i=1}^{3}\,[Z_{i},Z_{i}^{\dagger}]_{\star}=0\,,
\label{FDcons2}
\end{equation}
the commutator $\left[\cdot,\cdot\right]_{\star}$ is defined in eq. (\ref{beta_star_product}). Hence, the ansatz in eq. (\ref{arb}) is not a good solution anymore. A simple generalization of the latter is the following \cite{Dorey:2004xm}. Let us consider the three chirals as in eq. (\ref{arb}), but this time with the eigenvalues $x^{(i)}_{a}$ no longer taken as unconstrained: for each entrance “$a$” only one chiral can displays a non-zero $x^{(i)}_{a}$, while the others must have vanishing elements. This ensures that the F and D conditions in eq. (\ref{FDcons2}) are satisfied, leading to a Coulomb branch configuration for a generic value of the deformation parameter $\g$.\\
Moreover, following \cite{Lunin:2005jy,Berenstein:2000hy}, we notice that for specific values of  $\g$, namely
\ba
\label{rationalg}
\g = \frac{m}{n}\,, \quad m\,,n\,\in \mathbb{N}\,,\,\,\text{coprime}\,,
\ea
the F and D constraints in eq. (\ref{FDcons2}) admit another solution, which reads
\ba
\label{NCCoulomb}
Z_1 = \chi_1 U^m\,,\quad Z_2 = \chi_2 V\,,\quad  Z_3 = \chi_3 V^{-1}U^{-m}\,,
\ea
where here $\chi_i$ are complex numbers and the matrices $U$ and $V$ satisfying the algebra
\ba
\label{NCtorusAlgebra}
\left[e^{i\pi/n} UV -e^{-i\pi/n} VU\right] =0\,.
\ea
In the case in which $N=nk$, with $k \in \mathbb{N}$, $U$ and $V$ have a simple representation as $SU(N)$  matrices, \textit{i.e.}
\ba
&&U =  {\rm diag}\left[ 1,e^{-\frac{2\pi i}{n}},
\ldots,e^{-\frac{2\pi i(n-1)}{n}}\right]\otimes \mathbb{1}_k\,,\quad V = \hat V\otimes \mathbb{1}_k\,,
\label{UV}
\ea
where $\hat V$ is a $n\times n$ matrix with non-zero entrances given by 
\ba
\hat V_{i+1,i} = \hat V_{n,1} =1\,.
\ea
As a simple example, let us consider the case $k =1$, $N=3$: in this case 
\ba
U={\rm diag}\left[ 1,e^{-\frac{2\pi i}{3}},e^{-\frac{4\pi i}{3}}\right]\,,\quad V = \begin{pmatrix}
0&1&0\\0&0&1\\1&0&0
\end{pmatrix}\,.
\ea
Here, it easy to verify that $UV = e^{2\pi i/3}VU$ and so that the algebra in eq. (\ref{NCtorusAlgebra}) is satisfied.\\

In light of these results, let us move a step forward through the analysis of a possible presence of Coulomb branches in the beta deformed theory of Subsection \ref{sec:marginal_N_4}. Notice that after the 
compactification procedure along the $\phi$ direction the spectrum of the effective $3D$ theory is given by  eq. (\ref{Bspecturm}), were the KK modes associated to the chiral multiplets $Z_i$, acquire real masses.
Then, we expect the $\mathcal{N} = 4$ Coulomb branch to be lifted due to the latter masses for the chirals. \\
In the $3D$ $\mathcal{N}=2$ effective theory, a massless scalar is present, namely the zero mode of the “$\phi$” component of the vector multiplet gauge field, $\Theta^{(0)}$. In analogy to what is observed in \cite{Kumar:2024pcz}, $\Theta^{(0)}$ can acquire a VEV of the form
\ba
\label{theta_VEV}
\langle \Theta^{(0)}\rangle = {\rm diag}\left[x_1^{(0)},\ldots,x_N^{(0)}\right]\,,\quad x_i^{(0)} \in \mathbb{C}\,,
\ea
with $x_i^{(0)} \sim x_i^{(0)} + 1/R$.
The configuration in eq. (\ref{theta_VEV}) along with the choice for the (massive) complex scalars KK modes
\ba
\langle Z_i^{(n)}\rangle  = 0\,, \quad i= 1,2,3\,,
\ea
yield a breaking of gauge group from $SU(N)$ to $U(1)^{N-1}$, leading to a (classic) Coulomb branch for the $3D$ effective theory.
At the quantum level, however, it can be show that $\mathcal{N}=2$ vector multiplet zero modes gain masses via loop corrections. Hence, the Coulomb branch is completely lifted, as expected \cite{Kumar:2024pcz}.
\paragraph{Holographic point of view}\,\\
In the previous subsection, we have observed how it is expected to find a lifted Coulomb branch in the field theory dual to the beta-transformed $\widehat{AdS_5}\times \widehat{S^5}$ background of Subsection \ref{sec_beta_deformation}. Here, we explore this from the holographic point of view. In fact, the presence (or absence) of the special Coulomb branch appearing in the Lunin-Maldacena background \cite{Lunin:2005jy} for rational $\gamma$ as in eq.(\ref{rationalg}). In the original supergravity solution \cite{Lunin:2005jy},\footnote{Notice that the Lunin-Maldacena background can be recovered from the one in eq.(\ref{sol1}) by simply fixing $Q\to0$.} this Coulomb branch is probed by the use of a D5-brane extending along the Poincare slice of AdS and wrapping the TsT two torus. The DBI action for the D5-brane thus reads
\ba
S_{DBI} = T_{D5}\int_{\mathbb{R}^{1,3}\times T^2} \mathrm{d}^{5+1}\xi\, e^{-\Phi}\sqrt{-\det(P[g]_{ab} +F_{ab}+P[B]_{ab})}\,.
\ea
We have switched on a gauge field with field strength $F$ on the brane, and $P[\cdot]$ indicates the pullback on the D5-worldvolume of the various background fields. Moreover, the Wess-Zumino term is given by \footnote{Here and in what follows, for a lighter notation we have omitted the pull-back symbols for the $C_p$ and $B$ forms.}
\ba
\label{WZterm}
S_{WZ} = T_{D5} \int_{\mathbb{R}^{1,3}\times T^2} C_6 + C_4\wg (F+B)\,.
\ea
Following a similar argument to the one in \cite{Lunin:2005jy}, we  notice that
\ba
\mathrm{d}\left( C_6 + C_4\wg B\right) &=& \star \mathrm{d} C_2 + H\wg C_4 +\mathrm{d}C_4\wg B = -F_7+G_5\wg B=0\,,
\ea
where we have used the following relations for RR forms
\ba
G_5 = \mathrm{d} C_4\,,\quad F_7 = -\mathrm{d} C_6 - H\wg C_4\,,
\ea
and the specific expression for $F_7$ of solution in eq. (\ref{sol1}). Hence, $ C_6 + C_4\wg B $ is exact and the coupling of the brane with the form is constant.
We can chose a configuration for the D5-brane that shrinks to a D3, for which the coupling with $C_6 + C_4\wg B$ is not present \cite{Lunin:2005jy}. This argument allows us to set it to zero. Then, the WZ term in eq. (\ref{WZterm}) reduces to 
\ba
\label{WZterm2}
S_{WZ} = T_{D5} \int_{\mathbb{R}^{1,3}\times T^2} C_4\wg F\,.
\ea
We analyze the BPS condition for the probe D5-brane, \textit{i.e.} 
\ba
\label{BPS}
S_{DBI}+S_{WZ} =0\,.
\ea
Setting $F_2$ to extend along the two torus only, the lhs of eq. (\ref{BPS}) reads
\ba
\label{BPS2}
T_{D5}\int_{\mathbb{R}^{1,3}\times T^2} \mathrm{d}^{5+1}\xi\, \left[e^{-\Phi}r^4\sqrt{f(r)}\sqrt{G^2g_0 + (F_{12} + B_{12})^2} +F_{12}r^4\left(1-\frac{Q^4}{r^4}\right) \right]\,,
\ea
where we have use the expression for $C_4$\,, namely
\ba
C_4 = r^4 \left(1-\frac{Q^4}{r^4}\right)\mathrm{d}t\wg \mathrm{d}x_1\wg \mathrm{d}x_2\wg \mathrm{d}\phi\,. 
\ea
We set a special value for the field strength $F$ on the two torus, that is \footnote{The particular choice in eq.(\ref{D5field_strength}) follows the one in \cite{Lunin:2005jy}, where such a fine tuned D5-brane (and gauge field) setting allows to find a BPS configuration. Moreover, the latter, in the case in which $\gamma$ is taken as in eq.(\ref{rationalg}), represents the non-commutative Coulomb branch of eq.(\ref{NCCoulomb}). }
\ba
\label{D5field_strength}
F_{12} = -\frac{1}{\g}\,,
\ea
from which eq. (\ref{BPS2}) reduces to 
\ba
\label{BP3}
&&T_{D5}\int_{\mathbb{R}^{1,3}\times T^2} \mathrm{d}^{5+1}\xi\, \left[e^{-\Phi}r^4\sqrt{f(r)}\sqrt{G^2g_0 + \frac{1}{\g^2}(1-\g^2 g_0 G )^2} -\frac{1}{\g}r^4\left(1-\frac{Q^4}{r^4}\right) \right]\,,\nb\\
&&=T_{D5}\int_{\mathbb{R}^{1,3}\times T^2} \mathrm{d}^{5+1}\xi\, \frac{1}{\g}\left[e^{-\Phi}r^4\sqrt{f(r)}\sqrt{G}- r^4\left(1-\frac{Q^4}{r^4}\right) \right]\,,\nb\\
&&=T_{D5}\int_{\mathbb{R}^{1,3}\times T^2} \mathrm{d}^{5+1}\xi\, \frac{r^4}{\g}\left[\sqrt{\left(1-\frac{Q^6}{r^6}\right)}-\left(1-\frac{Q^4}{r^4}\right) \right]\geq 0\,.
\ea
From eq. (\ref{BP3}) we can observe that the D5 probe brane is attracted to the tip of the cigar at $r=Q$, and therefore this leads to a complete lifting of the Coulomb branch. Finally, notice that in the limit in which $Q\to 0$, we recover the D5-brane to be a good BPS configuration as in \cite{Lunin:2005jy}.

\paragraph{Comments on Chern-Simons terms}\,\\
According to the analysis in \cite{Kumar:2024pcz},\cite{Cassani:2021fyv}, the low energy regime of field theory dual to the background in eqs.(\ref{metric-ARxS5})-(\ref{RR-S5}) is an ${\cal N}=2$ with a Chern-Simons term of level $N$. 

It is natural to ask if something similar occurs for the background in eq.(\ref{sol1}).
To answer this, let us consider a D7-probe extending along the directions $\left[t,x_1,x_2, \theta, \varphi, \varphi_1,\varphi_2,\psi\right]$ in the beta-deformed $\widehat{AdS_5}\times \widehat{S^5}$ of eq.(\ref{sol1}). 
We are interested in the Wess-Zumino part of the action, that is used to read the Chern-Simons level. The Bon-Infeld-Wess-Zumino action reads,
\ba
S_{\text{BI}} + S_{WZ}&=& T_{D7}\int_{\mathbb{R}^{1,2}\times \widehat{S^5}}\mathrm{d}^8\xi \, e^{-\Phi} \sqrt{-\det\left(P[g]_{ab}+F_{ab}+ P[B]_{ab}\right)} \,+\nb\\
&&+ T_{D7} \int_{\mathbb{R}^{1,2}\times \widehat{S^5}} C_8 + C_6\wedge(F+B) + \frac{1}{2}C_4\wedge(F+B)\wedge(F+B)\,.
\ea
We have introduced a gauge field $A$, with field strength $F$, on the $\mathbb{R}^{1,2}$ part of the brane worldvolume.
In the background in eq. (\ref{sol1}) we have  $C_8 =0$ and the form $C_6 + C_4\wedge B$ is exact, as noticed in the Coulomb branch analysis. By using similar arguments we can set the constant coupling of the brane with $C_6 + C_4\wedge B$ to zero, implying a WZ term of the form
\ba
 S_{WZ}&=& T_{D7} \int_{\mathbb{R}^{1,2}\times \widehat{S^5}}  \frac{1}{2}C_4\wedge F\wedge F\, = -T_{D7} \int_{\mathbb{R}^{1,2}\times \widehat{S^5}}  F_5\wedge A\wedge F\nb\\
 &=& - \frac{N}{(2\pi)^3} \int_{\mathbb{R}^{1,2}} A\wedge F\,.\label{lionelandres}
\ea
Hence, we find a Chen-Simons term of level $N$ in the effective action for the gauge field $A$ on the D7 brane.
A similar calculation could be to switch on an axion field 
$C_0=  \frac{N}{2\pi R} \phi$, such that $F_1=  \frac{N}{2\pi R} d\phi$. Switching on a gauge field, the Wess-Zumino term for a D3 brane extended along $[t,x_1,x_2,\phi]$ reads,
\begin{equation}
S_{WZ}=T_{D3}\int_{\mathbb{R}^{1,2}\times S^1} C_0 F \wedge F= -T_{D3}\int_{S^1} F_1 \int_{\mathbb{R}^{1,2}} A\wedge F= - \frac{N}{(2\pi)^3}\int_{\mathbb{R}^{1,2}} A\wedge F\,.\label{angelito}
\end{equation}
The arguments above are indicative or intuitive. One can criticize the following: in the first argument, we are obtaining the action of an Abelian CS-term (the QFT is argued to be $SU(N)_N$ CS, see  \cite{Kumar:2024pcz}, \cite{Cassani:2021fyv}). Having $N$ D7 branes takes the system out of the probe approximation. Regarding the second argument leading to eq.(\ref{angelito}), one should be wary of the non-single valuedness of the field $C_0$. In fact, switching on a background  field with that strength should actually deform the background.
Another criticism can be phrased as follows: the Chern-Simons coefficients in eqs.(\ref{lionelandres})-(\ref{angelito}) do not depend on the parameters $Q,\gamma$. How come the case $Q=\gamma=0$ corresponding to ${\cal N}=4$ SYM does not display a Chern-Simons coefficient? To answer this, observe that the Born-Infeld term indicates a tension-full probe for our case, whilst in the case of $Q=0$ the tension vanishes. In fact, the brane sits at a value of the radial coordinate that minimises its action. This is the value $r=Q$ in our backgrounds (leading to a probe with tension proportional to $Q$) and $r=0$ in the fully symmetric case (leading to a tensionless object). Hence, there is not a domain wall sustaining the Chern-Simons action in the conformal case.

In the next section, we focus our attention to observables that are purely non-perturbative. We have access to them via the holographic backgrounds in eqs.(\ref{sol1}) and (\ref{S5dip}).

\subsection{Observables}
\label{sec:observables}

In this subsection, we analyze various observables of the field theories corresponding to the $\beta$-deformed $\widehat{AdS_5}\times \widehat{S^5}$ type IIB supergravity backgrounds introduced in Section \ref{section-geometry}.  We briefly review the main features of the accounted quantities.\\
We leave for future work \cite{Castellani:2024pmx} an extended study of observables of field theories dual to TsT transformed $\widehat{AdS_5}\times \widehat{T^{1,1}}$ solutions.

\subsubsection{Wilson loops and confinement properties}\label{section-wilson1}
The first feature of the TsT-transformed  $\widehat{AdS_5}\times \widehat{S^5}$ backgrounds in Section \ref{section-geometry}, on which we are interested in is the confining  behaviour. We focus on the marginal and dipole TsT-transformed solutions presented in Subsections \ref{sec_beta_deformation} and \ref{sec:dipole}.
To unravel the confining properties of these backgrounds, we study  the (non-dynamical) quark-anti-quark potential energy via the holographic computation of the field theory rectangular Wilson loop expectation value. The latter is identified with the worldsheet boundary of a probing string that explores the bulk geometry \cite{Rey:1998ik,Maldacena:1998im}.\\
More precisely, the classical string (of length $L$) embedding can be chosen as
\bea
\label{emb}
t = \tau \,, \hspace{1cm}  x^1 = \sigma\,, \hspace{1cm} r = r(\sigma)\,, \hspace{1cm} \s \in \left[-L/2,L/2\right]\,,
\eea
while taking as constants the remaining coordinates.
In the marginal and dipole $\beta$-transformations of subsections \ref{sec_beta_deformation} and \ref{sec:dipole}, the Minkowski and the radial coordinates are not involved in the TsT procedure. Hence, we expect the confining behavior of the deformed backgrounds to be equal to the corresponding $\widehat{AdS_5}\times \widehat{S^5}$ seed-solution. Indeed, the directions containing the new physical information of the deformed solutions are frozen in analyzing these types of observables, that result to be blind to the transformation.\\
In both the $\beta$-deformed backgrounds in eqs. (\ref{sol1}) and (\ref{S5dip}), the induced metric on the worldsheet following from the embedding in eq. (\ref{emb}) reads
\bea
\label{indmet}
\mathrm{d}s^2_{\text{ind}} = r^2\left[-\mathrm{d}\t^2 + \left(1+ \frac{r^\prime(\s)^2}{r^4f(r)}\right)\mathrm{d}\s^2 \right]\,,\quad r^\prime(\s)= \partial_\s r(\s)\,.
\eea
Using eq. (\ref{indmet}), the Nambu-Goto action for the probe string is,
\bea
\label{NGaction}
S_{NG} &=& \frac{1}{2\pi} \int{\mathrm{d}\s\mathrm{d}\t}\sqrt{F_w^2(r)+ G_w(r)^2r^\prime(\s)^2}= \frac{T}{2\pi} \int_{-L/2}^{L/2}\mathrm{d}\s\sqrt{F^2_w(r)+ G_w(r)^2r^\prime(\s)^2}\,.
\ea
In the notation of \cite{Chatzis:2024kdu,Nunez:2009da}, we have defined the functions
\ba
\label{WFG}
F_w(r) = r^2\,, \quad G_w(r) = \frac{1}{\sqrt{f(r)}}\,.
\ea
%
%
%
%
From the equation (\ref{NGaction}) we find the effective potential \cite{Nunez:2009da}
\ba
\label{Veff}
r^\prime(\s) = \pm V_{\text{eff}}(r) = \pm \frac{r^2(\s)}{r_0^2} \sqrt{f(r)\left(r^4(\s) -r_0^4\right)}\,,
\ea
with $r_0$ defined as the turning point of the string in the radial direction.
This suggests a confinement behavior for the backgrounds. Indeed, notice that $V_{\text{eff}}(r)$ diverges in the boundary limit of large $r$ (with power of $\sim \,r^4$), while goes to zero when we approach the turning radius $r\to r_0$, satisfying the necessary conditions for confinement \cite{Nunez:2009da}.\\
Using the expression in eq.(\ref{Veff}), we obtain that the quark-anti-quark length as function of the turning point $r_0$ is given by
\ba
\label{LQQ}
L_{QQ}(r_0)= 2 \int_{r_0}^{\infty} \frac{\mathrm{d}r }{V_{\text{eff}}(r)} =  2r_0^2  \int_{r_0}^{\infty}\mathrm{d}r \frac{r}{\sqrt{(r^6-Q^6)\left(r^4 -r_0^4\right)}}\,.
\ea
It is useful to introduce a combination of the functions $F_w(r)$ and $G_w(r)$ in eq. (\ref{WFG}), that well approximates the length $L_{QQ}(r_0)$ in eq.(\ref{LQQ}). This is given by \cite{Kol:2014nqa,Faedo:2014naa}
\bea
\label{tilde_LQQ}
\tilde L_{QQ}(r_0) \equiv \pi\frac{ G_w(r)}{\partial_r  F_w(r)}\bigg|_{r = r_0} = \frac{\pi}{2}\frac{r_0^2}{\sqrt{r_0^6-Q^6}}\,.
\eea
Notice that $\tilde L_{QQ}(r_0)$ in eq. (\ref{tilde_LQQ}) diverges in the limit in which the turning point radius approaches the minimal one, \textit{i.e.} $r_0\to Q$. This feature suggests a confining behavior for the present background. In fact, the deeper the string probes the IR region of the geometry, the more the quark-anti-quark pair are separated, reaching an infinite relative distance. A finite maximal separation, on the other hand, would indicate screening.

The on-shell action in eq.(\ref{NGaction}) can be written as
\ba
\label{Onshell_NGaction}
S_{NG}
&=& \frac{T}{2\pi}r_0^2 L_{QQ}(r_0)+ 2 \frac{T}{2\pi}\int_{r_0}^{\infty}\mathrm{d}r \,\frac{\sqrt{r^4-r_0^4}}{r^2 \sqrt{f(r)}}\,.
\ea
The (regularized) quark-anti-quark pair energy as a function of the turning point and $L_{QQ}(r_0)$ can be read from eq. (\ref{Onshell_NGaction}) and the relation 
\ba
S_{NG} \sim T\times E_{QQ}(r_0)\,.
\ea
Hence we obtain,
\ba
\label{EQQ}
E_{QQ}(r_0) = r_0^2 L_{QQ}(r_0)+ 2 \int_{r_0}^{\infty}\mathrm{d}r \,\frac{\sqrt{r^4-r_0^4}}{r^2 \sqrt{f(r)}}- 2 \int_{Q}^{\infty}\mathrm{d}r \,\frac{1}{ \sqrt{f(r)}}\,.
\ea
Notice then that the first term in  eq. (\ref{EQQ}) being linearly dependent on the pair separation length, underlines the confining behavior of the dual field theories.\\
Let us conclude this subsection by commenting on the stability of the probe string embedding in eq. (\ref{emb}) in the present confining backgrounds. 
We introduce a $r_0$-dependent quantity useful when analyzing the (perturbative) stability of the embedding \cite{Faedo:2013ota}, namely
\bea
\label{Z_W}
\mathcal{Z}_w(r_0) \equiv \diff{\tilde L_{QQ}(r_0)}{r_0} = -\pi r\frac{\left(2r^6+Q^6\right)}{\left(r^6-Q^6\right)^{\frac{3}{2}}}\,.
\ea
A classically stable quark-anti-quark solution is characterised by a force which is both attractive and non-increasing with increasing distance. These two conditions are,
\bea
\label{W_stability}
&&\diff{E_{QQ}}{L_{QQ}}  = F_w(r_0) >0\,,\quad \diff*[2]{E_{QQ}}{L_{QQ}}\,\sim \mathcal{Z}_w(r_0)^{-1}\,\diff{ F_w(r_0)}{r_0} <0\,.
\eea
Inspecting the expressions in eqs. (\ref{WFG}) and (\ref{Z_W}) we  observe that the conditions in eq. (\ref{W_stability}) are verified.
This strongly suggests that the embedding for the probe string is stable.
\\
In summary, for both backgrounds in eqs.(\ref{sol1}), (\ref{S5dip}) we find indications that the holographic dual QFTs are confining, as is the seed-background in eq.(\ref{metric-ARxS5})-(\ref{RR-S5}).

\subsubsection{'t Hooft loops}
Another interesting observable to analyze is the 't Hooft loop. As in the previous subsection, here we focus on the marginal and dipole deformed background of Subsections \ref{sec_beta_deformation} and \ref{sec:dipole}.\\ The effective string connecting a  monopole-anti-monopole pair is modelled by studying  a probe D3 brane extended in the $[t,x_1, \phi, \alpha]$ directions ($\alpha$ is an angular coordinate along the internal $\widehat{S^5}$ submanifold), with the radial direction taken as function of the spatial Minkowski direction, \textit{i.e.} $r(x_1)$. All the remaining coordinates are fix to be constant. \\
The D3-brane DBI action, 
\ba
\label{DBI}
S_{D_3}= \int{}\mathrm{d}^{4} \xi\,e^{-\Phi} \sqrt{-\det\left(P[g]_{ab} + P[B]_{ab} \right)}\,,
\ea
and integrating out the compact coordinates $[\phi,\a]$, we end up with a two-dimensional effective action.
In eq. (\ref{DBI}), $P[\cdot]$ represents the pull-back of the background fields on the worldvolume. We study this in the case of the backgrounds of eqs.(\ref{sol1}) and (\ref{S5dip}).
\paragraph{Marginal beta deformed solution}\, \\
Let us consider the TsT-transformed background in eq. (\ref{sol1}). We choose the embedding for the probe D3 brane along the $[t,x_1, \phi, \psi]$ directions. Furthermore, we set
%
$\theta = \frac{\pi}{4}\,, \quad \varphi = \frac{\pi}{2}\,$.
%
The induced metric is,
\ba
\label{D3ind}
\mathrm{d}s^2_{D3} = r^2\left[-\mathrm{d}t^2+  \left(1+ \frac{r^\prime(x_1)^2}{r^4f(r)}\right)\mathrm{d}x_1^2 + f(r)\mathrm{d}\phi^2  \right] + G\left(\mathrm{d}\psi + \mathcal{A}\right)^2\,,\quad r^\prime(x_1)= \partial_{x_1}r(x_1)\,,
\ea
while 
\ba
g_0 = \frac{1}{4}\, \quad \to \quad G^{-1} = 1+\frac{\g^2}{4}\,.
\ea
We also have a non-trivial dilaton given in eq. (\ref{sol1}). The $B$ field in eq. (\ref{sol1}) does not contribute to this particular D3-brane probe. \\ The DBI action for the probe brane in eq. (\ref{DBI}) reduces to
\ba
S_{D3} = \int{}\mathrm{d}^4 \xi\,e^{-\Phi} \sqrt{-\det g_{D3}}\,,
\ea
and after integrating-out the compact $[\phi,\psi]$ coordinates, we find a two-dimensional effective action for a magnetically charged probe, 
\begin{multline}
\label{SeffHL}
S_{\text{eff}} =e^{-\Phi}L_{\phi} L_{\psi} \int{}\mathrm{d}t\,\mathrm{d}x_1\,\sqrt{G} \sqrt{r^6f(r) + r^2 r^{\prime\,2}}\,= L_{\phi} L_{\psi} T \int_{-L/2}^{L/2}\mathrm{d}x_1\, \sqrt{F_t(r)^2 + G_t(r)^2 r^{\prime\,2}}\,.
\end{multline}
%
We have defined the functions
\ba
\label{HLFG}
F_t(r) = r^3 \sqrt{f(r)}\, \quad G_t(r) = r\,.
\ea
From eqs. (\ref{SeffHL}) and (\ref{HLFG}), we  observe that the effective string tension is vanishing in the limit $r\to Q$, because $F_t(r) \to 0$ in the same limit. This means that in the IR the magnetic monopole pair can be separated without any energy expense. The latter observation suggests a screening behavior of the dual field theory for magnetic charges. \\
Following a similar procedure to the Wilson loop one, we can derive the effective potential, the monopole-anti-monopole separation length $L_{MM}(r_0)$ and its approximate expression $\hat{L}_{MM}(r_0)$:
\ba
\label{VeffHL}
V_{\text{eff}}(r) &=& \frac{F_t(r)}{F_t(r_0)G_t(r)} \sqrt{F_t(r)^2 -F_t(r_0)^2}\,,
\ea
and
\ba
\label{LMM}
&&L_{MM}(r_0)= 2 \int_{r_0}^{\infty} \frac{\mathrm{d}r }{V_{\text{eff}}(r)} =  2F_t(r_0) \int_{r_0}^{\infty}\mathrm{d}r \frac{1}{r^2\sqrt{f(r)}}\frac{1}{\sqrt{r^6 f(r) -F_t(r_0)^2}}\,,\nb\\
&&\tilde L_{MM}(r_0) \equiv \pi\frac{ G_t(r)}{\partial_r  F_t(r)}\bigg|_{r = r_0} = \frac{\pi}{3r_0^4}\sqrt{r_0^6-Q^6}\,.
\ea
From eq. (\ref{LMM}), we observe that as $r_0\to Q$ the approximated length $\tilde L_{MM}(r_0)$ vanishes, suggesting a screening behavior for the monopole-anti-monopole pair.
Finally, the magnetic charge pair energy is given in terms of $L_{MM}(r_0)$ as it follows
\ba
\label{EMM}
E_{MM}(r_0) = F_t(r_0) L_{MM}(r_0)+ 2 \int_{r_0}^{\infty}\mathrm{d}r \,\frac{\sqrt{r^6f(r) -F_t(r_0)^2}}{r^2 \sqrt{f(r)}}- 2 \int_{Q}^{\infty}\mathrm{d}r \,r\,.
\ea
In the same way as in the Wilson loop case, we have regularized $E_{MM}$ adding a counterterm to the on-shell action.
In the expression in eq. (\ref{EMM}) the linear term in $L_{MM}$ is proportional to the function $F_t(r_0)$ which goes to zero as the turning point approaches the minimum radius $r=Q$. This feature underlines the screening property of the system. \\
Regarding the stability of the embedding, similarly to eq. (\ref{Z_W}), we consider the function
\bea
\label{Z_t}
\mathcal{Z}_t(r_0) \equiv \diff{\tilde L_{MM}(r_0)}{r_0} = -\frac{\pi}{3}\frac{r_0^6-4Q^6}{r_0^5\sqrt{r_0^6-Q^6}}\,,
\eea
which displays a sign transition at $r_0 = 2^{1/3}Q$. As in the Wilson loop case, the region in which $\mathcal{Z}_t(r_0)$ is negative, is understood as the physical and stable one \cite{Faedo:2013ota}. Otherwise, the probe is likely unstable indicating a transition to a possibly disconnected configuration.\\
Finally, notice that we have found the same expressions for the separation length and energy in eqs. (\ref{LMM}) and (\ref{EMM}) that we would have obtained in the original solution in eq. (\ref{metric-ARxS5}). In fact, the dilaton and the factor of $\sqrt{G}$ in eq.(\ref{SeffHL}) have cancelled out. The probe D3 displays the same behaviour  in the seed-background of eq.(\ref{metric-ARxS5}). We refer to \cite{Chatzis:2024top,Chatzis:2024kdu} for a detailed analysis.

\paragraph{Dipole-deformed solution}\,\\
Now, we study the expectation value of a 't Hooft loop in the dipole $\beta$-transformed solution in eq. (\ref{S5dip}). Choosing the embedding of the D3-brane along the $[t,x_1, \phi, \varphi]$ coordinates and $r=r(x_1)$,  the induced metric on the worldvolume reads
\ba
\label{D3indDipole}
\mathrm{d}s^2_{D3} = r^2\left[-\mathrm{d}t^2+  \left(1+ \frac{r^\prime(x_1)^2}{r^4f(r)}\right)\mathrm{d}x_1^2 + G f(r)\mathrm{d}\phi^2  \right] + G\left(\mathrm{d}\varphi + \mathcal{A}\right)^2\,,
\ea
with the $G$ function defined as in eq. (\ref{GDipole}). We stress that  the dilaton in eq. (\ref{S5dip}) is $r$ dependent, so it does not factorize out in the effective action. Moreover, in the present case the $B$ field in eq. (\ref{S5dip}) does contribute to the DBI action. Using eqs. (\ref{S5dip}) and (\ref{D3indDipole}) we have that
\ba
\label{pullbackgB}
-\det\left(P[g]_{ab} + P[B]_{ab} \right) = G(r)\left(r^6f(r)+ r^2 r^{\prime\,2}\right)\,.
\ea
After the integration of the compact coordinates $[\phi,\varphi]$, the 2-dimensional effective action reads
\begin{multline}
\label{SeffHLdipole}
S_{\text{eff}} =L_{\phi} L_{\psi} \int{}e^{-\Phi}\mathrm{d}t\,\mathrm{d}x_1\, \sqrt{G(r)
(r^6f(r) + r^2 r^{\prime\,2})}=L_{\phi} L_{\psi} T \int_{-L/2}^{L/2}\mathrm{d}x_1\, \sqrt{F_t(r)^2 + G_t(r)^2 r^{\prime\,2}}\,.
\end{multline}
The functions $F_t(r)$ and $G_t(r)$  are defined in eq. (\ref{HLFG}). Hence, eq. (\ref{SeffHLdipole}) coincides with the expression in eq. (\ref{SeffHL}). The same analysis suggests that  the dipole-transformed background displays a magnetic monopole-anti-monopole screening behavior. Notice  the interesting cancellation between the dilaton and the function $G$
in eq.(\ref{SeffHLdipole}).

\subsubsection{Entanglement Entropy}\label{sec:EE}
The next observable we study is the Entanglement Entropy (EE) on a strip. Following the standard holographic procedure (see \textit{e.g.} \cite{Ryu:2006bv,Klebanov:2007ws,Kol:2014nqa}), the EE for two  boundary regions can be accounted for by considering a codimension two (minimal-action) manifold $\Sigma_8$ that approaches the boundary of the entangling surface. As in \cite{Klebanov:2007ws,Kol:2014nqa} the entangling surface can be taken as a strip of length $L$. The action  to  be minimized is given by
\begin{equation}\label{S_EE}
    S _{\mathrm{EE}} = \frac{1}{4G_N} \int _{\Sigma _8}\mathrm{d}^8x \, e^{-2\Phi}\sqrt{\mathrm{det} g _{\Sigma_8} }
\end{equation}
where $\det g _{\Sigma_8}$ is the determinant of the induced metric over the codimension two surface and $G_N$ is the ten-dimensional Newton's constant.\\
The papers \cite{Klebanov:2007ws,Kol:2014nqa} point out that (typically) in a confining background, the EE exhibits a first-order phase transition along with the variation of the width $L$ of the entangled region boundary. Thus the EE provides a useful tool in analyzing the confining properties of a given background. 
In the present case, we take $\Sigma_8$ in the seed-background of eq.(\ref{metric-ARxS5}), as a fixed time surface extending along internal $\widehat{S^5}$ submanifold and the $[x_1,x_2,\phi]$ $AdS$ directions, with $r=r(x_1)$. We can compare this with the analogous $\Sigma_8$ calculated in the backgrounds of eqs.(\ref{sol1}),(\ref{S5dip}). As we discuss below, a different behaviour occurs for the non-commutative background of eq. (\ref{NCS5}).


The calculation shows that after the TsT-tranformation, in both the marginal and dipole deformed backgrounds, we have that $\mathrm{det} g _{\Sigma_8}$ is modified such that
\ba
\mathrm{det} g _{\Sigma_8}\quad \longrightarrow \quad G^2\mathrm{det} g _{\Sigma_8}\,.
\ea
Notice that in the transformed backgrounds, we found out that the dilaton is given in terms of the $G$ by the relation
\ba
\label{EEdilaton}
e^{2\Phi_0}\quad \longrightarrow \quad \e^{2\Phi_0}G\,.
\ea
We denoted by $\Phi_0$ the value of the dilaton in the seed background of eq.(\ref{metric-ARxS5}). All in all, we have that the EE action in eq. (\ref{S_EE}) is an invariant quantity under the TsT-transformations considered here, since
\ba
S_{EE} \quad \longrightarrow \quad \frac{1}{4G_N} \int _{\Sigma _8}\mathrm{d}^8x \, G^{-1}e^{-2\Phi_0}\sqrt{G^2 \mathrm{det} g _{\Sigma_8} } = S_{EE}\,.
\ea
We conclude that the analysis of the $S_{EE}$ behavior and its dependence on the strip length $L$ follows identically what is presented in \cite{Chatzis:2024top,Chatzis:2024kdu}, to which we refer for precise formulae and plots.\\
Let us give some details. Considering the surface $\Sigma_8$ to be
\begin{eqnarray}
& &\Sigma_{8,\beta}=[x_1,x_2,\phi,\theta,\varphi,\phi_1,\phi_2, \phi_3],~~\text{with}~~r(x_1),\nonumber\\
 & &\Sigma_{8,dip}=[x_1,x_2,\phi,\mathbb{C}P2,\varphi],~~~~~~~~~\text{with}~~r(x_1),
\end{eqnarray}
for the beta deformed and dipole deformed backgrounds respectively. Calculating the $S_{EE}$ in eq.(\ref{S_EE}), we find
\begin{eqnarray}
& & S_{EE,\beta}= {\cal N}_\beta \int_{-L/2}^{L/2}dx_1 \sqrt{r^6 f(r) + r^2 r'^2},~~S_{EE,dip}= {\cal N}_{dip} \int_{-L/2}^{L/2}dx_1 \sqrt{r^6 f(r) + r^2 r'^2}.\label{SEEbetadip}
\end{eqnarray}
For the beta deformed and dipole backgrounds respectively. The constants ${\cal N}_{\beta}$ and ${\cal N}_{dip}$ refer to the integration over the remaining seven coordinates. 

In sharp contrast, for the non-commutative background of eq.(\ref{NCS5}), we find a qualitatively different result.
Choosing the $\Sigma_{8,NC}=[x_1,x_2,\phi, S^5]$, with $r(x_1)$, we find after integration
\begin{equation}
 S_{EE,NC}={\cal N}_{NC} \int_{-L/2}^{L/2}dx_1 \sqrt{r^6 f(r) + r^2 G(r) r'^2}.  
\end{equation}
The presence of the $G(r)$-factor changes the behaviour qualitatively. {In the case of the backgrounds of eqs.(\ref{NCS5}),(\ref{NCdipole}) the dual QFT is highly non-local in the UV. This makes the criterium of \cite{Klebanov:2007ws} not applicable, as discussed in \cite{Kol:2014nqa}, \cite{Jokela:2020wgs}.} Interestingly, the dipole theory, being non-local, is not afflicted by this and the EE gives a field theoretical result (see the flow central charge in the next section, for a fingerprint of such non-localities). 

In particular, in each (beta deformed and dipole deformed) backgrounds, the EE of the strip as a function of its length $L$ displays a so-called “butterfly shape” first-order phase transition between the connected and the disconnected solutions. The latter is argued to be a signal for confinement. \\
Finally, there is an interesting feature to notice: in \cite{Chatzis:2024kdu} has been observed that in the $\widehat{AdS_5}\times \widehat{S^5}$ background (and further examples analyzed in that paper), once having integrated out all the directions aside from $x_1$, $S_{EE}$ in eq.  (\ref{S_EE}) reduces to a one-dimensional integral expression equivalent (up to overall factors) to the one that appears in the 't Hooft loops calculations. Hence, in those cases, the 't Hooft loop and EE display the same behavior with the monopole-anti-monopole/entangling strip lengths respectively. 
In fact, compare the expressions in eq.(\ref{SEEbetadip}) with those in eqs.(\ref{SeffHL})-(\ref{HLFG}), and  (\ref{SeffHLdipole})
%

In summary, the Wilson loop, the 't Hooft loop and the Entanglement Entropy point to the confining behaviour of the QFTs dual to the backgrounds in eqs.(\ref{sol1}),(\ref{S5dip}). Interestingly, all the particular details of the TsT transformed backgrounds cancel-out in the calculation. This points to a form of universality for these QFTs that would be difficult to guess pure on field theoretical grounds.

\subsubsection{Holographic flow central charge }\label{sec:Hol_central_charge}
In this subsection we analyse a quantity of particular relevance in conformal field theories (which is not the case of the QFT dual to the present backgrounds), the holographic central charge. In a CFT, its central charge encodes the number of degrees of freedom and is related to the system's free energy. The study of this observable can be extended in the context of holography (both in $AdS$ like and non $AdS$ “confining” background) to the analysis of the holographic central charge flow: a monotonic function describing the counting of the  degrees of freedom of a system, undergoing  a flow across dimensions \cite{Macpherson:2014eza,Bea:2015fja}. To define this flow central charge, it is useful to write the metric of a general background in the following form
\begin{eqnarray}
ds^{2} &=&-\alpha_0\mathrm{d}t^2+ \sum_{n =1}^{d} \alpha_n \mathrm{d}x_n^2 +\left(\prod_{n =1}^d \alpha_n \right)^{\frac{1}{d}}\beta \left( r\right)
dr^{2} +g_{ij} (dy^{i}-A^i)(dy^{j}-A^j)\,,
\label{conventions central charge}
\end{eqnarray}%
as well as the dilaton
\ba
\Phi = \Phi(r, y^i)\,.
\ea
In equation (\ref{conventions central charge}), the $y^{i}$ identify the coordinates of the internal manifold and we called $x_n$ the coordinates ''of the QFT''. We define
\ba
& &\hat g_{ab}\mathrm{d}\xi^a \mathrm{d}\xi^b \equiv \sum_{n =1}^{d} \alpha_n \mathrm{d}x_n^2 +g_{ij} dy^{i}dy^{j}\,.\nonumber\\
&  &
H^{\frac{1}{2}}(r) \equiv \int\mathrm{d}\xi^a \, e^{-2\Phi }\sqrt{\det \hat g_{ab}}\,.\label{Hcflow}
\ea
Using the above equations and following \cite{Macpherson:2014eza,Bea:2015fja}, the holographic central charge flow is defined as \footnote{Here the primes “ $^\prime$ ” stands for the derivative respect with the $r$ coordinate. }
\begin{equation}
c_{\text{flow}}=d^{d}\frac{\beta \left( r\right) ^{d/2}H^{\left( 2d+1\right)
/2}}{G^{(10)}_{N}\left( H^{\prime }\right) ^{d}}\ .\label{chol}
\end{equation}
For the backgrounds in this work, we have $d=3$, since the $x_n$ span the $\left[x_1,x_2,\phi\right]$ directions. Moreover,  the coordinates $ y^{i}$ describe the five-dimensional internal manifold.
The functions in eq. (\ref{conventions central charge}) and the dilaton depend on the specific choice of background. We  observe a common feature in all the present examples. As noticed in the study of the entanglement entropy in Subsection \ref{sec:EE}, in all the $\beta$-transformed backgrounds of Section \ref{section-geometry} the dilaton is related to its original value as in eq. (\ref{EEdilaton}). Furthermore, in this work, the TsT-transformations have been performed involving $U(1)$ isometries laying in the submanifold described by $\left[x_n,y^i\right]$. We then have 
\ba
\det \hat g_{ab}\quad \longrightarrow_{\beta}\quad G^2\det \hat g_{ab}\,.
\ea
As before, we focus on the backgrounds in eqs.(\ref{sol1}) and (\ref{S5dip}). The function $H(r)$ in eq. (\ref{Hcflow}) is  invariant  under TsT-transformations. This could drive us to the conclusion that the flow central charge in eq. (\ref{chol}) is invariant as well. This conclusion is not correct, see below for a careful study. 
\\
Let us summarise the analysis of $c_{\text{flow}}$ in the $\widehat{AdS_5}\times \widehat{S^5}$ background of eq.(\ref{metric-ARxS5}) \cite{Chatzis:2024kdu}. In this case, we have that the functions in eq. (\ref{conventions central charge}) reads
\ba
\a_{1,2} = r^2\,, \quad \a_{3} = r^2 f(r)\,,\quad \beta(r) =\frac{1}{r^4}f(r)^{-\frac{4}{3}}\,.\label{diego}
\ea
Moreover, using the metric in eq. (\ref{metric-ARxS5}) we can evaluate the function $H(r)$  in eq. (\ref{Hcflow}), obtaining
\ba
\label{H2cflow}
H^{\frac{1}{2}}(r)= \text{vol}_{S^5}r^3 \sqrt{f(r)}\,.\label{lionel}
\ea
We find  $c_{\text{flow}}$ in eq. (\ref{chol}) to be,
\ba
\label{c2hol}
c_{\text{flow}} 
&=& \frac{\text{vol}_{S^5}}{8G^{(10)}_{N}}\left(\frac{\sqrt{f(r)}}{f(r)+\frac{r}{6}f^\prime(r)}\right)^3 = \frac{\text{vol}_{S^5}}{8G^{(10)}_{N}}\left(1-\frac{Q^6}{r^6}\right)^{\frac{3}{2}}\,.
\ea
%
The quantity in eq. (\ref{c2hol}) is  defined all along the range of the radial coordinate $r\in \left[Q, \infty\right)$. In particular, in the limit of the deep IR, \textit{i.e.} $r\to Q$,  $c_{\text{flow}}$ approaches zero, suggesting the lack of dynamical degrees of freedom (a gapped system). On the other hand, in the UV regime $c_{\text{flow}}$ has a non-vanishing limit value given by 
\ba
\label{cUV}
c_{\text{UV}} = \frac{\text{vol}_{S^5}}{8G^{(10)}_{N}}\,.
\ea
We have a monotonic quantity for a quantity that describes a flow across dimensions. This interpolates between zero degrees of freedom in the IR, and the value in eq.(\ref{cUV}) for the UV. This value in eq.(\ref{cUV}) corresponds to the 4d SCFT point. The same result is obtained if we work with the background of eq.(\ref{sol1}). In fact, being a marginal deformation, we should expect that the holographic central charge at the UV fixed point is unchanged, and the same to occur along the flow.

The case of the dipole deformation--the background in eq.(\ref{S5dip}) is subtly different. In fact, whilst $\alpha_1,\alpha_2$ take the same values as in eq.(\ref{diego}),  we find that
\begin{equation}
\alpha_1=\alpha_2=r^2,~~\alpha_3=r^2 f(r)G(r),~~~\beta(r)= \frac{1}{r^4 f(r)^{4/3} G^{1/3}}.
\end{equation}
The change in $\alpha_3$ respect to eq.(\ref{diego}) is not so relevant as it is cancelled by the factor of the dilaton $e^{-2\Phi}=\frac{1}{G}$, leading to the same $H(r)$ as in eq.(\ref{lionel}). In contrast, the change in the function $\beta(r)$ implies that after calculating with eq.(\ref{chol}) we find,
\begin{equation}
 c_{flow}=   \frac{\text{vol}_{S^5}}{8G^{(10)}_{N}}\left(\frac{\sqrt{f(r)}}{f(r)+\frac{r}{6}f^\prime(r)}\right)^3 \times \frac{1}{\sqrt{G(r)}}.
\end{equation}
Analysing this expression we find still the tell-sign of  a gapped system (vanishing at $r=Q$), but the far UV asymptotics is not that corresponding to a CFT. In fact, we observe that the quantity grows unbounded $c_{flow}\sim r$. We anticipated this, given that the $r\to\infty$ asymptotics of the background is not $AdS_5$--see the comment around eq.(\ref{xxy}). This is the sign of a QFT in need of a UV completion. As we insisted, this dipole-deformed QFTs do not have a well defined UV, they are non-local field theories. The good definition of the system is (probably) in terms of a string theory. In some sense, the UV-behaviour is reminiscent of what occurs in the systems of
\cite{Nunez:2023xgl}, \cite{Nunez:2023nnl}. Whilst the behaviour of the EE, Wilson and 't Hooft loops was universal for backgrounds in eqs.(\ref{metric-ARxS5}),(\ref{sol1}) and (\ref{S5dip}), the behaviour of the flow central charge is not the same for the three backgrounds. This could have been anticipated given the non-local character of the dipole QFT. 
\\

\subsubsection{A Polyakov-like loop and symmetry breaking}
In this subsection, we  study the properties of the Wilson loop of the $SU(N)$ gauge field along the compactified $S^1$ cycle. The latter plays the role of order parameter for the spontaneous breaking of a $\mathbb{Z}_N$ symmetry of the dual field theory on the $\mathbb{R}^{1,2}\times S^1$. 
\\ The loop we study is not taken along a thermal circle as in \cite{Witten:1998zw}. Following the notation in \cite{Kumar:2024pcz}, we refer to this order parameter as \textit{Polyakov loop}.\footnote{In its original definition as temporal Wilson loop, the Polyakov loop  acts as an order parameter for confinement. It is related to the free energy of a static (infinitely) massive quark, $|P_t| \sim \exp{\left(-F/T\right)}$ \cite{tHooft:1977nqb,Witten:1998zw}. Notice that this is not the case of the expression in eq. (\ref{Polyakov_loop}), where the integral is taken over the spatial cycle $S^1_\phi$ and there are no proper notions of temperature and free energy.} In particular in the present case, this is given by
\begin{equation}
\label{Polyakov_loop}
    P_\phi =\frac1N{\rm Tr} \,\exp i\oint_{S^1_\phi}\, \Theta^{(0)}\,,
\end{equation}
where $\Theta^{(0)}$ is the zero-mode of the $\phi$-component of the $SU(N)$ gauge field, $A_\phi$ after the Kaluza-Klein reduction. Let us consider an extended (large) gauge transformation.
Then, the Polyakov loop in eq. (\ref{Polyakov_loop}) transforms under the latter as \cite{tHooft:1977nqb,Witten:1998zw}
\begin{equation}
\label{Polyakov_loop_gt}
    P_\phi \to h P_\phi\,,\quad h\in \mathbb{Z}_N = \left\{e^{2\pi ki/N}\,, k=0,...,N-1\right\}\,.
\end{equation}
From eq. (\ref{Polyakov_loop_gt}) is clear why the expectation value of the Polyakov loop is an order parameter for the $\mathbb{Z}_N$ spontaneous symmetry breaking. If $\langle P_\phi\rangle = 0$, the $\mathbb{Z}_N$ symmetry is unbroken. On the other hand when $\langle P_\phi\rangle \neq 0$, it reveals the spontaneously broken phase of the  symmetry.
Notice that we do not associate this discrete symmetry with the center of the electric gauge group.
\\
The authors of  \cite{Kumar:2024pcz} noted that the holographic computation for the (regularized) Polyakov loop in the $\widehat{AdS_5}\times \widehat{S^5}$ background leads to a non-vanishing expectation value proportional to the square root of the 't Hooft coupling $\l = g_{YM}^2N$, namely
\begin{equation}
\label{PLEV}
    \langle P_\phi\rangle \sim \,\exp\left(-S^{\text{reg}}_{NG}
\right)\sim \,e^{-a \sqrt \l}\neq 0\,,
\end{equation}
with $a$ being a constant.
This yields the conclusion that field theory dual to solution in (\ref{metric-ARxS5}) displays a spontaneous breaking of the $\mathbb{Z}_N$ symmetry.\\
Thus, it is very interesting to analyze the Polyakov loop behavior also in the field theories dual to the backgrounds introduced in Section \ref{section-geometry}, with special regard to the marginal and dipole deformed ones. Let us start by analyzing the holographic computation of the Polyakov loop in the $\beta$-transformed $\widehat{AdS_5}\times \widehat{S^5}$ solution in Subsection \ref{sec_beta_deformation}.
To do that we take the embedding for the  (Euclidean) string worldsheet as given by
\ba
\label{PLstring_marginal}
\phi(\t) = \t\,,\quad r(\s) = \s\,, \quad \m_i = \text{constant}\,,
\ea
from which the induced metric reads 
\ba
\mathrm{d}s^2_{\text{ind}} =\frac{1}{r^2f(r)} \mathrm{d}r^2 + \left[r^2f(r) + G\,\left(1+9\gamma^2\m_1^2\m_2^2\m_3^2\right)\,Q^6 \left(\frac{1}{r^2}-\frac{1}{Q^2}\right)^2\right]\mathrm{d}\phi^2\,.
\ea
We can immediately notice that given the embedding in eq. (\ref{PLstring_marginal}), this probe string configuration in thus not blind to the TsT-transformation, since the result is directly depending on the $\gamma$ parameter.\\
The on-shell Nambu-Goto action for the string is given by
\ba
\label{SF1_beta}
S^{\text{on-shell}}_{NG} &=& \frac{1}{2\pi}\int_{Q}^\infty\mathrm{d}r \int_0^{2\pi R}\mathrm{d}\phi \sqrt{\det g_{\text{ind}}} = R\int_{Q}^\infty\mathrm{d}r \sqrt{1+ G\, \left(1+9\gamma^2\m_1^2\m_2^2\m_3^2\right)\left(\frac{Q^2(r^2-Q^2)^2}{r^6 f(r)}\right)}\,\nb\\
&=&\sqrt{2\l}R Q\,\int_{1}^\infty\mathrm{d}\r \sqrt{1+ G\, \left(1+9\gamma^2\m_1^2\m_2^2\m_3^2\right)\frac{(\r^2-1)^2}{(\r^6-1)}}\,.
\ea
In the limit  $\gamma\to 0$, this expression reduces to the one for the seed-background $\widehat{AdS_5}\times \widehat{S^5}$ solution in eq. (\ref{PLEV}) \cite{Kumar:2024pcz}. In the last line we have introduced back the $AdS$ radius $\ell$ and $\alpha^\prime$, used the identification
\ba
\label{tHooft}
\ell^4/\a^{\prime\,2} = 2g_{YM}^2 N = 2\l\,.
\ea
and defined the dimensionless radial coordinate $\rho$ as
\bea
r = Q\ell^2 \rho.
\eea
Notice that similarly to what observed in \cite{Kumar:2024pcz}, the expression is regular at $\r = 1$ but has to be regularized at infinity adding a well-suited counterterm (in particular the integral in eq. (\ref{SF1}) is linearly divergent). Thus, following the standard procedure, we can write the UV regularized on-shell action as
\bea
\label{SF1reg_beta}
S^{\text{reg}}_{NG} 
=\lim_{\L\to \infty}\,\sqrt{2\l}RQ\left[\int_{1}^\L\mathrm{d}\r \sqrt{1+ G\, \left(1+9\gamma^2\m_1^2\m_2^2\m_3^2\right)\frac{(\r^2-1)^2}{(\r^6-1)}}- \L\right]\,,
\eea
which remarkably is finite and non-vanishing.

Interesting is also the case of the field theory dual to $\b$-dipole solution of Subsection \ref{sec:dipole}. Using the background in eq. (\ref{S5dip}), and fixing the probe string embedding as in eq. (\ref{PLstring_marginal}), we have that the induced metric on the worldsheet is
\ba
\mathrm{d}s^2_{\text{ind}} =\frac{1}{r^2f(r)} \mathrm{d}r^2 + G(r)\left[r^2f(r) + Q^6 \left(\frac{1}{r^2}-\frac{1}{Q^2}\right)^2\right]\mathrm{d}\phi^2\,,
\ea
and the on-shell string action reads
\ba
\label{SF1}
S^{\text{on-shell}}_{NG} &=& \frac{1}{2\pi}\int_{Q}^\infty\mathrm{d}r \int_0^{2\pi R}\mathrm{d}\phi \sqrt{\det g_{\text{ind}}} = R\int_{Q}^\infty\mathrm{d}r \sqrt{G(r)\left(1+ \frac{Q^2(r^2-Q^2)^2}{r^6 f(r)}\right)}\,\nb\\
&=&\sqrt{2\l}RQ\int_{1}^\infty\mathrm{d}\r \sqrt{\frac{\r^4}{\r^4 + \g^2 Q^2\ell^2 (\r^6-1)}\left(1+ \frac{(\r^2-1)^2}{(\r^6-1)}\right)}\,.
\ea
In the present case, the expression in eq. (\ref{SF1}) is regular in $\rho =1$, while the regularized action is given by
\begin{multline}
\label{SF1reg}
S^{\text{reg}}_{NG} 
=\lim_{\L\to \infty}\,\sqrt{2\l}RQ\left[\int_{1}^\L\mathrm{d}\r \sqrt{\frac{\r^4}{\r^4 + \g^2 Q^2\ell^4 (\r^6-1)}\left(1+ \frac{(\r^2-1)^2}{(\r^6-1)}\right)}-\frac{1}{\g Q\ell^2}\log \L\right]\,,
\end{multline}
which gives a ($\g$ and $Q$ depending) non-zero value .\\
Given the results in eqs. (\ref{SF1reg_beta}) and (\ref{SF1reg}), we  observe that after the marginal and dipole TsT-transformations, the field theories dual to the deformed backgrounds in Subsections \ref{sec_beta_deformation} and \ref{sec:dipole},  display a non-vanishing Polyakov loop expectation value as in eq. (\ref{PLEV}). From this feature we conclude that the present theories are characterized by a symmetry $\mathbb{Z}_N$ broken phase, with $N$ possible different vacua. Note also that in both the marginal and dipole deformed QFTs dual to the backgrounds in eqs.(\ref{sol1}) and (\ref{S5dip}), the VEV of the Polyakov loop is dependent on the deformation parameter $\gamma$. This is an observable whose dynamics is actually influenced by the KK-modes of the circle reduction. The KK modes do contribute to the VEV of the Polyakov-like loop.

\subsubsection{Semiclassical spinning  strings in beta-deformed \texorpdfstring{$\widehat{AdS_5}\times \widehat{S^5}$}{AdS5xS5}}
In this subsection, we  study a semiclassical spinning string in the marginal $\beta$-deformed solution of eq. (\ref{sol1}).
We consider the simple case of a string laying on the minimal radius surface $r= Q$ of the $AdS$-spacetime and probing the internal $\beta$-deformed $\widehat{S^5}$, spinning in two directions. Since in this case the string is not moving around the compact $\phi$ or the radial directions, we expect that the results will not differ so much from the one in \cite{Frolov:2005ty,Bobev:2005cz} for the Lunin-Maldacena $\beta$-transformed $AdS_5\times S^5$ solution.\\
We start displaying the bosonic part of the Polyakov classical string action in curved background,\footnote{We use  the convention $\e^{\t\s} = 1$. Also, to avoid confusion between target space and worldsheet quantities, we  use Greek and Latin letters for worldsheet and background indices, respectively. }
\bea
\label{Polyakov}
S_P = -\frac{1}{4\pi}\int\, \mathrm{d}^2\s\left[ \eta^{\a\b}\partial_\a X^m\partial_\b X^n\, g_{mn}
-\e^{\a\b}\partial_\a X^m\partial_\b X^n\, B_{mn}\right]\,.
\eea
Here $g_{mn}$ is the ten-dimensional target spacetime metric, $B_{mn}$ is the NS-NS 2-form field and $X^m(\t,\s)$ are the embedding functions. We have chosen  the conformal gauge fixing
\ba
\sqrt{-h}\, h^{\a\b} \equiv \eta^{\a\b}\,.
\ea
The Euler-Lagrange equations for the Polyakov action in eq.(\ref{Polyakov}) read
\begin{multline}
\label{EL}
2\p_\a\left(\eta^{\a\b}\p_\b X^m g_{ms}-\e^{\a\b}\p_\b X^m B_{sm}\right) =\eta^{\a\b}\p_\a X^m \p_\b X^n \p_s g_{mn} -\e^{\a\b}\p_\a X^m\p_\b X^n \p_s B_{mn}\,.
\end{multline}
or equivalently \footnote{Here, we  have use the identity: $\e^{\a\b}\p_\a X^m\p_\b X^n\left[\p_m B_{sn} -\frac{1}{2}\p_sB_{mn}\right] = -\frac{1}{2}\e^{\a\b}\p_\a X^m\p_\b X^n H_{smn}$. }
\begin{multline}
\label{EL2}
g_{sm }\eta^{\a\b}\p_\a\p_\b X^m + \frac{1}{2}\left[\p_n g_{ms} +\p_m g_{ns}- \p_s g_{mn}  \right]\eta^{\a\b}\p_\a X^m \p_\b X^n =-\frac{1}{2}\e^{\a\b}\p_\a X^m\p_\b X^n H_{smn}\,.
\end{multline}
Then, introducing the Christoffel symbols for background metric, $\Gamma^s_{mn}$, we can recast eq. (\ref{EL2}) in the form \cite{Callan:1989nz,Forini:2015mca}\footnote{In the case of vanishing Christoffel connection for the induced metric $P[g_{mn}]$ (as in the configurations analyzed in what follows), the rhs of eq.(\ref{EL3}) coincides with the trace of the so-called second fundamental form or extrinsic curvature  of the worldsheet $K^m_{\a\b}$ \cite{Forini:2015mca}.}
\ba
\label{EL3}
\eta^{\a\b}\left(\p_\a\p_\b X^s + \p_\a X^m\p_\b X^n \Gamma^s_{mn}\right)= -\frac{1}{2}\e^{\a\b}\p_\a X^m\p_\b X^n H^s\,_{mn}\,.
\ea
Equation (\ref{EL3}) has to be solved along with the Virasoro constraints
\ba
\label{Virasoro}
T_{\a\b} = \p_\a X^m \p_\b X^n g_{mn} -\frac{1}{2}\eta_{\a\b} \,
\eta^{\r\l}\p_\r X^m \p_\l X^n g_{mn} = 0\,.
\ea
In this subsection we  follow a similar approach to the one in
\cite{Frolov:2005ty,Bobev:2005cz,Frolov:2002av}. We refer to these works for more details of the semiclassical analysis of rotating and spinning strings in $AdS_5\times S^5$ and $\beta$-deformed $AdS_5\times S^5$.

Let us start our analysis by requiring the string to move on the surface described by
\ba
\label{surface}
\Sigma = \left\{\theta= \frac{\pi}{2}\,, \quad \m_1 = \sin\varphi\, \quad \m_2 = \cos\varphi\,, \quad \m_3 = 0\,,\quad x_1\,,x_2\,,\phi\,,\psi = \text{const}\right\}\,,
\ea
where the $\m_i$ coefficient are defined  in eq. (\ref{mu_i}).
Using the marginal deformed solution in eq. (\ref{sol1}) and angular coordinates in eq. (\ref{anglephi}),\footnote{Since the $\psi$-direction is frozen, the $\phi_i$ angles are defined as $\phi_1 = -\varphi_1$\, and $\phi_2 = \varphi_1 +\varphi_2$.} we can write down the induced metric on the surface in eq. (\ref{surface}) as
\ba
\label{metricSpinning}
\mathrm{d}s^2_{10}\big|_\Sigma&=& -r^2\mathrm{d}t^2 + \frac{\mathrm{d}r^2}{ r^2 f(r)}+\mathrm{d}\varphi^2 +  G\left(\sin^2\varphi \, \mathrm{d}\phi_1^2  + \cos^2\varphi \,\mathrm{d}\phi_2^2\right)\,,
\ea
where 
\ba
\label{Spinndef}
\quad f(r) = 1-\frac{Q^6}{r^6}\,, \quad G^{-1} = 1+ \frac{\g^2}{4}\sin^2 2\varphi\,,\quad g_0 = \sin^2\varphi  \cos^2 \varphi\,.
\ea
Moreover, for the presents ansatz the $B$-field is given by
\ba
\label{BSpinning}
B\big|_\Sigma&=& -\g g_0 G d\phi_1 \wedge d \phi_2\,.
\ea
Using the ansatz in eq. (\ref{surface}), we can derive the equations of motions for the fundamental string modes as in eq. (\ref{EL3})
\ba
\label{EOM}
&&\eta^{\a\b}\left(\p_\a\p_\b t +2\G^t_{tr} \p_\a t \p_\b r\right)=0\,,\nb\\
&&\eta^{\a\b}\left(\p_\a\p_\b r +\G^r_{rr} \p_\a r \p_\b r+\G^r_{tt} \p_\a t \p_\b t\right) =0\,,\nb\\
&&\eta^{\a\b}\left(\p_\a \p_\b \varphi +\G^\varphi_{\phi_1\phi_1} \,\p_\a \phi_1\p_\b \phi_1 + \G^\varphi_{\phi_2\phi_2}\,\p_\a \phi_2\p_\b \phi_2\right) +\p_\varphi B_{\phi_1\phi_2}\, \e^{\a\b}\p_\a \phi_1\p_\b \phi_2=0\,,\nb\\
&&\eta^{\a\b} \left(\p_\a\p_\b \phi_1 +2\G^{\phi_1}_{\phi_1 \varphi}\,\p_\a \phi_1\p_\b \varphi \right) - \e^{\a\b}\p_\a \left(B_{\phi_1\phi_2} \,\p_\b \phi_2 \right)g^{\phi_1\phi_1} = 0\,,\nb\\
&&\eta^{\a\b} \left(\p_\a\p_\b \phi_2 +2\G^{\phi_2}_{\phi_2 \varphi}\,\p_\a \phi_1\p_\b \varphi \right) + \e^{\a\b}\p_\a \left(B_{\phi_1\phi_2} \,\p_\b \phi_1 \right)g^{\phi_2\phi_2} = 0\,,
\ea
We further choose to place the string at the minimal radius $r=Q$ and with $\varphi = \frac{\pi}{4}$. We find \footnote{Notice that working around the cigar tip, we formally consider string worldsheet at constant $r= Q +\varepsilon$ and must take the $\varepsilon\to 0$ limit only at the end of the geometrical calculations in order to circumvent any coordinate singularities. See \cite{Bigazzi:2023oqm} for a similar worldsheet analysis. }
\ba
\G^r_{tt} = 0\,,\quad \G^{\phi_1}_{\phi_1\varphi} = -\G^{\phi_2}_{\phi_2\varphi} =1\,,\quad \G^\varphi_{\phi_1\phi_1} = -\G^\varphi_{\phi_2\phi_2} =-\frac{2}{4+\g^2}\,.
\ea
Hence, a solution for eq. (\ref{EOM}) is then given by 
\ba
\label{EOMsol}
t = \kappa\t\,, 
\quad \phi_i = \o_i\t +m_i \s \quad (i = 1,2)\,,\quad r = Q\,, \quad\varphi = \frac{\pi}{4}\,,
\ea
with the “$\varphi$” equation prescribing
\ba
\label{EOMsolC}
\quad \o_1^2 -m_1^2 = \o_2^2 -m_2^2\,.
\ea
Moreover, the solution in eq. (\ref{EOMsol}) has to satisfy the Virasoro constraints in eq. (\ref{Virasoro}), namely
\ba
&&T_{\t\t} = 
-\frac{ Q^2}{2}\kappa^2+ \frac{G}{4}\left(\o_1^2+\o_2^2 +m_1^2+m_2^2\right)\equiv 0\,,\nb\\
&&T_{\t\s} = \frac{G}{2} \left(\o_1\,m_1 +\o_2 \,m_2\right)\equiv 0\,,
\ea
which further fix
\ba
\label{Virasoro2}
&&\o_1 = \o_2 = \o\,,\quad m_1 = -m_2 = -m\,,\quad 
\kappa^2 Q^2 = 
G\left(\o^2+m^2\right)\,.
\ea
We can now account for the conserved charges corresponding to the $t$, and $\phi_1$ and $\phi_2$ variables, \textit{i.e.} the Energy and angular momenta $J_1$, $J_2$. The first step consists in writing down the canonical momentum density
\ba
\label{ConjMomenta}
\Pi^\tau_m = \frac{\d\mathcal{L}}{\d\p_\t X^m} = \frac{1}{2\pi}\left(\p_\t X^n g_{mn} +\p_\s X^n B_{mn}\right)\,,
\ea
Thus, using eq. (\ref{ConjMomenta}) we have that
\ba
\label{Pim}
&&\Pi^\t_t = -\frac{1}{2\pi}Q^2 \p_\t t = -\frac{\sqrt{2\l}}{2\pi}Q^2\kappa\,,\nb\\
&& \Pi^\t_{\phi_1} = \frac{1}{2\pi}\frac{G}{2} \left(\p_\t \phi_1 - \frac{\g}{2}\p_\s \phi_2 \right)  = \frac{\sqrt{2\l}}{2\pi} \frac{G}{2} \left(\o -\frac{\g}{2}m \right)\,,\nb\\
&& \Pi^\t_{\phi_2} = \frac{1}{2\pi} \frac{G}{2} \left(\p_\t \phi_2 +  \frac{\g}{2}\p_\s \phi_1 \right)  = \frac{\sqrt{2\l}}{2\pi}\frac{G}{2}  \left(\o -\frac{\g}{2}m \right)\,,
\ea
where we have restored the $AdS$ radius $\ell$ and $\alpha^\prime$, and used the relation in eq. (\ref{tHooft}) for the 't Hooft coupling $\l$.
 Hence, we can finally compute the conserved charges associated with the Poincaré invariance
 \ba
 P_m = \int_{0}^{2\pi} \mathrm{d}\s\, \Pi^\tau_m\,.
 \ea
In particular using eq. (\ref{Pim}) we have
\ba
\label{EJJ}
&&E= \sqrt{2\l} \, \mathcal{E}= -P_0 = \frac{\sqrt{2\l}}{2\pi}\int_{0}^{2\pi} \mathrm{d}\s Q^2\kappa = \sqrt{2\l} Q^2 \kappa\,,\nb\\
&& J_1 = J_2 = \sqrt{2\l}\, \mathcal{J}_1 = \frac{\sqrt{2\l}G}{4\pi}\int_{0}^{2\pi} \mathrm{d}\s \left(\o -\frac{\g}{2}m \right) =\sqrt{2\l}\,\frac{G}{2} \left(\o -\frac{\g}{2}m \right)\,,\nb\\
&&\mathcal{J} = \mathcal{J}_1+\mathcal{J}_2\,.
\ea
Using the Virasoro constraints in eq. (\ref{Virasoro2}) and expressions in eq. (\ref{EJJ}), we can then relate the string energy as
\ba
E = \sqrt{2\l}Q\sqrt{\mathcal{J}^2 + (m+\frac{\g}{2}\mathcal{J})^2}\,.\label{energy10}
\ea
This result is closely related to those in \cite{Frolov:2005ty} for the circular  string spinning in two directions  of $\beta$-deformed  $AdS_5\times S^5$ background.

Notice that the result above mixes the phenomena of confinement (characterised by the parameter $Q$) and the beta-deformation (characterised by the parameter $\gamma$). The spinning strings allow us to observe, using the energy in eq.(\ref{energy10}) a superposition between the two effects. One effect is typical of the IR of the QFT, the other is associated with the UV of it.

\subsection{Closing discussion on field theory and holography}
In this short section we  collect some comments and discuss aspects of the Physics of our models. Part of the discussion is direct consequence of the calculations in the previous sections. Some other comments apply to  the QFT dual to the backgrounds of eqs.(\ref{metric-ARxS5})-(\ref{RR-S5}), namely the QFT without marginal deformations of any kind.
\\
To begin with one may ask what is the deformation on the QFT side that triggers the RG-flow ending in a gapped and confining system.
In the dual of eqs.(\ref{metric-ARxS5})-(\ref{RR-S5}), we can perform holographic renormalisation. This shows that the deformation introduced by the one-form ${\cal A}$ in eq.(\ref{metric-ARxS5}) is dual to a VEV for an operator of dimension three (the needed UV-expansions in a related system are performed in \cite{Fatemiabhari:2024lct}). This operator needs to be a current as it is dual to a vector in the bulk. It also breaks $SO(6)_R$-symmetry into $U(1)^3$ R-symmetry. This deformation acts on the metric and on the five form. A Kaluza-Klein decomposition of Type IIB on the five sphere indicates that a possible operator with these characteristics is $O_3=$~Tr~$ \lambda\gamma_\mu\lambda$, a descendant from a chiral operator. This can be read for example, from Table 7, page 50 of the review \cite{DHoker:2002nbb}.
In other words, the deformation is via a VEV
of the current $J_\mu=$~Tr~$ \lambda\gamma_\mu\lambda $, choosing among the four  fermions a combination preserving $U(1)^3$ in $SU(4)_R$. 
\\
After the marginal deformation, characterised by the parameter $\gamma$ in eq.(\ref{sol1}), the dynamics is ruled by the Lagrangian in Section \ref{sec:marginal_N_4}, with the star-product between fields in eq.(\ref{beta_star_product}). This Lagrangian, in the presence of a VEV for the dimension-three operator drives the dynamics.
\\
The holographic approach is powerful and interesting because it shows the applicability of this logic to systems that are not described by a Lagrangian.  Indeed, as shown in \cite{Chatzis:2024top}, \cite{Chatzis:2024kdu}, \cite{Castellani:2024pmx}, the same deformation works for non-Lagrangian CFTs. After the marginal deformation (if possible), the dynamics works similarly.
\\
\\
Let us comment on the scales in our systems. From the field theory side we have a 4d SCFT, which implies (at least) a $U(1)_R$ symmetry. This theory is compactified on a  circle of fixed radius $R$. The radius is related to the gravity parameter  $R=\frac{1}{3Q}$, as expressed in eq.(\ref{RQ}). This introduces a scale in the SCFT, breaks conformality and an RG-flow ensues. Excitations of masses $m\sim\frac{1}{R}$ arise. Twisting, that is mixing the $U(1)_R$-symmetry with translations in the periodic dimension $\phi$ of size $R$, allows the QFT to preserve some amount of SUSY. The presence of massles fermions, completing the 3d ${\cal N}=2$ vector multiplet is the content of eq.(\ref{Bspecturm}). These display a SUSY spectrum (after spectral flow takes place). The massive modes have masses $m\sim \frac{n}{R}$. The marginal $\gamma$-deformation does not change this spectrum, but can change the interaction of the massive modes. The integration-out of the massive modes leads to a Chern-Simons term. This procedure is subtle, involving a  tricky regularisation--see \cite{Cassani:2021fyv},\cite{Kumar:2024pcz}--and relies heavily on the amount of SUSY preserved. In the case for which the Chern-Simons coefficient $k=N$, the QFT confines and has a unique vacuum.
\\
The holographic approach to these systems is again very powerful. On the one hand, the integration-out of modes is reflected by the shrinking-character of the $\phi$-direction. In fact $g_{\phi\phi}\sim r^2 f(r)+{\cal A}^2$. This vanishes at $r=Q$, indicating that as we move towards the end of the space, excitations on the circle become heavier and difficult to produce. This process does not depend on SUSY.
The integration out of modes is subtle. The lightest states (associated with glueballs of the QFT) have a mass degenerate wit the confinement scale $\Lambda_{YM}\sim Q$. The lack of separation between the IR QFT and the UV completion (KK modes) is a feature afflicting the supergravity approximation to the dynamics.
\\
\\
The confining properties of the solution are analysed in Section \ref{section-wilson1}. With little effort one can extend this to the case of SUSY breaking solutions. In fact, choosing the function $f(r)= 1-\frac{Q^6}{r^6} -\frac{\hat{\mu}}{r^4}$, with $\hat{\mu}= r_*^4 -\frac{Q^2}{r_*^2}$, the space ends smoothly at $r=r_*$. In this case the one-form ${\cal A}=Q^3(\frac{1}{r^2}-\frac{1}{r_*^2})$. Various calculations in the paper follow similarly. The field theory approach is not possible (as it relies on SUSY), but the holographic dual indicates that the IR Physics is qualitatively similar.
\\
An interesting question is if it would be possible to retain gapped properties (smooth ending of the space at $r=r_*$), but in a non-confining situation. We believe this is not possible in the present system. A singular behaviour, either at small or large values of the radial coordinate would be needed. As discussed in \cite{Faedo:2017fbv}, the desired phenomenon occurs (gapped spectrum with perimeter law for Wilson loops). This is a consequence of the need to lift the system to eleven dimensions. We leave for future studies this kind of dynamics. Let us now present some conclusions. 

\section{Conclusions and future lines of study}\label{concl}
The conclusions drawn from this work span both  geometrical aspects and quantum field theory (QFT) observables, providing new connections between string theory and gauge theories using holography. Through the two main parts, Geometry and Quantum Field Theory duals—this study contributes to the understanding of simple background geometries and their field theory counterparts.

A summary of the paper goes as follows: in the first part we revisited the seed background originally written by Anabal\'on and Ross \cite{Anabalon:2021tua}, which serves as the seed for the new solutions generated using TsT (T-duality, shift, and T-duality) transformations. These transformations, applied with different choices of $U(1)$ symmetries in the five-sphere and in the compact direction of the QFT, have led to some new type IIB backgrounds. One is the $\beta$-deformed background, where the $U(1)$ symmetries are embedded as R-symmetries inside $S^5$. The second is the dipole background, where one of the $U(1)$ symmetries is inside the five-sphere and the other is associated with the compact direction of the UV-QFT. This distinction is central to the structure of the resulting quantum field theories. 

In the Appendix, we also introduced two additional backgrounds: a non-commutative deformation and its dipole counterpart. While these solutions are presented with detailed calculations of the associated quantised charges and geometric invariants, their full exploration is reserved for future work, as they lie outside the primary focus of this study.

In the second part, we turned to the dual QFTs associated with these geometric backgrounds, shedding light on how these deformations influence observables in the quantum theory. A perturbative analysis of the compactified QFT on $S^1$ with an R-symmetry twist reveals interesting behaviours through the use of the $\star$-product in the $\beta$-deformed and dipole-deformed field theories. These deformations introduce a  parameter, $\gamma$, which  alters observables involving KK modes. Our analysis emphasizes the impact of this parameter on the dynamics.

By analyzing the Wilson, 't Hooft, and Polyakov loops, we demonstrate that electric quarks in these backgrounds exhibit confinement, with a breaking of the $\mathbb{Z}_N$ symmetry. This would have been difficult to uncover without the use of holographic dual backgrounds. Moreover, the calculation of the entanglement entropy in the deformed backgrounds, supports the argument for confinement.

Of  particular note is the result of the flow central charge, which serves as an energy-dependent measure of the degrees of freedom in the dual quantum field theories. This quantity provides an insightful window into the  flow across dimensions of these QFTs, capturing the transition from zero degrees of freedom in the low-energy, gapped phase to the degrees of freedom corresponding to the UV conformal field theory. In the case of the dipole-deformed background, the central charge diverges in the UV, signaling the presence of a non-local theory that may not be describable by traditional field-theoretic means. This suggests the existence of more exotic UV behavior, possibly involving stringy or other non-local effects.

 The analysis of semiclassical strings further supports these findings. By studying the behavior of heavy operators in the deformed QFTs, we find that these strings exhibit characteristics  that mix  both confinement and the effects of marginal deformations. 

Let us propose some topics for future investigation: \begin{itemize} 
\item{An intriguing direction would be to apply TsT transformations (or more general Yang-Baxter deformations in a similar fashion to \textit{e.g.} \cite{vanTongeren:2015uha,vanTongeren:2016eeb,Araujo:2017jkb}) to other related systems, further exploring the interaction between massless modes and KK-modes in the context of dynamical observables. This could improve the understanding of the role of KK-modes in holographic setups.} 
\item{It would also be valuable to explore one-loop corrections to the Wilson loops computed in our background. Additionally, investigating the potential Hagedorn behavior of our QFTs would be of interest, building on the insights from recent studies \cite{Bigazzi:2023oqm, Bigazzi:2023hxt,  Bigazzi:2024biz, Bigazzi:2024sjy}. This could provide further information on the thermal properties and phase transitions of these theories.} 
\item{Another worthwhile pursuit would be the construction of Anabal\'on-Ross-like deformations of $AdS$ geometries in different dimensions. For instance, similar deformations could be considered in families of $AdS_7$ \cite{Cremonesi:2015bld, Filippas:2019puw}, $AdS_6$ \cite{DHoker:2017zwj, Legramandi:2021uds}, AdS$_5$ \cite{Chatzis:2024top, Chatzis:2024kdu, Nunez:2019gbg,Itsios:2017cew}, and $AdS_4$ \cite{Assel:2011xz, Akhond:2021ffz}. These generalizations could shed light on the RG flows and dual QFTs in various dimensions.} \end{itemize}

In conclusion, this work offers an exploration of both geometric and field-theoretic aspects of TsT-deformed backgrounds, with insights into the behavior of quantum field theories under such deformations. The holographic duality provides a powerful framework for analyzing these systems, allowing us to compute a wide range of observables that shed light on the underlying physics. The study of confinement, symmetry breaking, and entanglement entropy underscores the rich structure of these QFTs, while the flow central charge and semiclassical string analysis point towards interesting connections between geometry and field theory. 
Future work may also explore the non-commutative and dipole-deformed solutions in greater depth, potentially giving new understanding of non-local theories and their holographic duals.

\section*{Acknowledgments} For discussions, comments on the manuscript and for sharing their ideas with us, we wish to thank: 
Francesco Bigazzi, Riccardo Borsato, Aldo Cotrone, S. Prem Kumar, Massimo Porrati, Ricardo Stuardo, Stijn van Tongeren.
F.C. wants to thank the Swansea University for the hospitality during the research period that led to this work.
We are supported by the grants ST/Y509644-1, ST/X000648/1 and ST/T000813/1.


{\bf Open Access Statement}---For the purpose of open access, the authors have applied a Creative Commons 
Attribution (CC BY) licence to any Author Accepted Manuscript version arising.

{\bf Data Statement}---no new data were generated in this work.

\appendix
\section{TsT: details of the generating technique}\label{TsTdetails}
In this appendix we clearly derive some of the formulas obtained in the main body of the paper.
\subsection{Abelian T-duality}
\label{Tstdetails_Tdual}
Following a similar approach to the one in \cite{Kelekci:2014ima}, it is also useful to recall Abelian T-duality transformation rules in the flat-index formalism. Let us consider a general ten-dimensional metric with a $U(1)$ isometry, then the latter can be written in the form
\bea
\mathrm{d}s^2_{10} = \mathrm{d}s^2_{9} + e^{2C}(\mathrm{d}z + A_1)^2\,.
\eea
Moreover, we can decompose further the fields of the related type II supergravity solution as
\bea
B = B_2 + B_1\wg \mathrm{d}z\,,\qquad F = F_{\perp} + F_{\parallel}\wg e^{C}(\mathrm{d}z + A_1)\,,
\eea
and in addition, consider a non-zero dilaton field $\Phi$. Then, following the conventions for Abelian T-duality transformations in \cite{Kelekci:2014ima}, we get the T-dual solution will be given by
\bea
\label{Tdual}
&&\mathrm{d}s^2_{10}\,^{(2)} = \mathrm{d}s^2_{9} + e^{-2C}(\mathrm{d}z - B_1)^2\,,\nonumber\\
&&B^{(2)} = B_2 - A_1\wg (\mathrm{d}z - B_1)\,, \qquad \Phi^{(2)} = \Phi-C\,,\nonumber\\
&&F^{(2)}  = F^{(2)}_{\perp} + F^{(2)} _{\parallel}\wg e^{-C}(\mathrm{d}z - B_1)\,,\quad  F^{(2)}_{\perp} =  e^{C}F_{\parallel}\,,\quad F^{(2)}_{\parallel} =  e^{C}F_{\perp}\,.
\eea
The results in (\ref{Tdual}) turn out to be very helpful in performing the TsT transformation of type II supergravity analyzed in this work.

\subsection{TsT transformation on \texorpdfstring{$AdS_5\times S^5$}{AdS5xS5}}\label{TsTdetails_LM}
In this appendix, referring closely to \cite{Lunin:2005jy}, we want to briefly review the original Lunin-Maldacena $\beta$-transformation in $AdS_5\times S^5$. Let us start by displaying the field content of the $AdS_5\times S^5$ type IIB supergravity solution, which reads
\bea
\label{ds2LMAdS}
&&\mathrm{d}s^2_{10}=\mathrm{d}s^2_{AdS_5}
+\sum_{i=1}^3 \mathrm{d}\mu_i^2 + 9\frac{\m_1^2\m_2^2\m_3^2}{g_0}\mathrm{d}\psi^2+(\m_1^2+\m_2^2)\left( D\varphi_1 + \frac{\m_2^2}{\m_1^2+\m_2^2}D\varphi_2\right)^2+\frac{g_0}{\m_1^2+\m_2^2} D\varphi_2^2\,,\nonumber\\
&&e^{2\Phi}= 1\,,\qquad B = 0\,,\nonumber\\
&&F_5 = 4 \left(\omega_{AdS_5}+\omega_{S^5} \right)\,, \quad \omega_{S^5}= 3 \mathrm{d}\omega_1\wg  \mathrm{d}\psi \wg \mathrm{d}\varphi_2 \wg\mathrm{d}\varphi_1\,,\quad  \mathrm{d}\omega_1 =  \sqrt{\m_1^2+\m_2^2}\,\m_1\m_2\m_3\, \mathrm{d}\theta\wg \mathrm{d}\varphi\,, \nonumber\\
\eea
where we the $\m_i$ are defined as in eq.(\ref{mu_i}) and
\bea
g_0 &=& \m_1^2\m_2^2+ \m_1^2\m_3^2+\m_2^2\m_3^2\,,\nonumber\\
 D\varphi_1 &=& \mathrm{d}\varphi_1 +\left(3\frac{\m_2^2\m_3^2}{g_0}-1\right)\mathrm{d}\psi\,,\quad D\varphi_2 = \mathrm{d}\varphi_2 +\left(3\frac{\m_1^2\m_2^2}{g_0}-1\right)\mathrm{d}\psi \,.
\eea

Now, looking at the prescription for Abelian T-dualities in \ref{Tstdetails_Tdual}, we can apply step by step the TsT-transformation algorithm on the $AdS_5\times S^5$ solution as in \cite{Lunin:2005jy}. Let us start by performing a T-duality with respect to one of the $U(1)$ isometries of the metric in (\ref{ds2LMAdS}), say associated with the $\varphi_1$ direction. In particular, before applying the T-transformation, it is useful to look at the expressions in (\ref{Tdual}), and recognize the following quantities
\bea
&&\mathrm{d}z = \mathrm{d}\varphi_1\,\quad A_1 = \left(3\frac{\m_2^2\m_3^2}{g_0}-1\right)\mathrm{d}\psi+ \frac{\m_2^2}{\m_1^2+\m_2^2}D\varphi_2 \,,\nonumber\\
&&\mathrm{d}s^2_{9}= \mathrm{d}s^2_{AdS_5}+\sum_{i=1}^3 \mathrm{d}\mu_i^2 + 9\frac{\m_1^2\m_2^2\m_3^2}{g_0}\mathrm{d}\psi^2+\frac{g_0}{\m_1^2+\m_2^2} D\varphi_2^2\,,\nonumber\\
&&e^{2C} =(\m_1^2+\m_2^2)\,,\quad B_1 = B_2 = 0\,, \quad F_{\perp} =4\omega_{AdS_5}\,,\quad  \quad F_{\perp} =4e^{-C}i_{\varphi_1}\omega_{S^5}\,,
\eea
where $i_{\varphi_1}\cdot$ is the interior product respect with the $\varphi_1$ direction.
Then, we can perform the T-duality along the $\varphi_1$ direction, obtaining
\bea
\label{T1}
&&\mathrm{d}s^2_{10}\,^{(2)} = \mathrm{d}s^2_{9} + e^{-2C}\mathrm{d}\varphi_1^2\,,\nonumber\\
&&B^{(2)} = - A_1\wg \mathrm{d}\varphi_1\,, \qquad \Phi^{(2)} = \Phi-C\,,\nonumber\\
&&F^{(2)}  =   F^{(2)}_{\perp} +F^{(2)} _{\parallel}\wg e^{-C}\mathrm{d}\varphi_1\,,\quad F^{(2)}_{\parallel} =  e^{C}4\, \omega_{AdS_5}\,,\quad F^{(2)}_{\perp} =  4\, i_{\varphi_1}\omega_{S^5}\,,
\eea
After that, we can perform the second step of the $\beta$-transformation by shifting the $\varphi_2$ as
\bea
\label{shift}
\varphi_2 \to \varphi_2 + \gamma\varphi_1\,,
\eea
Hence under this shift (\ref{shift}) the metric in (\ref{T1}) transforms to 
\bea
\label{Ts}
&&\mathrm{d}s^2_{10}\,^{(3)} = \mathrm{d}s^2_{AdS_5}
+\sum_{i=1}^3 \mathrm{d}\mu_i^2 + 9\frac{\m_1^2\m_2^2\m_3^2}{g_0}\mathrm{d}\psi^2+ \frac{g_0}{\m_1^2+\m_2^2} (D\varphi_2+\gamma\mathrm{d}\varphi_1)^2
 + \frac{1}{\m_1^2+\m_2^2}\mathrm{d}\varphi_1^2\,,\nonumber\\
 \eea
and the solution (\ref{T1}) can be recast as
\bea
\label{Ts2}
&&\mathrm{d}s^2_{10}\,^{(3)} = \mathrm{d}s^2_{AdS_5}
+\sum_{i=1}^3 \mathrm{d}\mu_i^2 + 9\frac{\m_1^2\m_2^2\m_3^2}{g_0}\mathrm{d}\psi^2+ \frac{g_0\,G}{\m_1^2+\m_2^2} D\varphi_2^2
 + \frac{G^{-1}}{\m_1^2+\m_2^2}\left(\mathrm{d}\varphi_1 +A_1^{(3)} \right)^2\,,\nonumber\\
 && A_1^{(3)} = \gamma g_0 G\, D\varphi_2\,,\qquad G^{-1} = 1+\gamma^2 g_0\,,\quad e^{2C^{(3)}} =\frac{G^{-1}}{\m_1^2+\m_2^2}\,, \nonumber\\
&&B^{(3)} = - A_1\wg \mathrm{d}\varphi_1\,, \qquad \Phi^{(3)} = \Phi-C\,,\nonumber\\
&&F^{(3)}  = F^{(3)} _{\perp}+ F^{(3)} _{\parallel}\wg e^{C^{(3)}}\left(\mathrm{d}\varphi_1 +A_1^{(3)} \right)\,,\nonumber\\
&&F^{(3)}_{\parallel} =  e^{-C^{(3)}}4\,\left(\omega_{AdS_5} +\gamma\, i_{\varphi_2}i_{\varphi_1}\omega_{S^5}\right)\,,\quad F^{(3)}_{\perp} = 4
\left(G\,i_{\varphi_1}\omega_{S^5}-\omega_{AdS_5}\wg A_1^{(3)}\right)\,.
\eea
Finally, using again the prescription in (\ref{Tdual}), we have to T-dualize back (\ref{Ts2}) along the $\varphi_1$ direction, and we end up on the type IIB solution given by
\bea
\label{TsT}
&&\mathrm{d}s^2_{10}\,^{(4)} = \mathrm{d}s^2_{AdS_5}
+\sum_{i=1}^3 \mathrm{d}\mu_i^2 + 9\frac{\m_1^2\m_2^2\m_3^2}{g_0}\mathrm{d}\psi^2+ \frac{g_0\,G}{\m_1^2+\m_2^2} D\varphi_2^2
 + G(\m_1^2+\m_2^2)\left(\mathrm{d}\varphi_1 +A_1 \right)^2\,,\nonumber\\
&&B^{(4)} =-A_1^{(3)}\wg (\mathrm{d}\varphi_1+A_1)\,, \qquad \Phi^{(4)} =\Phi+\frac12 \log G\,,\nonumber\\
&&F^{(4)}  = F^{(4)} _{\perp}+ F^{(4)} _{\parallel}\wg e^{-C^{(3)}}\left(\mathrm{d}\varphi_1 +A_1 \right)\,,\nonumber\\
&&F^{(4)}_{\perp} = 4\,\left(\omega_{AdS_5} + \gamma\, i_{\varphi_2}i_{\varphi_1}\omega_{S^5}\right) \,,\quad F^{(4)}_{\parallel} = 4e^{C^{(3)}}
\left(G\,\omega_{S^5}-\omega_{AdS_5}\wg A_1^{(3)}\right)\,,
\eea
that can be expressed as
\bea
\label{TsT2}
&&\mathrm{d}s^2_{10} = \mathrm{d}s^2_{AdS_5}
+\sum_{i=1}^3 \mathrm{d}\mu_i^2 + 9\frac{\m_1^2\m_2^2\m_3^2}{g_0}\mathrm{d}\psi^2+ \frac{g_0\,G}{\m_1^2+\m_2^2} D\varphi_2^2
 + G(\m_1^2+\m_2^2)\left( D\varphi_1 + \frac{\m_2^2}{\m_1^2+\m_2^2}D\varphi_2\right)^2\,,\nonumber\\
&&B =\gamma g_0\,G D\varphi_1\wg D\varphi_2\,, \qquad \Phi^{(4)} =\Phi+\frac12 \log G\,,\nonumber\\
&&F_5  = 4\left(\omega_{AdS_5} + G\,\omega_{S^5}\right)\,,\quad F_3 = 4 \gamma \,\,i_{\varphi_2}i_{\varphi_1}\omega_{S^5}\,, \quad F_7 = 4 \omega_{AdS_5} \wg B\,,
\eea
which coincides with the Lunin-Maldacena solution for the $\beta$-deformed $AdS_5\times S^5$ \cite{Lunin:2005jy}.

\subsection{Details on the \texorpdfstring{$\beta$}{beta}-transformed RR forms in \texorpdfstring{$\widehat{AdS_5}\times \widehat{S^5}$}{AdS5xS5}}
\label{RR_form_TsT_details}
Here, in this appendix, we give some details on the $\beta$-transformed RR forms in the $\widehat{AdS_5}\times \widehat{S^5}$ background in eq. (\ref{RR-S5}). In particular, we provide a step by step calculation of the TsT transformations acting on the electric part of the five-form $G_5$ defined in eq. (\ref{RR-S5}). The result will be easily extendable to the magnetic part $\star_{10}G_5$. Using the definitions in eq. (\ref{RR-S5}), $G_5$ can be written as
\begin{multline}
\label{G5}
G_5= -4 \text{vol}_5-2Q^3 J_2\wedge \mathrm{d}t\wedge \mathrm{d}x_1 \wedge \mathrm{d}x_2\,,\\
= -4 \text{vol}_5 -2Q^3  \bigg[ \sum_{i=1}^3 \mu_i \mathrm{d}\mu_i \wg D\psi +\left(\mu_2 \mathrm{d}\mu_2-\mu_3 \mathrm{d}\mu_3\right)\wg \mathrm{d}\varphi_2 + \left(\mu_2 \mathrm{d}\mu_2-\mu_1 \mathrm{d}\mu_1\right) \wg \mathrm{d}\varphi_1\bigg]\wg\mathrm{d}t\wedge \mathrm{d}x_1 \wedge \mathrm{d}x_2\,,\\
= -4 \text{vol}_5 + \b_4 \wg D\psi +\g_4 \wg \mathrm{d}\varphi_2 + \a_4 \wg \mathrm{d}\varphi_1\,,
\end{multline}
where, for a shorter notation, we have introduced the following definition
\bea
 &&\text{vol}_5 \equiv \mathrm{d}t\wedge \mathrm{d}x_1\wedge \mathrm{d}x_2\wedge\mathrm{d}r\wedge \mathrm{d}\phi\,,\quad \a_4 \equiv 2Q^3 \left(\mu_2 \mathrm{d}\mu_2-\mu_1 \mathrm{d}\mu_1\right)\wg \mathrm{d}t\wedge \mathrm{d}x_1 \wedge \mathrm{d}x_2\,, \nb\\
 &&\b_4 \equiv 2Q^3 \sum_{i=1}^3 \mu_i \mathrm{d}\mu_i \wg \mathrm{d}t\wedge \mathrm{d}x_1 \wedge \mathrm{d}x_2\,,\quad \g_4 \equiv 2Q^3\left(\mu_2 \mathrm{d}\mu_2-\mu_3\mathrm{d}\mu_3\right)\wg\mathrm{d}t\wedge \mathrm{d}x_1 \wedge \mathrm{d}x_2\,.\nb\\
\eea
Moreover, having in mind the expressions in eq. (\ref{Tdual}), we can massage $G_5$ in eq. (\ref{G5}) as
\ba
\label{G52}
&&G_5 = G_{5\,\perp}+ \a_4 \wg (\mathrm{d}\varphi_1+ A_1)\,,\nb\\
&&G_{5\,\perp} =-4 \text{vol}_5 + \b_4 \wg D\psi +\g_4 \wg \mathrm{d}\varphi_2 -\a_4 \wg A_1\,,\nb\\
&&A_1 = \left(3\frac{\m_2^2\m_3^2}{g_0}-1\right)D\psi + \frac{\m_2^2}{\m_1^2+\m_2^2}D\tilde\varphi_2\,.
\ea
Then, at this point we can proceed with the T-transformation respect with $\varphi_1$, obtaining a four-form and a six-form
\ba
&&G_4^{(2)} = \a_4\,,\quad G_6^{(2)} =  G_{5\,\perp}\wg\mathrm{d}\varphi_1\,,
\ea
and after the shift $\varphi_2 \to \varphi_2 +\g \varphi_1$, these read
\ba
&&G_4^{(3)} = \a_4\,,\nb\\
&&G_6^{(3)} = G_{5\,\perp} \wg\left(\mathrm{d}\varphi_1+A_1^{(3)}\right) - G_{5\,\perp} \wg A_1^{(3)}\,,
\ea
where, in analogy with eq. (\ref{Ts2}),  $A_1^{(3)}$ is defined as 
\ba
A_1^{(3)} = \gamma g_0 G\, D\tilde\varphi_2\,.
\ea
Finally, we can T-dualize again along the $\varphi_1$ direction. The electric form content of the TsT-transformed type IIB theory is then given by
\ba
&&G_5^{(4)} = G_5\,,\nb\\
&&G_7^{(4)} = - G_{5\,\perp} \wg A_1^{(3)}\wg  (\mathrm{d}\varphi_1+ A_1)\,.
\ea
Now, let us look closer at the above seven-form: indeed, using eq. (\ref{G5}) and eq. (\ref{G52}), the latter can be written as
\ba
G_7^{(4)} &= &- \gamma g_0 G\, G_{5\,\perp} \wg D\tilde\varphi_2\wg  \left(D\tilde\varphi_1 + \frac{\m_2^2}{\m_1^2+\m_2^2}D\tilde\varphi_2\right)\nb\\
&=&- \gamma g_0 G\, G_{5\,\perp} \wg D\tilde\varphi_2\w D\tilde\varphi_1 \nb\\
&=&  \gamma g_0 G\, \left(-4\text{vol}_5 +\b_4 \wg D\psi +\g_4 \wg \mathrm{d}\varphi_2 -\a_4 \wg A_1 \right) \wg D\tilde\varphi_1\wg D\tilde\varphi_2 \nb\\
&=&  \gamma g_0 G\, \left(-4\text{vol}_5 +\b_4 \wg D\psi +\g_4 \wg \mathrm{d}\varphi_2 + \a_4 \wg \varphi_1 \right) \wg D\tilde\varphi_1\wg D\tilde\varphi_2 \nb\\
&=&  \gamma g_0 G\, G_5 \wg D\tilde\varphi_1\wg D\tilde\varphi_2 \,,
\ea
obtaining the result in eq. (\ref{TsT2}). Notice that, performing a similar analysis for the magnetic RR-forms, we expect the total RR form content of the $\b$-dual theory to be
\ba
&&F_3 = \gamma\, i_{\varphi_2}i_{\varphi_1}\star_{10}G_5 = \frac{\g}{G}\, i_{\varphi_2}i_{\varphi_1}\star_{10\beta}G_5\,,\nb\\
&&F_5 = (1 + G \star_{10}) G_5 = (1 + \star_{10\beta}) G_5\,,\nb\\
&&F_7 = \gamma g_0 G\, G_5 \wg D\tilde\varphi_1\wg D\tilde\varphi_2 =G_5\wg B\,.
\ea

 \section{{Non-commutative deformation}}\label{sec-non-commutative}
{In this appendix,} we are interested in transformations of the $\widehat{AdS_5}\times \widehat{S^5}$ in eq. (\ref{metric-ARxS5}) involving a TsT performed along  two Minkowski directions. The latter will lead to a type IIB supergravity solution dual to a \textit{non commutative} gauge theory \cite{Maldacena:1999mh}. A simple motivation for this argument can be given as follows \cite{Lunin:2005jy, Seiberg:1999vs}. As we have observed in the examples in Section \ref{section-geometry}, the TsT procedure introduces for the deformed solution a $B$ field along the two $U(1)_{\Theta_i}$ directions (see \textit{e.g.} eqs. (\ref{sol1}) and (\ref{S5dip})). Hence, in this case, we expect a $B$-field extending along the $AdS$ directions of the deformed background. Then, the presence of the latter intuitively yields a non-commutativity (anti-symmetric) coupling modifying the effective metric for an open string theory as
\bea
g^{ij} \to G^{ij} +\theta^{ij}\,,\quad G^{ij} = \left(\left(g +B\right)^{-1}\right)^{(ij)}\,,\quad \theta^{ij} = \left(\left(g +B\right)^{-1}\right)^{[ij]}\,.
\eea
In what follows, we focus on two different possible non-commutative $\beta$ transformations of $\widehat{AdS_5}\times \widehat{S^5}$. First, we deform  choosing the TsT directions, $\Theta_i$, as two Minkowski coordinates, say $x_1$ and $x_2$. In the second case, we perform the TsT along the shrinking circle $\phi$ and the $x_1$ direction.

\subsection{First non-commutative  solution}\label{sec:Non_commutative_proper}
Let us start by expressing the metric in eq.(\ref{metric-ARxS5}) in the following useful way
\bea
\label{dsNC}
 &&\mathrm{d}s^2 _{10} =  \mathrm{d}s^2_{8} + r^2(\mathrm{d}x_1^2 + \mathrm{d}x_2^2)\,,\nb\\
 &&\mathrm{d}s^2_{8}=r^2 (-\mathrm{d}t^2 +  f(r)\mathrm{d}\phi^2) + \frac{\mathrm{d}r^2}{ r^2 f(r)} +  \sum_{i=1}^3 \mathrm{d}\mu_i^2 + \mu_i^2 \left( \mathrm{d}\phi_i + \mathcal{A} \right) ^2\,,
\eea
where the function $f(r)$ and the one-form $\mathcal{A}$ are given as in eqs.(\ref{metric-ARxS5}). Now, having the metric in the form of (\ref{dsNC}), we can identify $(x_1,x_2)$ as the two torus subspace along with to carry out the TsT transformation. Hence, let us perform the first T-duality along the $x_1$ direction, followed by the shift
\bea
x_2 \to x_2 + \g x_1\,,
\eea
and then T-dualize back on $x_1$. This process returns a deformed solution of the form
\bea
\label{NCS5}
& &\mathrm{d}s^2_{10}=\mathrm{d}s^2_{8} + G r^2\left(\mathrm{d}x_1^2+  \mathrm{d}x_2^2\right)\,,\nb\\
&&B =\gamma r^4 G\mathrm{d}x_1 \wg \mathrm{d}x_2\,, \qquad  \quad e^{2\Phi} = G\,,\nb\\
&&F_5  = \left(1 + \star_{10\beta}\right)G\,G_5\,,\quad F_3 =  \g \,\,i_{x_2}i_{x_1}G_5\,,\quad F_7 = \star_{10\beta}G_5 \wg B\,,
\eea
where, here we have defined the $G$-factor as
\bea
\label{NCG}
&& G^{-1} = 1 + \g^2r^4\,.
\eea
Moreover the Hodge-dual operator $\star_{10\beta}$ refers to the TsT-transformed metric in eq.(\ref{NCS5}), while the five-form $G_5$ is given as in eq.(\ref{RR-S5}) 
\bea
&&G_5 = -4r^3 \mathrm{d}t\wg \mathrm{d}x_1 \wg\mathrm{d}x_2\wg \mathrm{d}r\wg \mathrm{d}\phi\,- 2Q^3
J_2\wedge \mathrm{d}t\wedge \mathrm{d}x_1 \wedge \mathrm{d}x_2\,, \nb\\
 & & J_2=  \sum_{i=1}^3 \mu_i \mathrm{d}\mu_i \wedge \left(\mathrm{d}\phi_i +\mathcal{A}\right)\,.
\eea
In analogy with what has been done {in Section \ref{section-geometry}}, here we provide the charge quantization conditions for the possible emerging $D_p$-branes. In particular these read\footnote{Notice that in this case, there is not a $B\wg F_3$ term in eq.(\ref{QD3prime_NC}) since both $B$ and $F_3$ extend along non-compact directions.}
\bea
\label{QD3prime_NC}
Q_{D3}^\prime =\frac{1}{(2\pi)^4}\int_{\Sigma_5} \star_{10\beta}\, G\,G_5 =\frac{1}{(2\pi)^4}\int_{\Sigma_5} \star_{10} G_5 =  N\,,
\eea
where $\Sigma_5 = \left[\theta,\varphi, \phi_i\right]$, and 
\bea
\label{QD1prime_NC}
Q_{D1}^\prime =\frac{1}{(2\pi)^6}\int_{\Sigma_7}  \star_{10\beta} F_3 + B\wg \star_{10\beta} G_5 =\frac{1}{(2\pi)^6}\int_{\Sigma_7}  B\wg \star_{10\beta} G_5 - F_7 = 0\,.
\eea
%
Here we have $Q_{D5}^\prime =0$, as F$_3$ is electric. 
{An analogous argument can be given regarding the NS5-brane charge: as there is no a compact three cycle supporting $H_3$ (see the expression for $B$ in eq.(\ref{NCS5})), we can conclude that $Q_{\text{NS}5}^\prime =0$. }\\
In order to analyze the smoothness properties of the background in eq. (\ref{NCS5}), let us show the expression for the Ricci scalar
\bea
&& R=\frac{8 \gamma ^2 \left(Q^6 \left(4 \gamma ^2 r^4-3\right)+2 \gamma ^2 r^{10}+9 r^6\right)}{r^2\left(\gamma ^2 r^4+1\right)^2}\,,
\eea
which is regular at all $r$ finite values. The asymptotic IR and UV limits of the latter are given by
\bea
&&R\big|_{r=Q}=\frac{48 \gamma ^2 Q^4}{\gamma ^2 Q^4+1}\,,\quad R\big|_{r\to \infty} \sim 16 +O\left(r^{-3}\right)\,.
\eea

Below we display also other curvature invariants, such as the contractions of two Ricci tensors and the Kretschmann scalar evaluated at the minimal radius $r = Q$ and in the boundary limit. In particular these read respectively
\bea
&& R^{\m\n}R_{\m\n}\big|_{r=Q} = \frac{24 \left(17 \gamma ^4 Q^8+2 \gamma ^2 Q^4+9\right)}{\left(\gamma ^2 Q^4+1\right)^2}\,, \quad R^{\m\n}R_{\m\n}\big|_{r\to \infty}\sim 96 +O\left(r^{-3}\right)\,, \nb\\
&&R^{\m\n\r\s}R_{\m\n\r\s}\big|_{r=Q}=\frac{72 \left(7 \gamma ^4 Q^8+6 \gamma ^2 Q^4+7\right)}{\left(\gamma ^2 Q^4+1\right)^2}\,,\quad R^{\m\n\r\s}R_{\m\n\r\s}\big|_{r\to \infty} \sim 80 +O\left(r^{-3}\right)\,.\nb\\
\eea
The invariants in Einstein frame have asymptotics
\begin{eqnarray}
& & R\vert_{r\to \infty}\sim -\frac{2}{\sqrt{\g}r} +O\left(r^{-3}\right)\,,\\
& & R_{\mu\nu}R^{\mu\nu}\vert_{r\to \infty}\sim \frac{116}{\g r^2} +O\left(r^{-3}\right)\,,\\
& & R_{\mu\nu\rho\sigma}R^{\mu\nu\rho\sigma}\vert_{r\to \infty}\sim \frac{97}{\g r^2} +O\left(r^{-3}\right)\,.\label{ivarinatseinsteinNC}
\end{eqnarray}
Notice that also in this frame none of these diverge.\\
Let us also notice that since $G$ factor in eq. (\ref{NCG}) scales as $G\sim (\g^2 r^4)^{-1}$ in the large $r$ limit, it spoils the $AdS_5$ UV asymptotic behavior of the background. The latter feature is understandable by the fact that the present non-commutative TsT-transformation is equivalent to the introduction of an irrelevant operator in the dual field theory.

For the present deformed solution in eq.(\ref{NCS5}) we verified on Mathematica the type IIB Einstein, Maxwell and dilaton equations, as well as the Bianchi identities for the NS and RR fields.

\subsection{Second non-commutative solution}\label{sec:Non_commutative_dipole}
Finally, we focus on an intermediate  example, between the dipole and non commutative $\beta$-transformed solution in Subsections \ref{sec:dipole} and \ref{sec:Non_commutative_proper}. In doing that we  analyze a TsT of the background in eq.(\ref{metric-ARxS5}), where the T-duality transformations are performed along the shrinking compact direction $\phi$ and the shift along the spatial $x_1$ coordinate as
\bea
\label{NCshift}
x_1 \longrightarrow x_1 + \g \phi\,.
\eea
In the present case, before to T-dualize along the $\phi$ direction it is useful to rewrite the metric in eq.(\ref{metric-ARxS5}) as
\bea
\label{CP2NC}
& &\mathrm{d}s^2_{10}=\mathrm{d}s^2_{8}+ r^2 \mathrm{d}x_1^2 + \left(r^2f(r)+ \hat{\mathcal{A}}^2\right)\left( \mathrm{d}\phi +\frac{\hat{\mathcal{A}}}{r^2f(r)+ \hat{\mathcal{A}}^2}D\varphi\right)^2\,,\nb\\
&& \mathrm{d}s^2_{8} = r^2 (-\mathrm{d}t^2 + \mathrm{d}x_2^2 ) + \frac{\mathrm{d}r^2}{ r^2 f(r)} + \mathrm{d}s^2_{\mathbb{C}P^2} + \frac{r^2f(r)}{r^2f(r)+ \hat{\mathcal{A}}^2}D\varphi^2\,,\nb\\
&& \mathcal{D}\varphi = \mathrm{d}\varphi +\eta\,,
\eea
where again the $\mathrm{d}s^2_{\mathbb{C}P^2}$ is the Fubini-Study metric given as in eq.(\ref{CP2metric}) \cite{Pope:1980ub,Herrero:2011bk} and 
\bea
\label{hatA}
\hat{\mathcal{A}}  =  Q^3\left( \frac{1}{r^2}- \frac{1}{Q^2}\right)\,.
\eea
Now we can apply the T-dualities along $\phi$, interspersing them with the shift in eq.(\ref{NCshift}). Hence this procedure leads to the solution
\bea
\label{NCdipole}
& &\mathrm{d}s^2_{10}=\mathrm{d}s^2_{8}+ Gr^2 \mathrm{d}x_1^2 + G\left(r^2f(r)+ \hat{\mathcal{A}}^2\right)\left( \mathrm{d}\phi +\frac{\hat{\mathcal{A}}}{r^2f(r)+ \hat{\mathcal{A}}^2}D\varphi\right)^2\,,\nb\\
&&B =-\gamma G r^2\left(r^2f(r)+ \hat{\mathcal{A}}^2\right)\mathrm{d}x_1 \wg \left( \mathrm{d}\phi +\frac{\hat{\mathcal{A}}}{r^2f(r)+ \hat{\mathcal{A}}^2}D\varphi\right)\,, \quad e^{2\Phi} = G\,,
\eea
along with the RR-forms content given by
\bea
\label{RR_NC_dipole}
&&F_5  = \left( 1+ \star_{10\beta}\right)\tilde G_5\,,\nb\\
&& \tilde G_5 = -4r^3 \mathrm{d}t\wg \mathrm{d}x_1 \wg\mathrm{d}x_2\wg \mathrm{d}r\wg \mathrm{d}\phi\,- Q^3\mathrm{d}\eta
\wedge \mathrm{d}t\wedge \mathrm{d}x_1 \wedge \mathrm{d}x_2 + B\wg F_3\,,\nb\\
 &&F_3 =  -4\g r^3\mathrm{d}t\wg \mathrm{d}x_2\wg \mathrm{d}r\,,\quad F_7 = -4\g r^4f(r)G \mathrm{d}x_1\wg\mathrm{d}\phi \wg\text{vol}_{\mathbb{C}P^2}\wg D\varphi\,.\,
\eea
We have defined the $G$-function as
\bea
\label{NCdipole2}
&& G^{-1} = 1 + \g^2r^2\left(r^2f(r) + \hat{\mathcal{A}}^2\right) \,.
\eea
Let us now analyze the various D$_p$-brane charge quantization conditions. Since the fluxes $F_3$, $F_7$ and the $B$ field in eq. (\ref{RR_NC_dipole}) have at least one leg on non-compact cycle, the only non trivial D$_p$ (NS) brane charge is the one related to D3 branes. Indeed, setting the five-cycle $\Sigma_5 = \left[\mathbb{C}P^2,\varphi\right]$, the latter is given by
\bea
\label{QD3prime_NC_dipole}
Q_{D3}^\prime =\frac{1}{(2\pi)^4}\int_{\Sigma_5} \star_{10\beta}\,\tilde G_5 =\frac{1}{(2\pi)^4}\int_{\Sigma_5}4\text{vol}_{\mathbb{C}P^2} \mathrm{d}\varphi =  N\,.
\eea

The Ricci scalar reads
\bea
R =&&\frac{2 \gamma ^2}{r^2 \left( 2 \gamma ^2Q^4-\gamma ^2 Q^2 r^2-\gamma ^2 r^4-1\right)^2} \bigg[19 \gamma ^2 Q^{10}+22 \gamma ^2 Q^8 r^2-8 \gamma ^2 Q^6 r^4+\nb\\&&-6 Q^6-64 \gamma ^2 Q^4 r^6+23 \gamma ^2 Q^2 r^8+15 Q^2 r^4+8 \gamma ^2 r^{10}+36 r^6\bigg]\,,
\eea
which displays the following IR and UV asymptotic limits
\bea
R\big|_{r=Q} = 90 \gamma ^2 Q^4\,,\quad R\big|_{r\to \infty}\sim 16 + 14 \frac{Q^2}{r^2} +O\left(r^{-3}\right)\,.
\eea
Moreover, we show here also the asymptotic behavior of other geometric scalars, namely
\bea
&& R^{\m\n}R_{\m\n}\big|_{r=Q} = 36 \left(81 \gamma ^4 Q^8-14 \gamma ^2 Q^4+6\right)\,, \quad R^{\m\n}R_{\m\n}\big|_{r\to \infty}\sim 96 +O\left(r^{-3}\right)\,, \nb\\
&&R^{\m\n\r\s}R_{\m\n\r\s}\big|_{r=Q}=36 \left(99 \gamma ^4 Q^8+24 \gamma ^2 Q^4+14\right)\,,\quad R^{\m\n\r\s}R_{\m\n\r\s}\big|_{r\to \infty} \sim 80 -40\frac{Q^2}{r^2} +O\left(r^{-3}\right)\,.\nb\\
\eea
%
The invariants in Einstein frame have the same UV asymptotics as the first non-commutative solution, namenly
\begin{eqnarray}
& & R\vert_{r\to \infty}\sim -\frac{2}{\sqrt{\g}r} +O\left(r^{-3}\right)\,,\\
& & R_{\mu\nu}R^{\mu\nu}\vert_{r\to \infty}\sim \frac{116}{\g r^2} +O\left(r^{-3}\right)\,,\\
& & R_{\mu\nu\rho\sigma}R^{\mu\nu\rho\sigma}\vert_{r\to \infty}\sim \frac{97}{\g r^2} +O\left(r^{-3}\right)\,.\label{ivarinatseinsteinNC2}
\end{eqnarray}
Remarkably, all these quantities are regular and vanishing in the large $r$ limit.\\
In the $r\to \infty$ limit, the $G$-function in eq. (\ref{NCdipole2}) behaves as
\bea
G\big|_{r\to \infty} \sim \frac{1}{\g^2r^4}\,,
\eea
preventing the background in eq. (\ref{NCdipole}) from having an $AdS_5$ UV asymptote as the original $\widehat{AdS_5}$ solution.\\
Let us conclude by emphasizing  that we have verified (using Mathematica) the Einstein, Maxwell, dilaton equations of motion for type IIB background in eqs.(\ref{NCdipole})-(\ref{NCdipole2}) as well as Bianchi identities for the RR and $H_3$ fields.

\bibliographystyle{JHEP}
\bibliography{Ref.bib}

\end{document}